\documentclass[aps, prb, reprint, showpacs, floatfix, citeautoscript, longbibliography, superscriptaddress]{revtex4-1}

 \pdfoutput=1

\usepackage[utf8]{inputenc}

\usepackage{color}
\usepackage{xcolor}
\usepackage{subfigure}
\definecolor{myblue}{rgb}{.8, .8, 1}

\usepackage{amsmath}
\usepackage{amssymb}
\usepackage{gensymb}
\usepackage{empheq}

\usepackage{listings}

\newlength\mytemplen
\newsavebox\mytempbox

\makeatletter
\newcommand\mybluebox{%
    \@ifnextchar[
       {\@mybluebox}%
       {\@mybluebox[0pt]}}

\def\@mybluebox[#1]{%
    \@ifnextchar[
       {\@@mybluebox[#1]}%
       {\@@mybluebox[#1][0pt]}}

\def\@@mybluebox[#1][#2]#3{
    \sbox\mytempbox{#3}%
    \mytemplen\ht\mytempbox
    \advance\mytemplen #1\relax
    \ht\mytempbox\mytemplen
    \mytemplen\dp\mytempbox
    \advance\mytemplen #2\relax
    \dp\mytempbox\mytemplen
    \colorbox{myblue}{\hspace{1em}\usebox{\mytempbox}\hspace{1em}}}

\usepackage{natbib}
\usepackage{graphicx}
\usepackage{amsmath}

\lstdefinestyle{mystyle}{
    commentstyle=\color{codegreen},
    keywordstyle=\color{magenta},
    numberstyle=\tiny\color{gray},
    stringstyle=\color{codepurple},
    basicstyle=\footnotesize,
    breakatwhitespace=false,         
    breaklines=true,                 
    captionpos=b,                    
    keepspaces=true,                 
    numbers=left,                    
    numbersep=5pt,                  
    showspaces=false,                
    showstringspaces=false,
    showtabs=false,                  
    tabsize=2
}
 
\begin{document}

\title{Probing charge carrier movement in organic semiconductor thin films via nanowire conductance spectroscopy}

\author{M. V. Klymenko}
\email[Email: ]{mike.klymenko@rmit.edu.au}
\affiliation{ARC Centre of Excellence in Exciton Science, RMIT University, Melbourne, Victoria 3001, Australia }
\affiliation{Chemical and Quantum Physics, School of Science, RMIT University, Melbourne, Victoria 3001, Australia}
\author{J. A. Vaitkus}
\affiliation{Chemical and Quantum Physics, School of Science, RMIT University, Melbourne, Victoria 3001, Australia}
\author{J. H. Cole}
\affiliation{ARC Centre of Excellence in Exciton Science, RMIT University, Melbourne, Victoria 3001, Australia }
\affiliation{Chemical and Quantum Physics, School of Science, RMIT University, Melbourne, Victoria 3001, Australia}

\begin{abstract}

Understanding the movement of charge within organic semiconducting films 
is crucial for applications in photo-voltaics and flexible electronics.
We study the sensitivity of the electrical conductance of a silicon 
nanowire to changes of charge states within an organic semiconductor 
physisorbed on the surface of the nanowire. Elastic scattering caused by 
motion of charge carriers near the nanowire modifies the mean-free path 
for backscattering of electrons propagating within it, which we have mathematically expressed in terms of 
the causal Green's functions. The scattering potential has been computed 
using a combination  of  the  polarizable  continuum  model and density 
functional theory with the range-separated exchange-correlation 
functional for organic molecules and the semi-empirical tight-binding 
model for silicon. As an example, the sensitivity to charge state 
changes in tetracene is computed as a function of operating temperature 
and geometrical parameters of a nanowire. For a single molecule, 
ultra-thin silicon nanowires with characteristic sizes of the 
cross-section below 2 nm produce a detectable conductance change at room 
temperature. For larger nanowires the sensitivity is reduced, however 
the conductance change grows with the number of charged molecules: with 
sub-4 nm nanowires being sensitive enough to detect several tens of 
charge carriers. We propose using noise spectroscopy to access the 
temporal evolution of the charge states. Information regarding the 
spatial distribution of charge carries in organic thin films can be 
obtained using a grid of nanowire resistors and electric impedance 
tomography.

\end{abstract}

\maketitle

\section{Introduction}

A general property of the semiconductors that makes them so widely used in electronics and optoelectronics is a strong dependence of the concentration of free charge carriers on various parameters such as temperature, concentration of dopants, stress and strain and applied electro-magnetic fields. Organic semiconductors are no exception\cite{Moses}.
Understanding the distribution and transport of charge carriers under non-equilibrium conditions in organic semiconductor films is crucial for designing new organic field-effect transistors \cite{OFET} and hybrid organic-inorganic photo-voltaic devices \cite{MacQueen}. Several different types of charge and excitonic excitations can be generated in such films, via electrical or photoexcitation. Each type of excitation will undergo different dynamics within the film and will induce both spatial and temporal changes in the local electrostatic environment. A local non-invasive charge sensing probe of the free carrier density that is induced by a spatially localized photo-excitation is of particular interest since it may give information regarding the anisotropic diffusion, quantum efficiency, mobility and lifetimes of charge carriers, similarly to the spatially resolved optical pump-probe experiments \cite{Ruzicka, Kumar}. Measuring a spatially resolved net stationary charge distribution in an organic thin films reveals details of the crystal morphology near the interface as well as the distribution of trap states and impurities. 

\begin{figure}[t]
    \centering
    \includegraphics[width=0.9\linewidth]{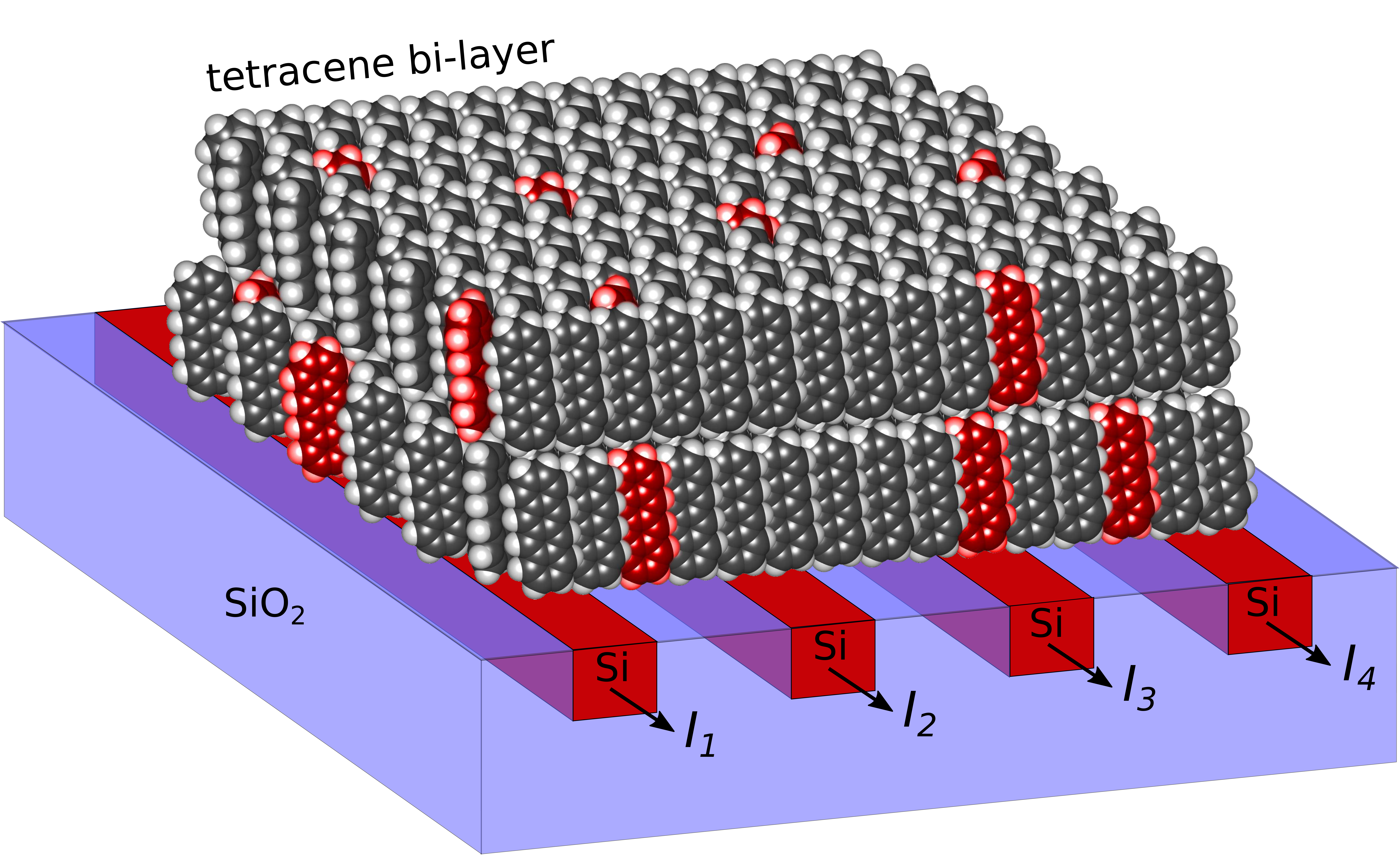}
    \caption{Linear array of silicon nanowires for probing charge states of organic molecules physisorbed on their surfaces. The charge carriers within the organic semiconductor (denoted by red shaded molecules) modify the conductance of the nanowires. The movement and position of the charges can be probed by fluctuations of currents $I_{j = {1...4}}$ in the linear transport regime.}
    \label{fig:intro}
\end{figure}

In this work we consider an array of silicon nanowires (NWs) as a measurement setup to access spatial and temporal information on the free charge carriers distribution in organic semiconductor thin films, illustrated in Fig.~\ref{fig:intro}. The goal is to estimate the sensitivity of the electron transport in silicon NWs to changes of the charge states of molecules in organic semiconductors. Previously, NWs have established themselves as highly sensitive chemical sensors \cite{Cui1289}. For instance, they are able to detect extremely small charge transferred from the ammonia molecules physisorbed at their surfaces in both gas and liquid environments \cite{ammonia1,ammonia2}. The possibility of non-invasive charge sensing by electrical current placed in close proximity to a confined charge has been also proven with quantum-point contact charge sensors \cite{Elzerman2004, QPC}.

The problem of the conductance change caused by motion of elastic scatterers, fluctuation of chemical potential or magnetic fields as well as their implications for $1/f$-noise had been intensively studied in the 80's and early 90's for metallic wires both numerically and analytically \cite{PhysRevLett.56.1960, PhysRevB.37.8557, PhysRevB.39.9929, Washburn_1992}. The theoretical models developed during that time concerned mostly the interference and localization effects in systems with relatively simple electronic structure (a single parabolic energy band) and lack detail on the scattering potential required for a quantitative treatment \cite{PhysRevB.37.8557}. Nevertheless, those works have produced the upper estimates for the conductance change and useful relationships between conductance sensitivity and dimensionality of the systems which are consistent with the results obtained in this work. In this paper we address the limit when the mean distance between scattering centers is large, and the scattering potential is weak, so the interference between scattering events may be neglected. At the same time the system itself is characterized via a realistic band structure with a scattering potential known in detail.

In this work, the conductance change caused by scattering from the potential induced by physisorbed charged molecules have been computed using the relationship between the conductance and mean-free path \cite{Lundstrom} which may be used in conjunction with the Matthiessen's rule \cite{Jauho} for various types of scatterer. We have expressed the mean free path in terms of the causal Green's functions using the Caroli formula for the transmission probability. The Green's function are computed using the semi-empirical tight-binding Hamiltonian for silicon NW with the reduced mode space transformation \cite{PhysRevB.85.035317, Huang_Klimeck}. In order to obtain the scattering potential generated by a physisorbed molecule  with an excess charge in a crystalline environment, we use a recently developed technique representing a combination of the polarizable continuum model and density functional theory with the range-separated exchange-correlation functional $\omega$B97XD with an adjustable parameter $\omega$ \cite{pcm2}.

We apply our model to a range of NWs with widths from 2 to 6 lattice constant (1.38 - 3.49 nm including hydrogen passivation layers). Sub 2-nm NWs has been fabricated by several authors using the self-limiting oxidation \cite{doi:10.1063/1.111914, 7998896}, the synthesis of vertically assembled
nanocomposites \cite{PhysRevMaterials.2.106003}, epitaxial Au-catalyzed vertical growth \cite{Zhou_2015} and scanning tunneling microscope directed self-assembly of dopants in silicon \cite{Weber64}.  Based on the obtained results, we propose a concept for the 2D spatially and temporary resolved charge sensing via noise spectroscopy and electric impedance tomography using a 2D grid of silicon NWs. 

The paper is organized as follows. In Sec. II we develop a computational model to simulate the effect of charge states of physisorbed organic molecules on the electron transport in nanowires. In Sec III, we compute the electrostatic fields of the neutral tetracene molecules as well as tetracene anions and cations in crystalline environment taking into account polarization effects. In Sec. IV we analyze computed conductance change caused by charge states of organic molecules physisorbed at the silicon surface. In Sec. V we show that the temporal information about ultra-fast electron transport processes can be extracted from the noise spectroscopy while 2D spatial information can be obtained using the electrical impedance tomography with a grid of NW resistors.

\section{Single nanowire as a charge sensor}

In this work we consider the problem of the surface roughness scattering for electrons propagating in NW. The surface roughness is caused by the electrostatic field of organic molecules with excess charge physisorbed on the surface of the NW.

We restrict our treatment to the case when organic semiconductor is in contact with one side of a rectangular silicon nanowire as is shown in Fig. \ref{fig:intro}. Moreover, the organic semiconductor is physisorbed at the hydrogen-passivated (100) silicon surface to ensure minimal deformations caused by the silicon crystal lattice. Although our treatment is applicable for any organic semiconductor held at the surface of NWs by van der Waals forces, here we consider the thin film of tetracene crystal as an example. 

\begin{figure}[t]
    \centering
    \includegraphics[width=0.8\linewidth]{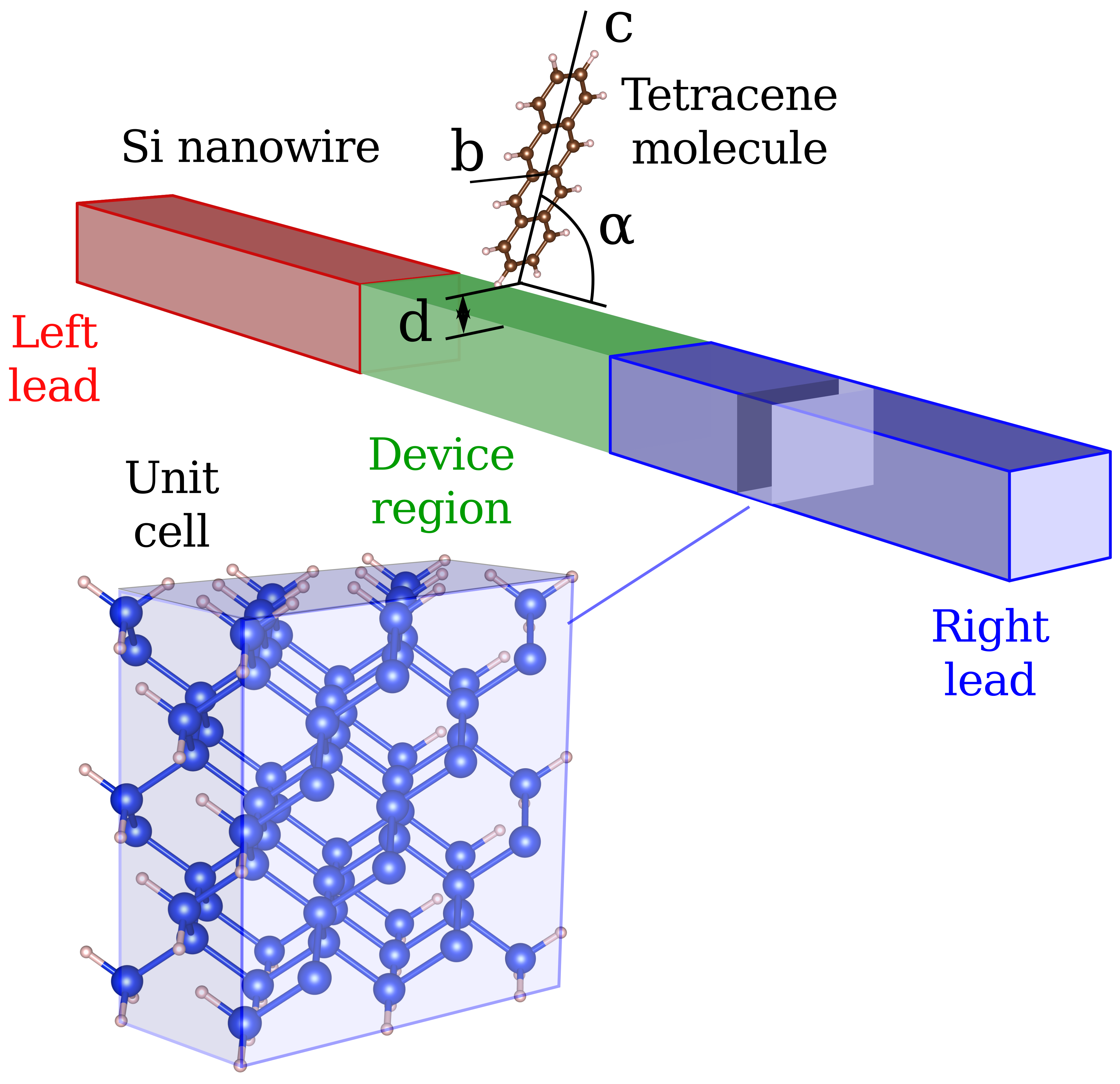}
    \caption{A simplified model of the system for the conductance change computations. The model consists of a single hydrogen passivated rectangular silicon NW with a single physisorbed tetracene molecule. The molecule is surrounded by other tetracene molecules in the crystal lattice (not shown in the figure). The effect of the tetracene crystal lattice is taken into account through the effective polarizable continuum media model (see discussion in Sec. III C).}
    \label{fig:intro1}
\end{figure}

The conductance sensitivity to a single scattering event is estimated using a model system represented by a single silicon NW and single molecule possessing different charge states (see Fig. \ref{fig:intro1}). If the interference between electrons scattered on neighboring charged molecules is negligibly small, the information on single molecule scattering can be easily generalized to the case of multiple scatterers as is shown in the following section.

\subsection{Finite temperature linear transport regime model for dilute elastic scattering centers}

For the linear transport regime close to the equilibrium the conductance of a NW can be computed using the Landauer formula for a two-terminal device (see Fig. \ref{fig:intro1}) modified to take into account interactions in the device region \cite{Landauer}:

\begin{equation}
    G = -\frac{2q^2}{\hbar} \int d\varepsilon  \left. \frac{d f(\varepsilon)}{d\varepsilon}  \right|_{\varepsilon_F} \text{tr} \left( \mathbf{G}^a \mathbf{\Gamma}^R \mathbf{G}^r \mathbf{\Gamma}^L \mathbf{\Sigma}_0^{-1} \mathbf{\Sigma}  \right)
    \label{eq:landauer}
\end{equation}
where: $\mathbf{G}^r$ and $\mathbf{G}^a$ are retarded and advanced Green's functions expressed in the matrix form, $\mathbf{\Gamma}^{L,R}=i\left(\mathbf{\Sigma}_{L,R}^r - \mathbf{\Sigma}_{L,R}^a \right)$, $\mathbf{\Sigma}_{L,R}^r$ and $\mathbf{\Sigma}_{L,R}^a$ are the retarded and advanced self-energies describing coupling to semi-infinite left and right leads correspondingly, $\mathbf{\Sigma}_0=-i\left(\mathbf{\Gamma}^L + \mathbf{\Gamma}^R \right)$, and $\mathbf{\Sigma}$ is a total self-energy describing all elastic and inelastic scattering in the system. In the case when only elastic scattering is present, $\mathbf{\Sigma} = \mathbf{\Sigma}_0$ (see Ref. \onlinecite{Landauer} for more details) and we obtain the Caroli formula for the transmission $T\left( \varepsilon \right) = \text{tr} \left( \mathbf{G}^a \mathbf{\Gamma}^R \mathbf{G}^r \mathbf{\Gamma}^L  \right)$ in Eq. (\ref{eq:landauer}).

Alternatively, scattering in the device region can be phenomenologically expressed as \cite{Lundstrom}:

\begin{equation}
    G = -\frac{2q^2}{\hbar} \int d\varepsilon \left. \frac{d f(\varepsilon)}{d\varepsilon}   \right|_{\varepsilon_F} T(\varepsilon) M(\varepsilon)
    \label{eq:lundstrom}
\end{equation}
where $M(\varepsilon) = \text{tr} \left( \mathbf{G}_0^a \mathbf{\Gamma}^R \mathbf{G}_0^r \mathbf{\Gamma}^L \right)$ is the number of modes defined in terms of the Green's function of the infinite uniform NW, $\mathbf{G}_0^r$, and $T(\varepsilon)$ is the transmission probability given by the formula\cite{Lundstrom}: 

\begin{equation}
T(\varepsilon) = \frac{\lambda(\varepsilon)}{\lambda(\varepsilon) + L}
\label{eq:trans}
\end{equation}
where $\lambda(\varepsilon)$ is the energy-dependent mean free path for backscattering and $L$ is the length of the device region. Comparing Eq. (\ref{eq:landauer}) and Eq. (\ref{eq:lundstrom}), we can establish the relationship between the mean free path and the causal Green's functions:

\begin{equation}
 \frac{\lambda(\varepsilon)}{\lambda(\varepsilon) + L} = \frac{\text{tr} \left( \mathbf{G}^a \mathbf{\Gamma}^R \mathbf{G}^r \mathbf{\Gamma}^L \mathbf{\Sigma}_0^{-1} \mathbf{\Sigma}  \right)}{\text{tr} \left( \mathbf{G}_0^a \mathbf{\Gamma}^R \mathbf{G}_0^r \mathbf{\Gamma}^L \right)} 
  \label{eq:trans1}
\end{equation}

If the system is represented by a few sets of indistinguishable scatterers and scattering events are independent, it is reasonable to derive the mean free path for a scattering center of a particular type  $j$ and then sum up contributions from each scattering center. In this case, the length of device region $L$ in Eq. (\ref{eq:trans1}) has to be chosen such that the device region contains only a single scattering center. Due to the random distribution of the scattering centers, this length varies along the nanowire. Averaging over all device regions of different lengths gives the expression for the mean-free path for backscattering: 

\begin{equation}
\lambda_j(\varepsilon) = \langle l \rangle_j \left[\frac{\text{tr} \left( \mathbf{G}_0^a \mathbf{\Gamma}^R \mathbf{G}_0^r \mathbf{\Gamma}^L \right)}{\text{tr} \left( \mathbf{G}_j^a \mathbf{\Gamma}^R \mathbf{G}_j^r \mathbf{\Gamma}^L \right)} - 1 \right]^{-1}
\label{eq:mfp}
\end{equation}
where $\langle l \rangle_{j=+, -}$ is the average distance between positively (index $+$) or negatively (index $-$) charged scattering centers. 

According to Eq. (\ref{eq:trans}), the resistance depends linearly on the length of the device $L$. This implies that we work in the Ohmic regime neglecting localization effects \cite{Todorov}. For our system this is a good approximation since the volume of the device affected by a single molecule has a characteristic size less than the total mean free path, and, at the same time, the average distance between molecules with excess charge is much larger than this number. This is usually the case for organic films as they contain relatively low free carriers concentrations at all temperatures. In this case, the electron scattering by phonons and by each charged molecule are independent events, and the quantum interference between electrons scattered at different molecules is negligibly small. As a result of those approximations, the mean free path can be computed independently for each kind of scattering and then summed up using Matthiessen's rule \cite{Jauho}:   

\begin{equation}
    \lambda^{-1}(\varepsilon) = \lambda_+^{-1}(\varepsilon) + \lambda_-^{-1}(\varepsilon) + \lambda_{\rm ph}^{-1}(\varepsilon)
    \label{eq:Matthiessen}
\end{equation}

The mean free path determined by the phonon scattering, $\lambda_{\rm ph}(\varepsilon)$, exceeds 500 nm in ultra-thin nanowires \cite{Jauho, Lu10046}, so the ballistic transport can be observed even at room temperature. Therefore, we neglect the electron-phonon contribution in this work. Each term in Eq. (\ref{eq:Matthiessen}) is computed independently using a formula derived from Eq. (\ref{eq:trans}). For instance in the case of elastic scattering, the mean free path reads:

\begin{equation}
\begin{split}
\lambda_{\rm{elastic}}^{-1}(\varepsilon) =  & \lambda_+^{-1}(\varepsilon) + \lambda_-^{-1}(\varepsilon) =  \\ & \sum_{j=+, -} \langle l \rangle_j^{-1} \left[\frac{\text{tr} \left( \mathbf{G}_0^a \mathbf{\Gamma}^R \mathbf{G}_0^r \mathbf{\Gamma}^L \right)}{\text{tr} \left( \mathbf{G}_j^a \mathbf{\Gamma}^R \mathbf{G}_j^r \mathbf{\Gamma}^L \right)} - 1 \right]
\end{split}
\label{eq:mfp1}
\end{equation}

In the case when a single type of elastic scattering dominates, the transmission probability in Eq. (\ref{eq:trans}) reads:

\begin{equation}
    T(\varepsilon) = \left\{1 + nL \left[\frac{\text{tr} \left( \mathbf{G}_0^a \mathbf{\Gamma}^R \mathbf{G}_0^r \mathbf{\Gamma}^L \right)}{\text{tr} \left( \mathbf{G}_{+, -}^a \mathbf{\Gamma}^R \mathbf{G}_{+, -}^r \mathbf{\Gamma}^L \right)} - 1 \right] \right\}^{-1}
    \label{eq:trans_final}
\end{equation}
where $n = \langle l \rangle_{+, -}^{-1}$ is the linear density of either positively or negatively charged scattering centers.

At zero temperature, the conductance change $\Delta G = |G_0-G_{+,-}|$ is proportional to the transmission change $\Delta T = 1 - T(\varepsilon)$. When $nL \ll 1$, according to Eq. \ref{eq:trans_final}, $\Delta T \propto nL$. In the limit when $nL \rightarrow \infty$, $\Delta T \rightarrow 1$. The observed dependence of the conductance change on the concentration of scattering centers and length of the device region are consistent with the results published in Ref. \onlinecite{PhysRevLett.56.1960}.

\subsection{Green's functions and band structure of the silicon nanowire}

A part of the NW where scattering due to the electrostatic potential of charged molecules occurs represents a device region (see Fig. \ref{fig:intro1}). The retarded Green's function for the system reads:

\begin{equation}
    \mathbf{G}_{+, -}^r = \left[ (\varepsilon + i\eta)\mathbf{I} - \mathbf{H}_0 - \mathbf{V}_{+, -} - \mathbf{\Sigma}_L^r - \mathbf{\Sigma}_R^r \right]^{-1}
    \label{eq:gf}
\end{equation}
where $\mathbf{H}_0$ is the tight-binding Hamiltonian of the device region without scatters, and $\mathbf{V}_{+, -}$ is the contribution from the electrostatic field of the molecular cation or anion respectively. A scalar potential $\phi(\mathbf{r})$ modifies only diagonal elements of the tight-binding matrix \cite{Graf} representing a contribution to the on-site energies that equals $-e\phi(\mathbf{r})$. Therefore, matrices $\mathbf{V}_{+, -}$ are diagonal matrices. The elements of those matrices are obtained by interpolating the computed electrostatic field of the molecules at the atomic sites of the nanowire for which the tight-binding model has been built. The electrostatic fields are computed for the configuration of the tetracene molecule that is discussed in the next section. The Green's function $\mathbf{G}_0^r$ in Eq. (\ref{eq:trans1}) and (\ref{eq:mfp}) are defined by Eq. (\ref{eq:gf}) for which $\mathbf{V}_{+, -} = 0$.

The Hamiltonian matrix $\mathbf{H}$ of silicon NW has been computed using the semi-empirical tight-binding sp$^3$d$^5$s$^*$ model \cite{PhysRevB.57.6493, PhysRevB.74.205323, PhysRevB.69.233101} neglecting spin-obit coupling. This method has been implemented in a custom open-source software package \cite{nanonet}. Since we probe molecular states via linear electron transport (not hole transport), we are interested in a relatively narrow energy interval around the conduction band edge. 

In order to reduce the computational burden, instead of direct numerical inversion of the matrix in Eq. (\ref{eq:gf}) we use the recursive algorithm \cite{gf_recursive} to compute the causal Green's functions. Also, we approximate the Hamiltonian matrices $\mathbf{H}_0$ and $\mathbf{V}_{+, -}$ that enters  Eq. (\ref{eq:gf}) by matrices of smaller dimensions having the same eigenvalues in a certain energy range. Finding those matrices is essentially building an equivalent reduced mode space for the original problem \cite{PhysRevB.85.035317, Huang_Klimeck}. The reduced matrices are obtained from the original ones by the following transformation:

\begin{equation}
    \mathbf{h} = \mathbf{\Phi}^{T}\mathbf{H}\mathbf{\Phi}
\end{equation}

The transformation matrices $\mathbf{\Phi}$ are build from a small set of so-called representative Bloch states being eigenvectors of the original matrix $\mathbf{H}$ taken at specific wave vectors and energies. The tight-binding matrix can be diagonalized in two ways: either one sets a wave vector as a parameter and finds energies as eigenvalues or one sets the energy as a parameter and finds eigenvectors as eigenvalues. In this work we implement the second method since it is more convenient for computing Green's functions that depend on energy as a parameter. The computation procedure for both the Hamiltonian $\mathbf{H}$ and self-energies $\mathbf{\Sigma}_L^r$, $\mathbf{\Sigma}_R^r$ is described in Ref. \onlinecite{Wimmer} and in Appendix A in more detail. 

The representative eigenvectors are collected in a matrix $\Tilde{\mathbf{\Phi}}$. In the general case, the eigenvectors are not orthogonal and form the overlap matrix which can be diagonalized by solving the eigenproblem $\Tilde{\mathbf{\Phi}}^T \Tilde{\mathbf{\Phi}} =  \mathbf{c} \mathbf{\Lambda} \mathbf{c}^T$ in order to obtain the matrix  $\mathbf{\Phi} = \Tilde{\mathbf{\Phi}} \mathbf{c} \mathbf{\Lambda} ^ {-1/2}$. In a relatively narrow energy window, matrix $\mathbf{h}$ gives a good approximation of the original matrix $\mathbf{H}$ and does not require any additional operations. However, when the energy window is extended, spurious solutions may appear in the spectrum along with the approximated energy bands. In this case one has to apply an algorithm eliminating spurious solutions; such an algorithm has been proposed by several authors in different modifications. In this work, we have found that for the range of energies up to 0.4 eV above the conduction band edge the spurious solutions do not appear.

\begin{figure}[t]
    \centering
    \subfigure[]{\includegraphics[height=4.7 cm]{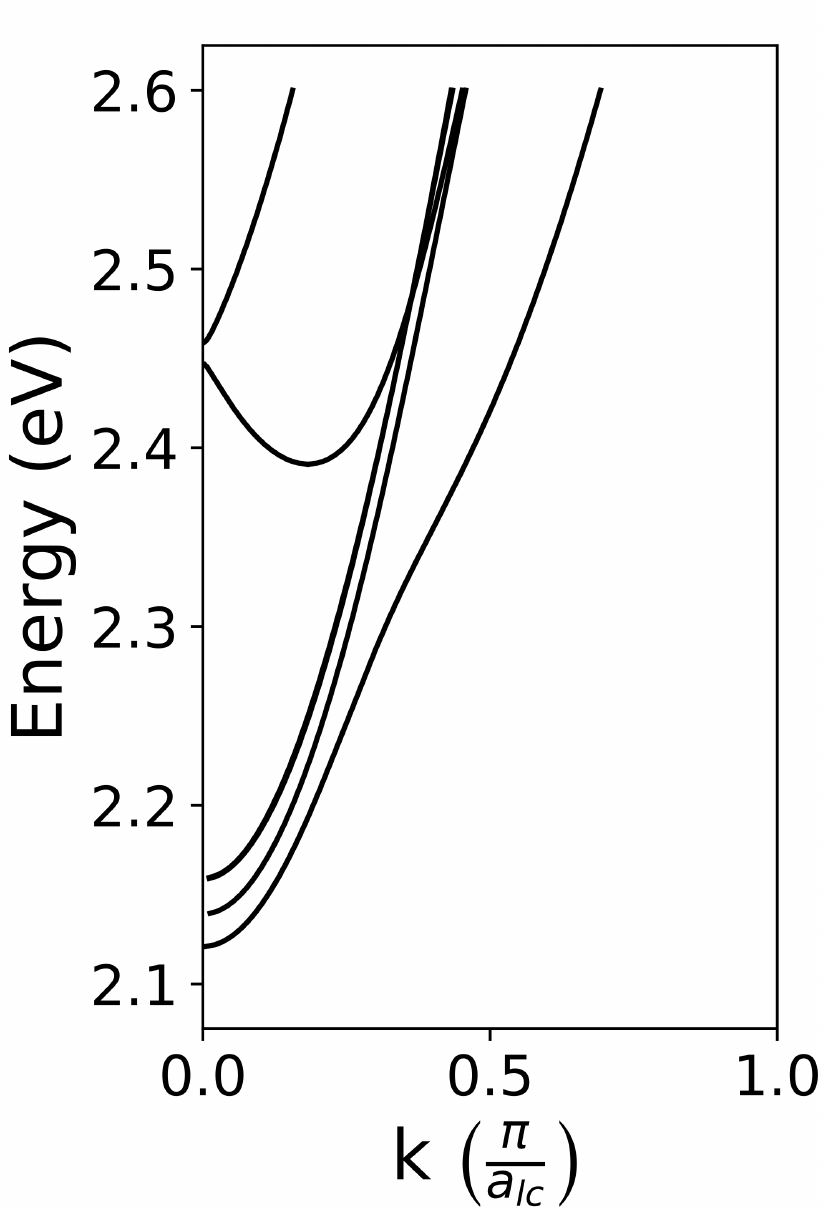}} 
    \subfigure[]{\includegraphics[height=4.7 cm]{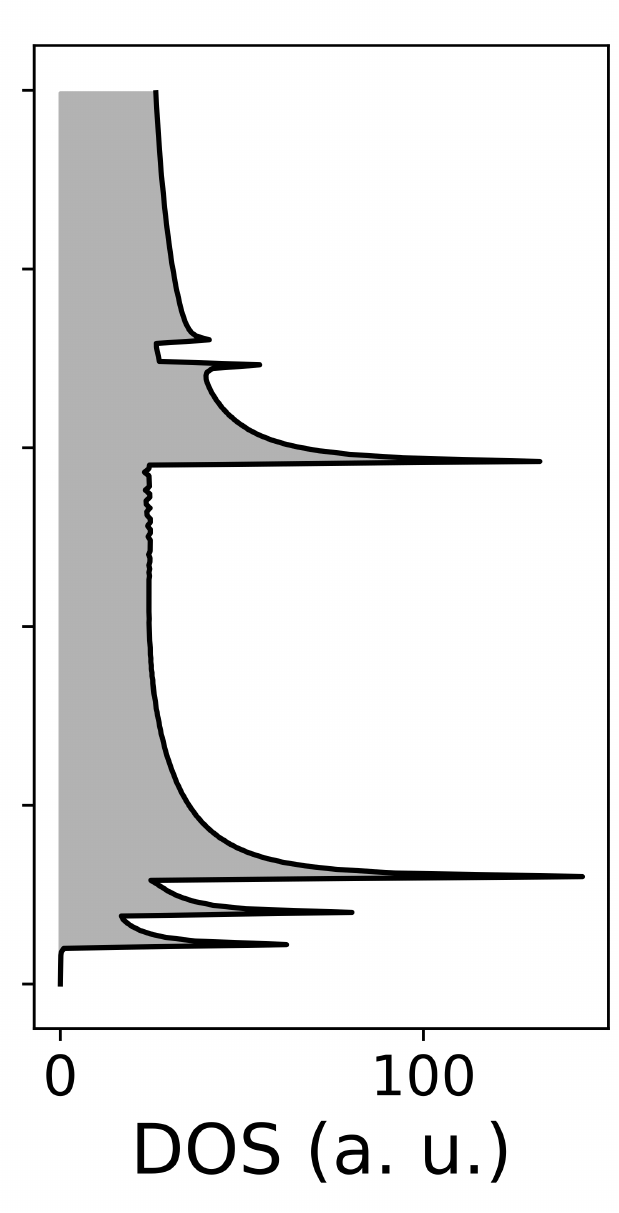}}
    \subfigure[]{\includegraphics[height=4.7 cm]{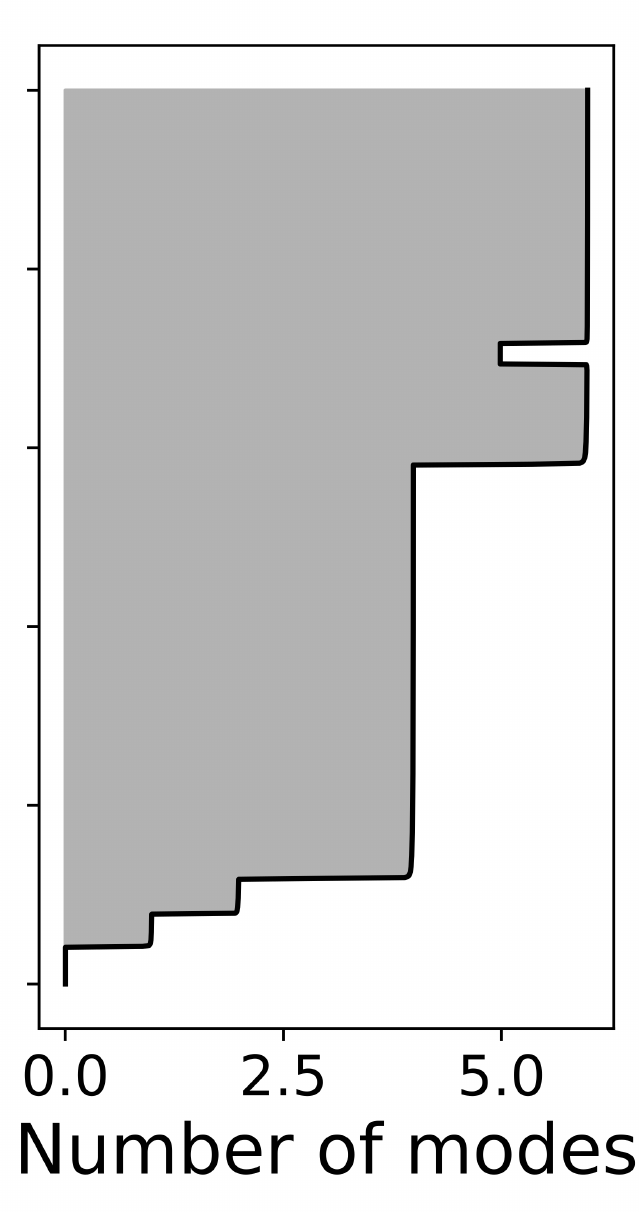}}
    \caption{a) The conduction band structure, b) density of states and c) number of modes of the pristine silicon NW with the width of 1.38 nm (2 lattice constants).}
    \label{fig:sinw}
\end{figure}

In Fig. \ref{fig:sinw} we show results of computations of the Green's functions $\mathbf{G}_0^r$ using Eq. (\ref{eq:gf}). The band structure of the pristine nanowire is given by the position of poles of $\mathbf{G}_0^r$ in the energy-momentum space; the density of states equals $\rm DOS = -2 \text{Im} \left( \mathbf{G}_0^r \right)$ and the number of modes is given by the Caroli formula discussed above.


\section{Electrostatic fields of molecules with excess charges in a crystalline environment}

\subsection{Orientation of molecules at the silicon substrate}

For solving a scattering problem it is crucial to determine the scattering potential that is, in our case, the electrostatic fields of a tetracene molecule physisorbed at the NW surface.

\begin{figure}[t]
    \centering
    \subfigure[]{\includegraphics[width=0.47\linewidth]{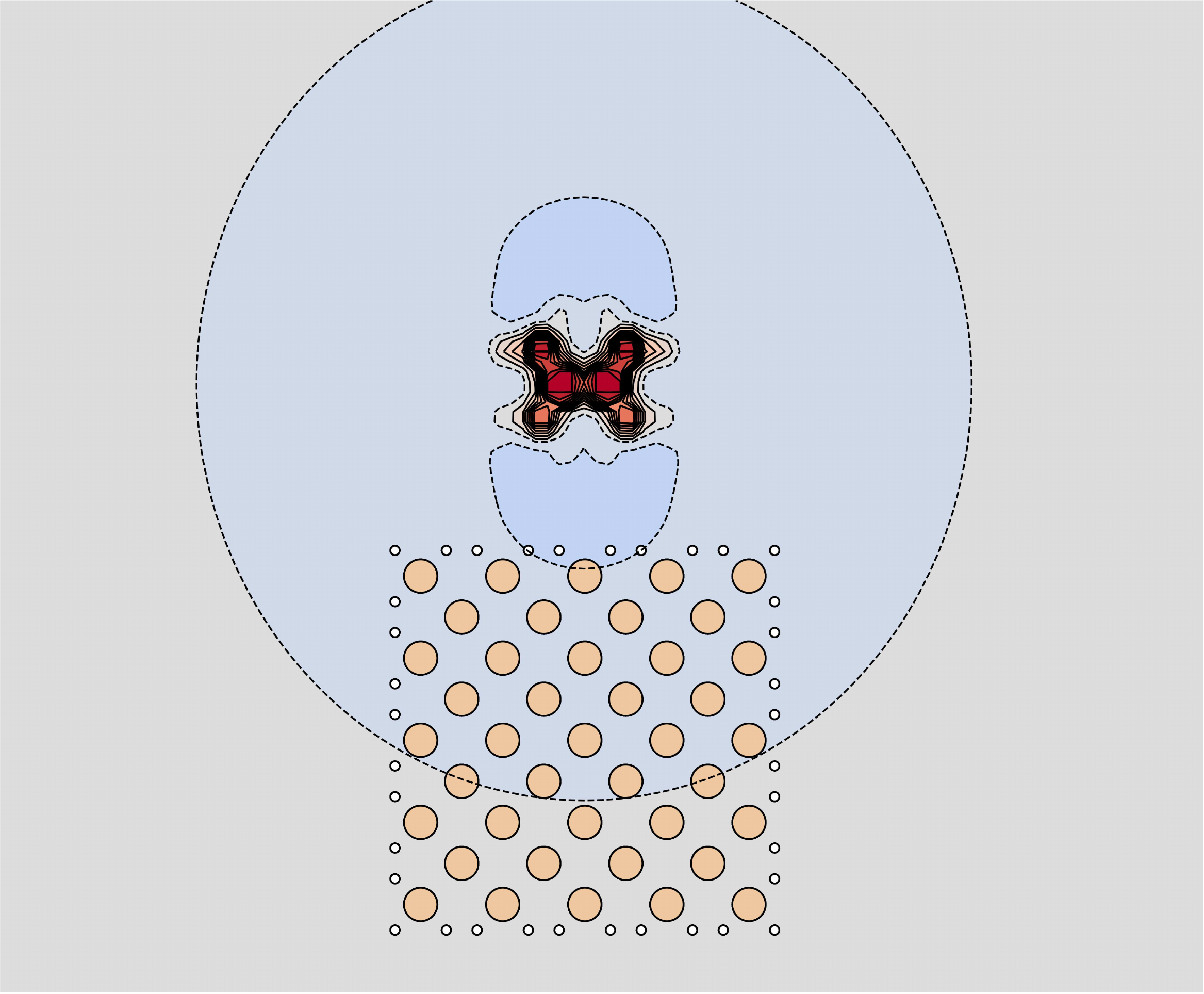}} \hspace{6 pt}
    \subfigure[]{\includegraphics[width=0.47\linewidth]{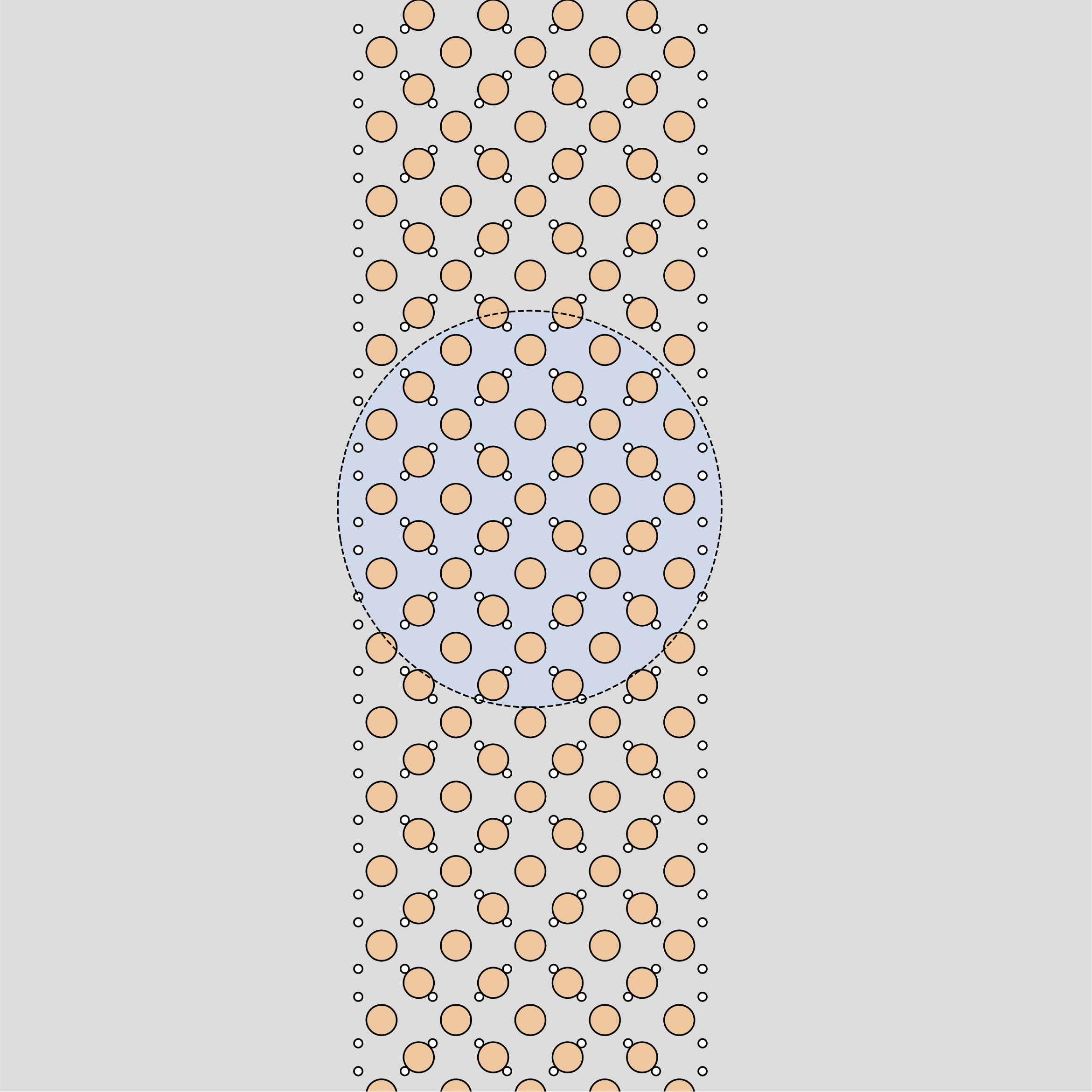}}
    \subfigure[]{\includegraphics[width=0.47\linewidth]{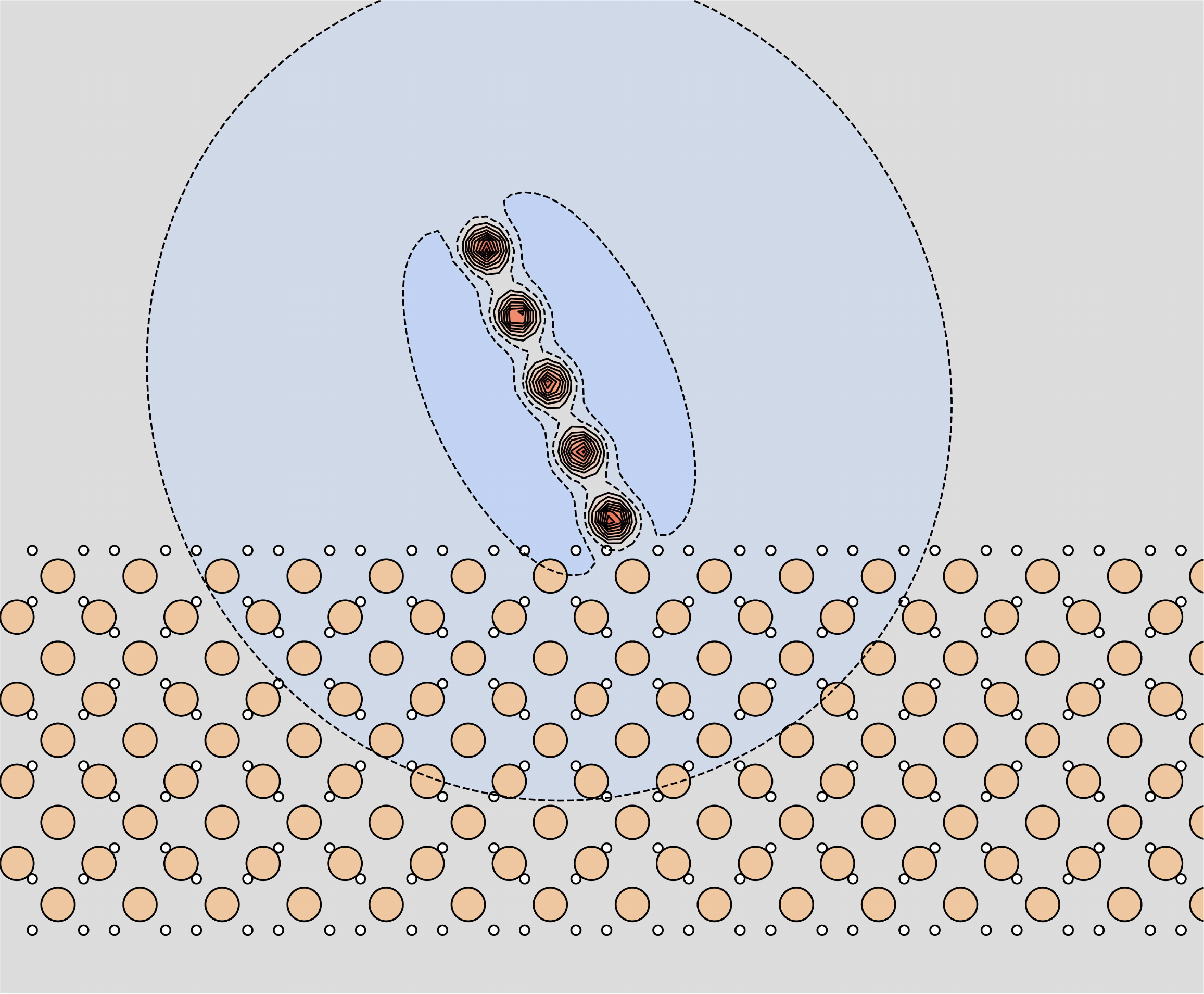}}
    \subfigure[]{\includegraphics[width=0.51\linewidth]{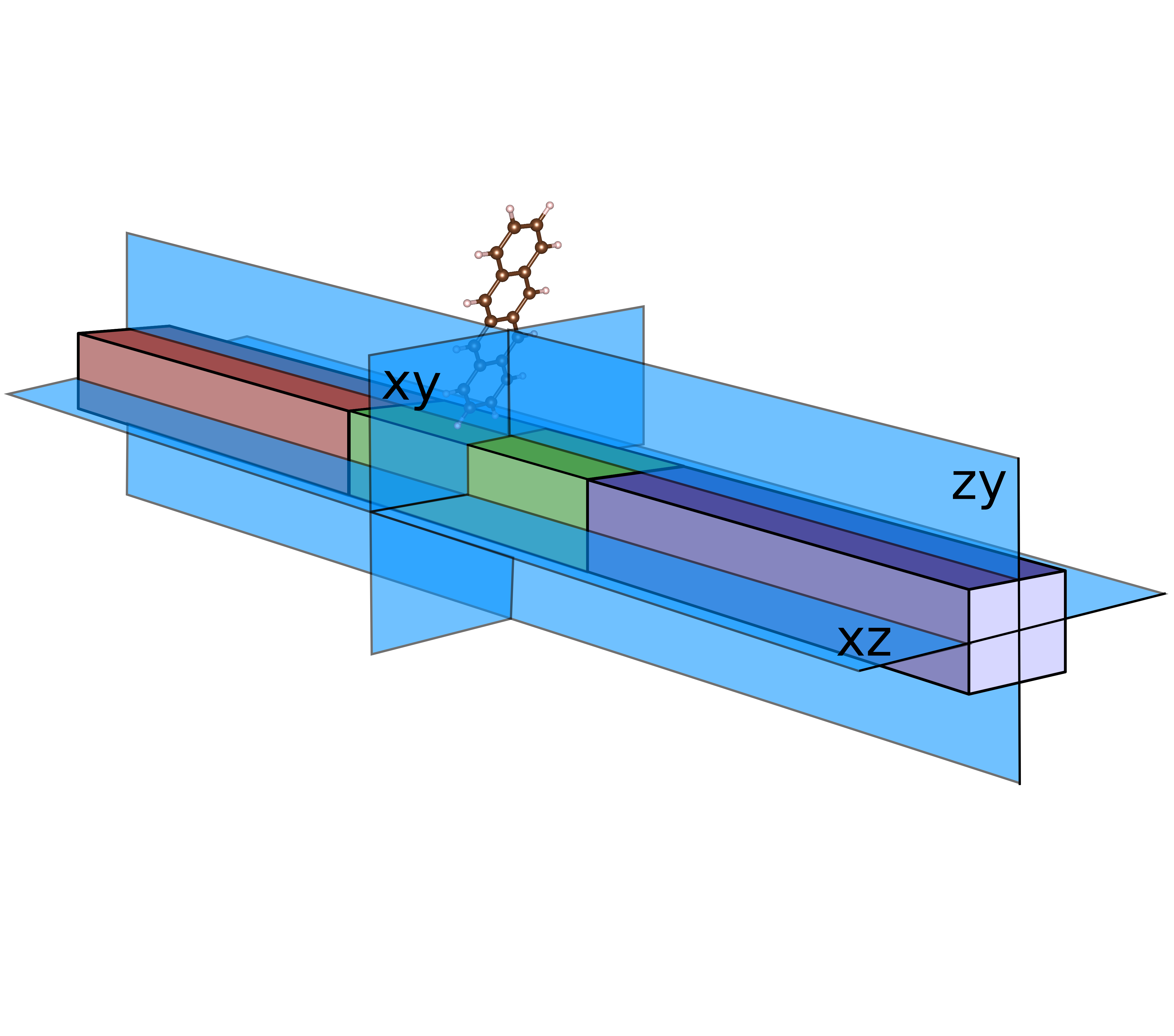}}
    \caption{Contribution to on-site energies of the NW tight-binding model from the electrostatic field of the tetracene radical anion physisorbed at the (100) surface. The plots show the magnitude of the electrostatic filed across the NW in a) xy plane, b) zy plane and c) xz plane. The orientations of those planes relative to the NW is illustrated in the panel d).}
    \label{fig:projections}
\end{figure}

For metallic substrates, the most energetically favorable geometrical configuration of the interface is such that the the molecular axis $c$ of tetracene (see Fig. \ref{fig:intro1}) is parallel to the interface, while the axis $b$ forms some angle with the surface due to the herringbone structure of the organic crystal. For insulators and semiconductors with passivated surfaces, the angle $\alpha$ between the axis $c$ and the surface is rather large and, unlike for metal substrates, the herringbone structure is formed in the direction parallel to the interface. Detailed information on the crystalline tetracene orientation relative to the silicon (100) surface is reported in Ref. \onlinecite{structure} where the authors have measured the averaged angle $\alpha=65 \degree \pm 3 \degree$ for the ultra-thin tetracene films using the near-edge X-ray absorption
fine structure. We use that average value in our calculations (see Fig. \ref{fig:projections}). To the best of our knowledge, the spacing $d$ between physisorbed tetracene molecules and silicon surface is not known so we perform simulations for a set of values of the parameter $d$. 

\subsection{Localized vs delocalized charge}

The electron transport in organic semiconductors is characterized by an interplay between coherent transport described by band theory and incoherent polaronic transport \cite{PhysRevB.79.235206}. At low temperatures, the coherent transport may dominate in the crystalline organic semiconductors resulting in delocalization of charge carriers over large length scales. As the temperature increases the electron-phonon scattering become more intense leading to so-called polaron bandwidth narrowing \cite{PhysRevB.69.075211}. At room temperature the incoherent transport dominates and the electron transport becomes hopping between first nearest neighbours, one at a time, with a charge localized mostly on a single molecule. In the case of organic semiconductor films with an amorphous structure, typically the charge carriers are localized within a single molecule even at low temperatures.\cite{PhysRevB.75.113203}.

\subsection{Effect of the polarizable environment on molecular charge states}

Comparing to the gas-phase, the ionization energy of an organic molecule in the crystalline phase is decreased by some value $P_h$ \cite{book:pope}. If we neglect a small hybridization of molecular orbitals \cite{genome} of adjacent molecules in the crystal, this energy corresponds to the polarization energy resulting from intermolecular interactions via the electrostatic fields and long-range dispersion forces. However even in this case, computing the electronic structure of molecular crystals is a challenging problem, especially for the case of the excess charge states, due to long range interaction between molecules. Such a multi-scale problem is often treated using a hybrid QM/MM/Continuum or QM/QM'/Continuum computational techniques \cite{pcm}. The simplest flavor of those hybrid methods exploited in this work is the QM/Continuum approach with the polarizable continuum model (PCM) well known from computational chemistry of solutions. 

In order to compute electrostatic fields of the tetracene molecules in their crystalline environment we employ a method proposed in Refs. \onlinecite{pcm1, pcm2}. The method treats the effect of environment on the electronic structure of a single molecule using PCM and tuned range-separated hybrid exchange-correlation potentials. The model has two parameters: the static dielectric constant, $\epsilon$, and the range-separation parameter, $\omega$.  For the example of the crystalline pentacene, it has been shown in Ref. \onlinecite{pcm2} that a proper choice of those parameters can reproduce energies of frontier orbitals with a qualitative accuracy. For the case of tetracene, we take the static dielectric constant of 3.7 from the literature\cite{tet_eps}. The range-separation parameter omega $\omega$ have been optimized to reproduce the ionization potential and electron affinity for bulk crystalline tetracene, following the procedure described in Ref. \onlinecite{pcm2} results in the optimal value of $\omega=0.032 \text{ Bohr}^{-1}$. 

\subsection{Electrostatic fields of tetracene anions and cations}

With the approach described above, the charge density and electrostatic fields of the molecular anion and cation as well as the neutral tetracene have been computed using Gaussian09 \cite{g09} with the density-functional method, using unrestricted-shell orbitals and basis set 6-31++g(d,p) including PCM and range-separated exchange-correlation functional $\omega$B97XD \cite{wB97XD}. Our results can be reproduced using input files for Gaussian09 listed in the Supplementary Information. 

The method described above gives the charge density and electrostatic potential neglecting an inhomogeneity of the environment, ie.\ assuming that tetracene is surrounded by other tetracene molecules only. The environment of a molecule physisorbed at the silicon surface includes not only crystalline tetracene, but also silicon NWs and SiO$_2$ substrate as is illustrated in Fig.~\ref{fig:intro}. A rough estimate of the static dielectric constant for the silicon NW can be computed using a simple analytic expression based on the Penn's model \cite{Penn} adapted to quantum-confined systems \cite{Zunger, Tsu}: $\epsilon = 1 + \left(\epsilon_B-1 \right)/\left( 1+\Delta E^2 / E_g^2 \right)$, where $\epsilon_B=11.3$ is the static dielectric constant for the bulk silicon, $E_g=4$ eV is the direct band-gap for the the single-oscillator model,  $\Delta E=\pi \hbar \sqrt{E_F} / \sqrt{2m}a$. All parameters are taken from Ref. \onlinecite{Tsu}. As a result, we have an effective dielectric constant in the range of 8.11-10.62 for the NWs we consider here. 

While the dielectric constant of 3.9 for SiO$_2$ \cite{handbook} is close to the corresponding value of tetracene, the dielectric mismatch between a silicon NW and tetracene is rather large. To obtain corrections to the electrostatic potential caused by the dielectric mismatch, we solve the Poisson equation with the tetracene charge density, $\rho_0$, obtained with neglecting the dielectric mismatch, and with a spatially dependent dielectric permittivity, $\epsilon(\mathbf{r})$: $ \nabla \left[ \epsilon(\mathbf{r}) \nabla  \phi(\mathbf{r}) \right] = - 4 \pi \rho_0$. The Poisson equation has been solved numerically using the induced charge computation method \cite{doi:10.1021/acsami.5b01606} with a self-consistent iteration loop. This method and the role of dielectric mismatch is discussed in detail in Appendix B.  The results of the electrostatic field computations are shown in Fig. \ref{fig:projections} for the case of the tetracene molecular anion.

\begin{figure}[t]
    \centering
    \includegraphics[width=\linewidth]{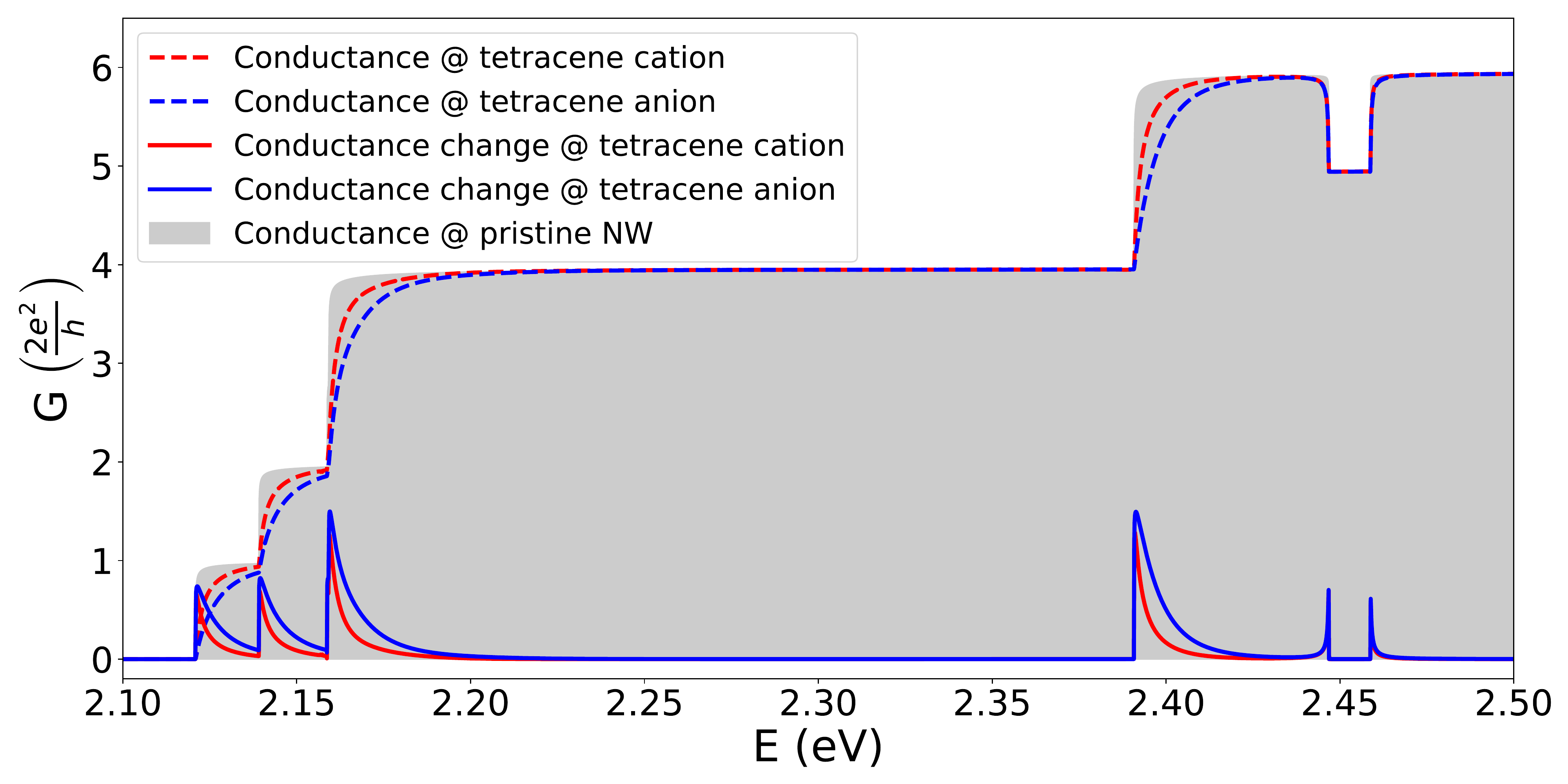}
    \caption{The conductance of a silicon NW having rectangular cross-section with a side length of 1.38 nm.}
    \label{fig:c11}
\end{figure}

\section{Conductance analysis}

\subsection{Conductance spectroscopy: static analysis}

We start our analysis with considering scattering at the single molecule limit for which $nL=1$ in Eq. (\ref{eq:trans_final}). The dependence of the conductance on the Fermi energy, shown in Fig. \ref{fig:c11}, demonstrates that the elastic scattering by the electrostatic fields of tetracene molecules with excess charge suppresses the ballistic electron transport around the subband edges. The electron wave packets with smaller group velocity have larger probability to reflect at the molecular potential. The only detectable difference between positively and negatively charged molecules is the overall suppression of the conductance change in the case of radical cation comparing to the radical anion. For both charge signs, the shapes of the conductance change energy spectra are similar.

\begin{figure}[t]
    \centering
    \subfigure[]{\includegraphics[width=\linewidth]{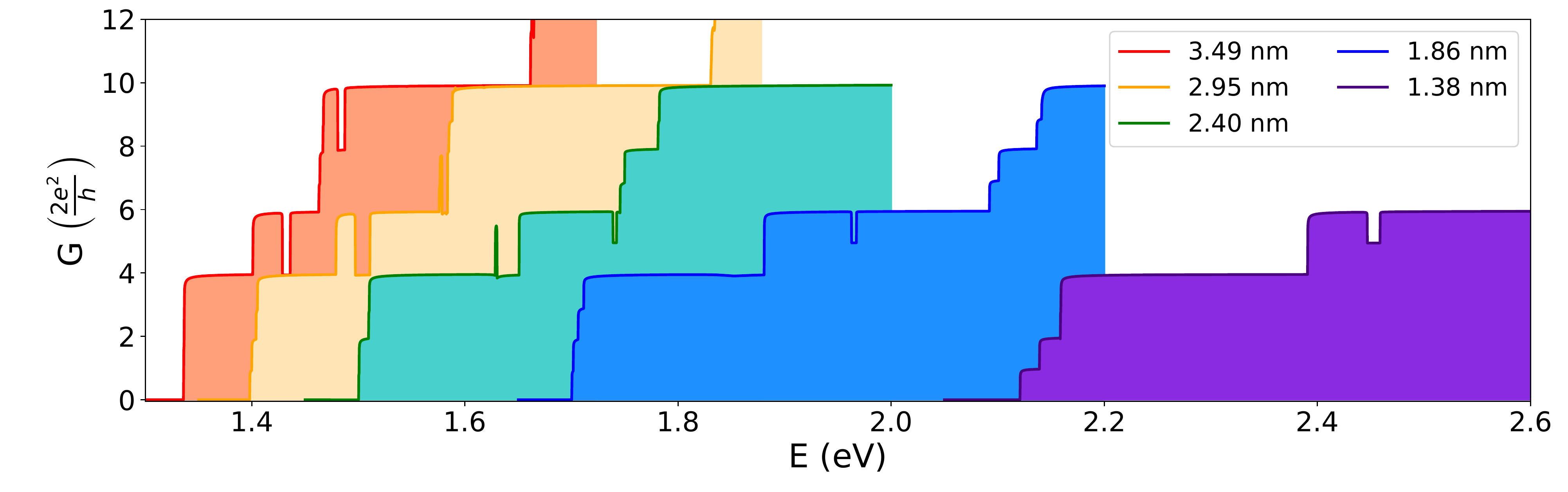}} 
    \subfigure[]{\includegraphics[width=\linewidth]{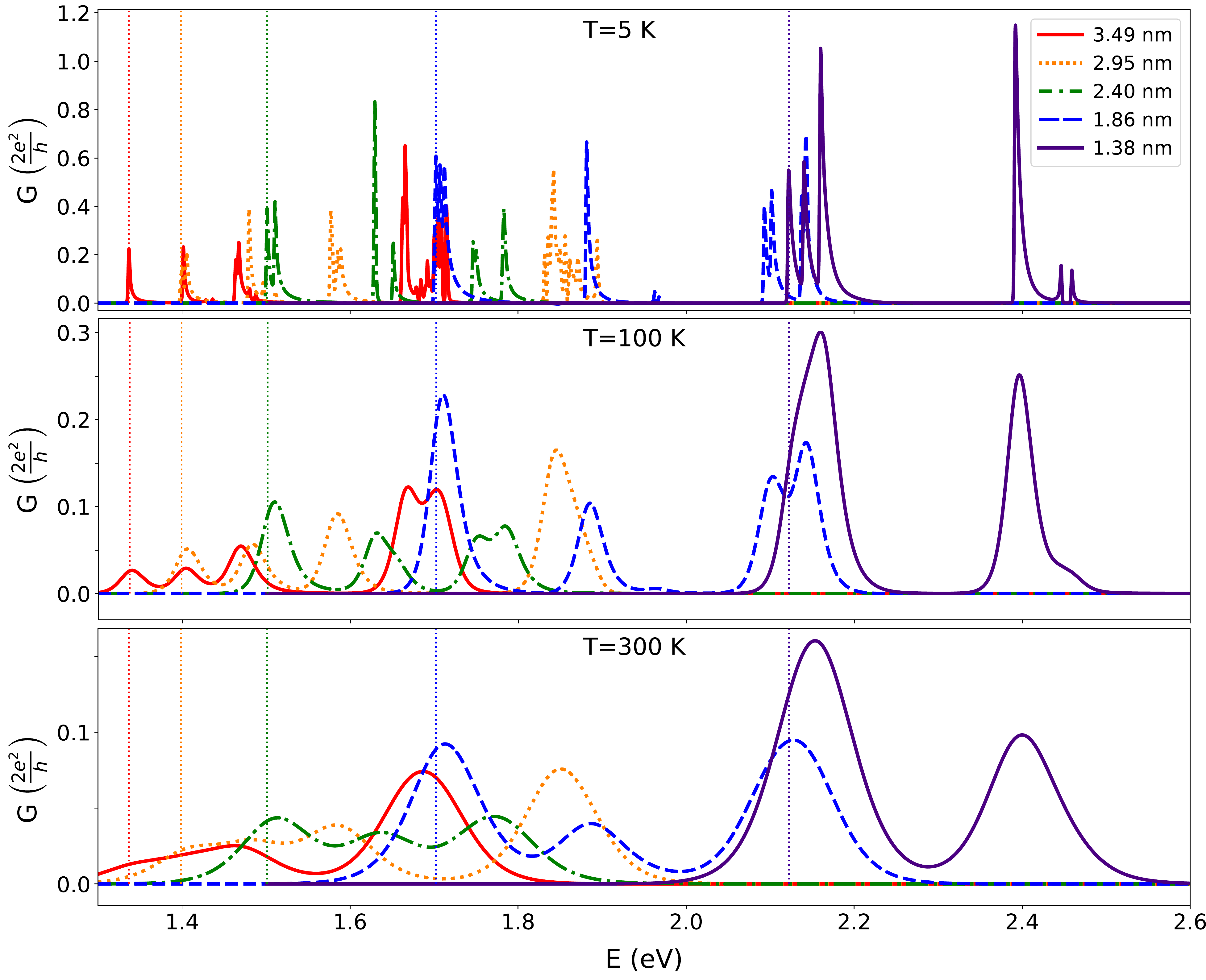}}
    \caption{a) Fermi energy dependence of the pristine nanowire conductance at $T=5$ K, and b) the conductance change determined by a scattering on the electrostatic field of the tetracene cation physisorbed at the surface at $T=$5, 100 and 300 K. The spacing between silicon surface and molecules is taken to be 5 \AA. The series of rectangular NWs are characterized by the length of the side of 1.38, 1.86, 2.40, 2.95 and 3.49 nm.}
    \label{fig:sinw1}
\end{figure}

In order to estimate the conductance sensitivity as a function of the NW widths and temperature, we have computed the conductance change for the case of the molecular cation for a series of rectangular NWs of various sizes at different temperatures (see Fig. \ref{fig:sinw1}). Increasing the width of NWs leads to the overall shift of the conductance spectra and to a reduction of spacing between propagating modes implying enhancement of the density of states. 

\begin{figure}[t]
    \centering
    \includegraphics[width=0.7 \linewidth]{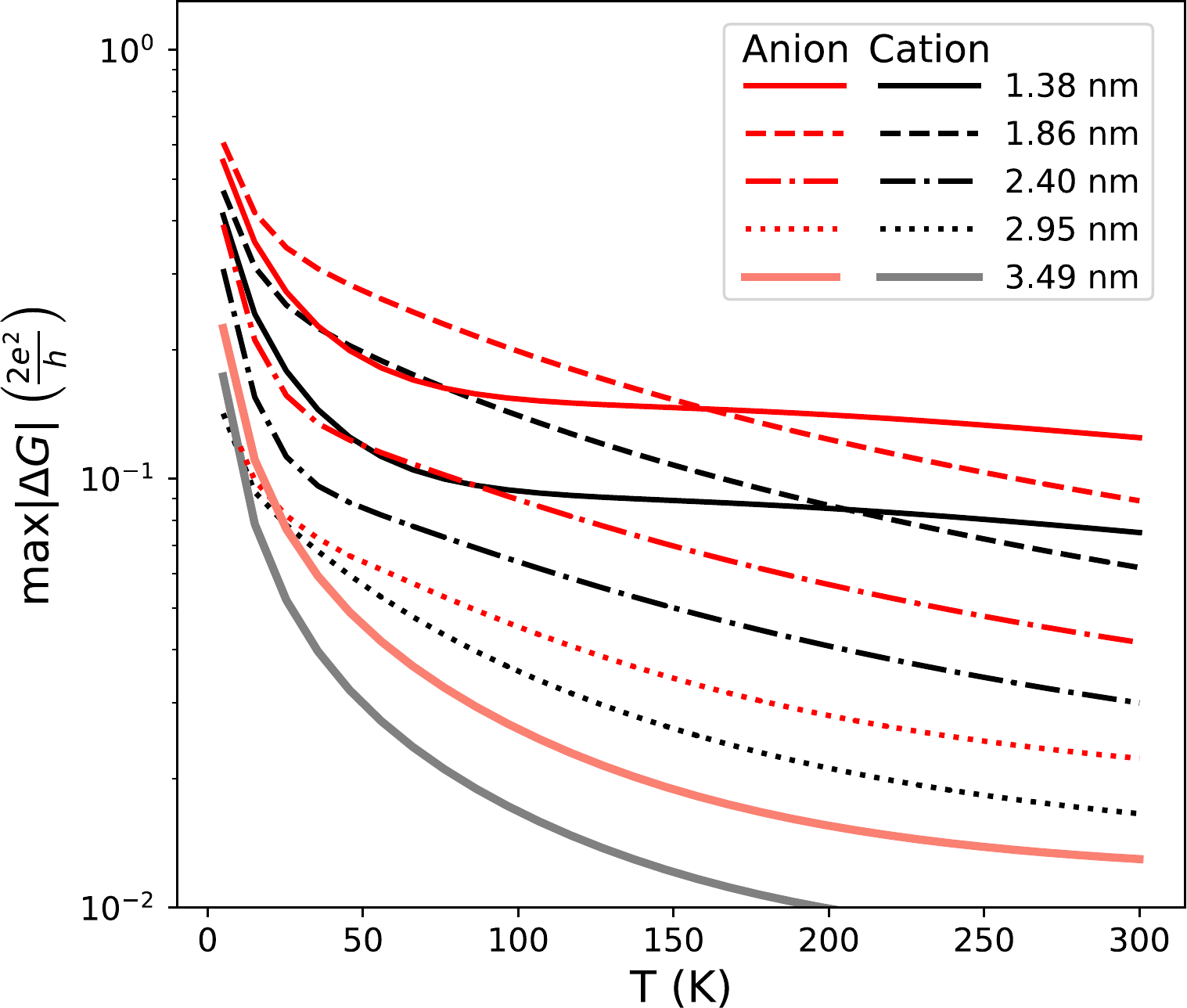}
    \caption{Temperature dependence of the peak conductance change for a series of rectangular NWs with a side length of the side of 1.38, 1.86, 2.40, 2.95 and 3.49 nm.}
    \label{fig:t_dep}
\end{figure}

At zero temperature, for each propagating mode the conductance change caused by the flip of the charge state of a molecule decreases with larger widths of NWs. However at higher temperatures, the Fermi window, defined by the expression $\left. d f(\varepsilon) / d\varepsilon \right|_{\varepsilon_F}$ in  Eq. \ref{eq:landauer}, becomes wider and the conductance is determined not by a single mode but by a contribution from several modes within the Fermi window. Therefore, the conductance change computed at a certain Fermi energy is defined by both the energy distribution of modes and conductance change per a mode. As a result, the interplay between the density of states and sensitivity per state determines a complicated temperature dependence of the peak conductance change (see Fig. \ref{fig:t_dep}). We define the peak conductance change as the conductance change at the first peak of the dependence of the conductance change on the Fermi energy computed at $T=5$ K. The Fermi energies corresponding to those peaks are denoted in Fig. \ref{fig:sinw1}b by the vertical thin lines. In Fig. \ref{fig:t_dep}, due to the above mentioned interplay, we see lots of intersections of the curves at low temperatures for the wider NWs and at relatively high temperatures for the smallest NW of 1.38 nm. The non-monotonic dependence of the conductance sensitivity on the NW size can be explained by the fact that the density of states in the smallest NW is so small that, despite its high sensitivity per mode, the maximal conductance change at low temperatures is observed in a slightly larger NW. At the temperatures above 200 K, the peak sensitivity of NWs depends monotonically on their widths.

\begin{figure}[t]
    \centering
    \includegraphics[width=0.8\linewidth]{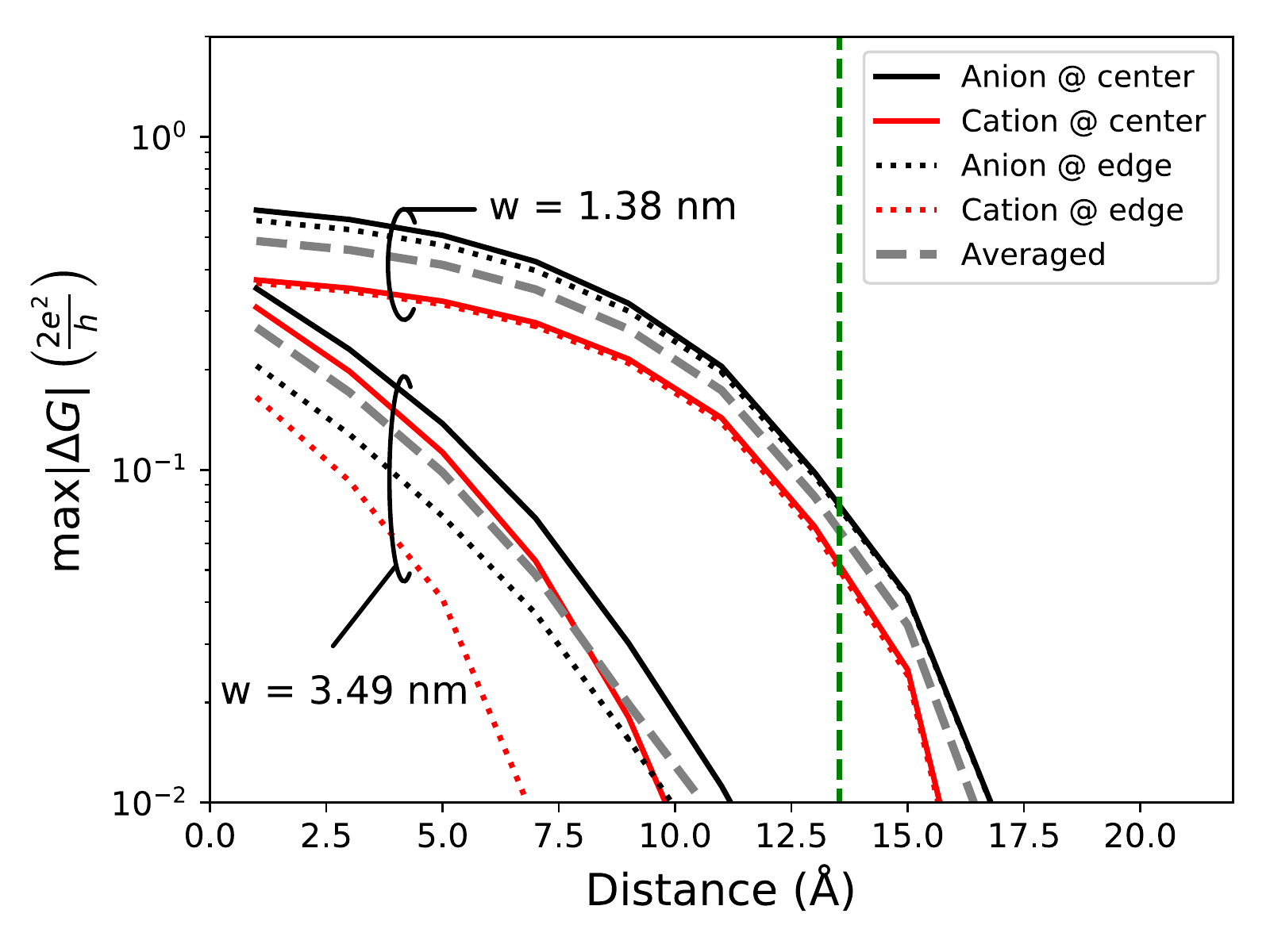}
    \caption{Dependence of the peak conductance change on the spacing between tetracene molecules and NW surface. The green dashed line denotes the width of one monolayer of the crystalline tetracene (13.53 \AA~according to Ref. \onlinecite{Campbell:a03426}).}
    \label{fig:spacing}
\end{figure}

Dependence of the peak conductance change on the spacing between organic molecules and silicon surface is shown in Fig. \ref{fig:spacing} for two NWs of different sizes, computed for both cation and anions placed at the center of NW as well as displaced to its edge. The obtained results show that the conductance change is very sensitive to the width of NWs while it is less sensitive to changes of the lateral coordinates of molecules within the nanowire width and the sign of their charge. Therefore we may average computed conductance over lateral displacement of molecules and charge signs (see grey thick curves in Fig. \ref{fig:spacing}).

\begin{figure}[t]
    \centering
    \subfigure[]{\includegraphics[width=0.8\linewidth]{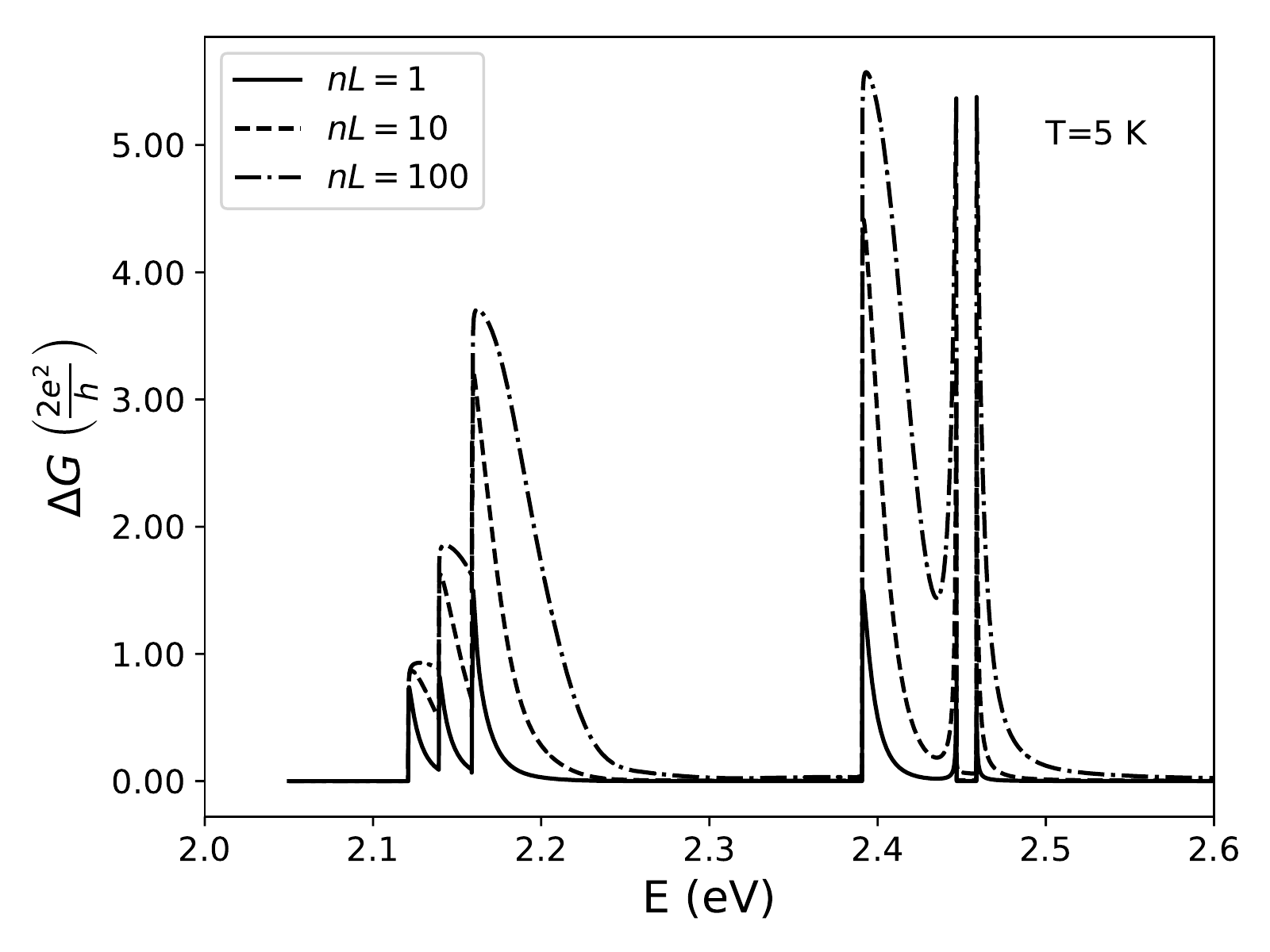}}
    \subfigure[]{\includegraphics[width=0.8\linewidth]{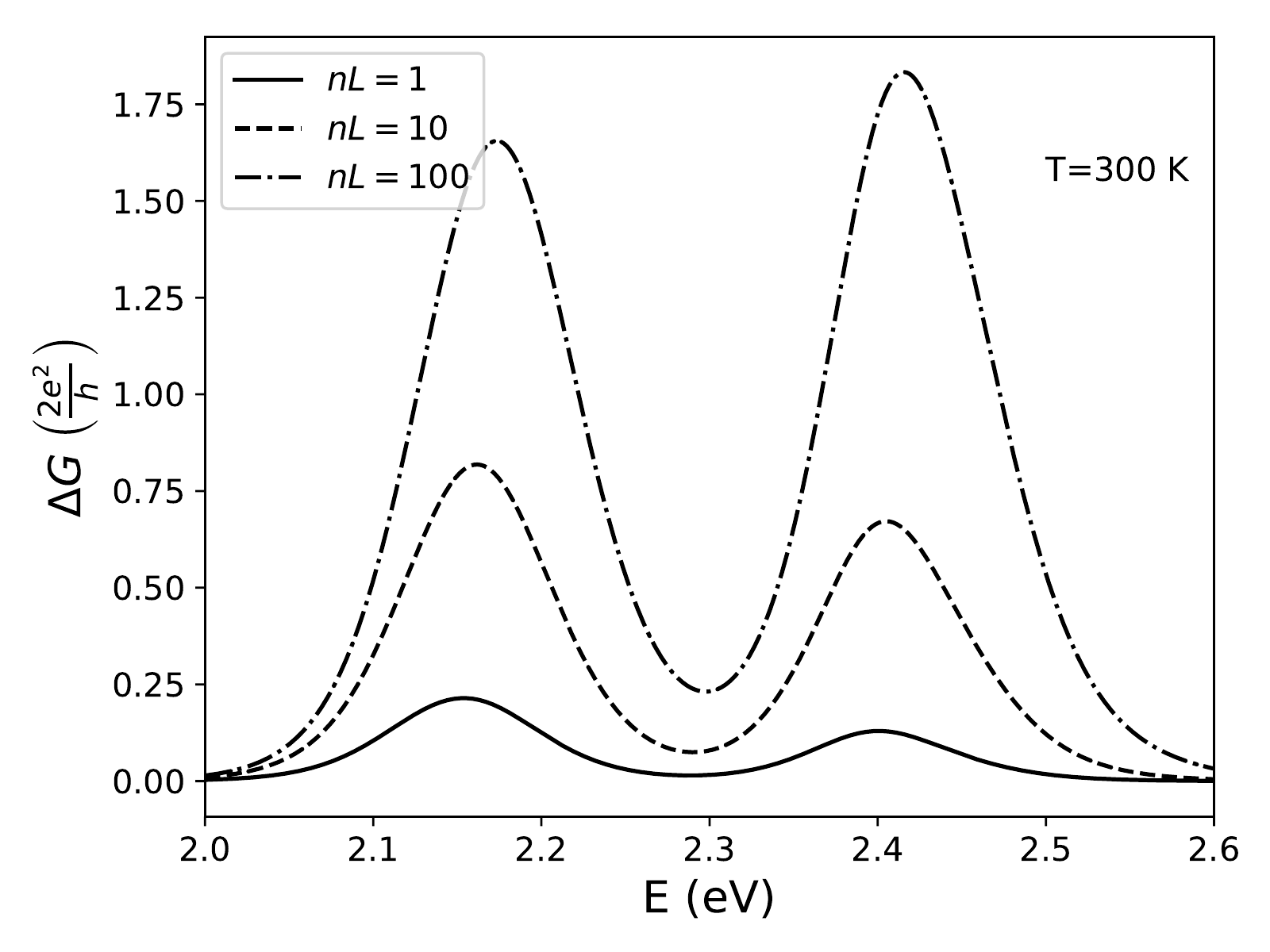}}
    \caption{Fermi energy dependence of the conductance change caused by scattering on the electrostatic field of the tetracene cation for different values of $nL$, effective number of molecules, and for a) T=5 K and b) T=300 K.}
    \label{fig:ensemble}
\end{figure}

In the smallest NW at T=5 K, the conductance is sensitive to the charge states in the first and second molecular monolayers (see Fig. \ref{fig:spacing}). For larger NWs, charges in the first monolayer are only detectable. The sensitivity significantly degrades at higher temperatures resulting in the maximal conductance change of several percent from $2e^2/ \hbar$ in the NW with the side length of 3.49 nm (see Fig. \ref{fig:sinw1}b). However, such conductance change is caused by elastic scattering on a single molecule only. Going beyond the single molecule limit by increasing the concentration of scattering centers and the length of the device region enhances the conductance change as is illustrated in Fig. \ref{fig:ensemble}. The conductance change grows with the product $nL$ and saturates at the conductance of the pristine nanowire. This effect has been first discussed in Ref. \onlinecite{PhysRevLett.56.1960}, showing results similar to ours.

\subsection{Noise spectroscopy: dynamic analysis}

Besides the static charge distribution sensitivity, information about the dynamic evolution of charge carriers can also be obtained. The time evolution of molecular charge states leads to fluctuations of the NW conductance which, in turn, can be probed by noise spectroscopy of the electric current in the NW \cite{Vitusevich}.

The noise in ultra-thin wires consists of shot noise and thermal noise that can be expressed in terms of the transmission probabilities:

\begin{equation}
\begin{split}
    S = & 2qV \coth \left( \frac{qV}{2k_BT} \right) \frac{2q^2}{\hbar} \sum_j T_j \left(1 - T_j \right) + \\ 
                & 4 k_BT  \frac{2q^2}{\hbar} \sum_j T_j
\end{split}
\label{noise}
\end{equation}

We assume that the charge state fluctuations of tetracene molecules represent a discrete Markov process like telegraph noise or generation-recombination noise. This kind of noise is described by the Lorentzian power spectrum \cite{kogan2008, PhysRevLett.99.207001}:

\begin{equation}
    S \left( \omega \right) = \frac{4 V^2 \Delta G_s^2 P_{+,-} P_0 \tau}{1+\omega^2 \tau^2}
    \label{noise01}
\end{equation}
where: $V$ is the applied electrical bias, $\Delta G_s$ is the conductance change caused by charge state switching of a single molecule, $P_{+,-}$ and $P_0$ are probabilities of finding molecule in one of positive or negative charge states and in the neutral state correspondingly. These probabilities can be expressed as: $P_{+,-} = nL / N$ and $P_0 = 1 - nL / N$, where $n$ is the mean linear density of molecules with excess charge, $L$ is the length of NW, and $N$ is the overall number of tetracene molecules covering NW. If we assume that the number of charged molecules is much less than the overall number of molecules, the resulted expression for the noise power spectral density reads:

\begin{equation}
    S \left( \omega \right) = \frac{nL}{N}\frac{4V^2 \Delta G_s^2 \tau}{1+\omega^2 \tau^2}
    \label{noise1}
\end{equation}

Eq. (\ref{noise1}) together with Eq. (\ref{noise}) can be used to fit the measured noise power density. As a result of the fitting procedure, one obtains the value of the characteristic time $\tau$. The values of $n$ and $\Delta G_s$ can be obtained from measurements of the mean conductance change discussed in the previous section.

It has been suggested in Ref. \onlinecite{Watanabe} that, in order to reduce the contributions of thermal noise and shot noise, the measurements of the noise power density can be performed in the frequency range up to 100 GHz where telegraph noise dominates.

The proposed method can be used to measure fluctuations of the intrinsic charge carriers as well as the photo-induced charge carriers. In the later case ultra-fast pulses combined with phase locked detection allow the charge movement within the film to be distinguished from the inherent photoconductance response of the nanowires. In addition, comparison of neighbouring NWs in an array (as visualized in Fig.1) provides a method of common mode rejection.

\subsection{Spatially resolved measurements: electrical impedance tomography}

\begin{figure}[t]
    \centering
    \subfigure[]{\includegraphics[width=0.45\linewidth]{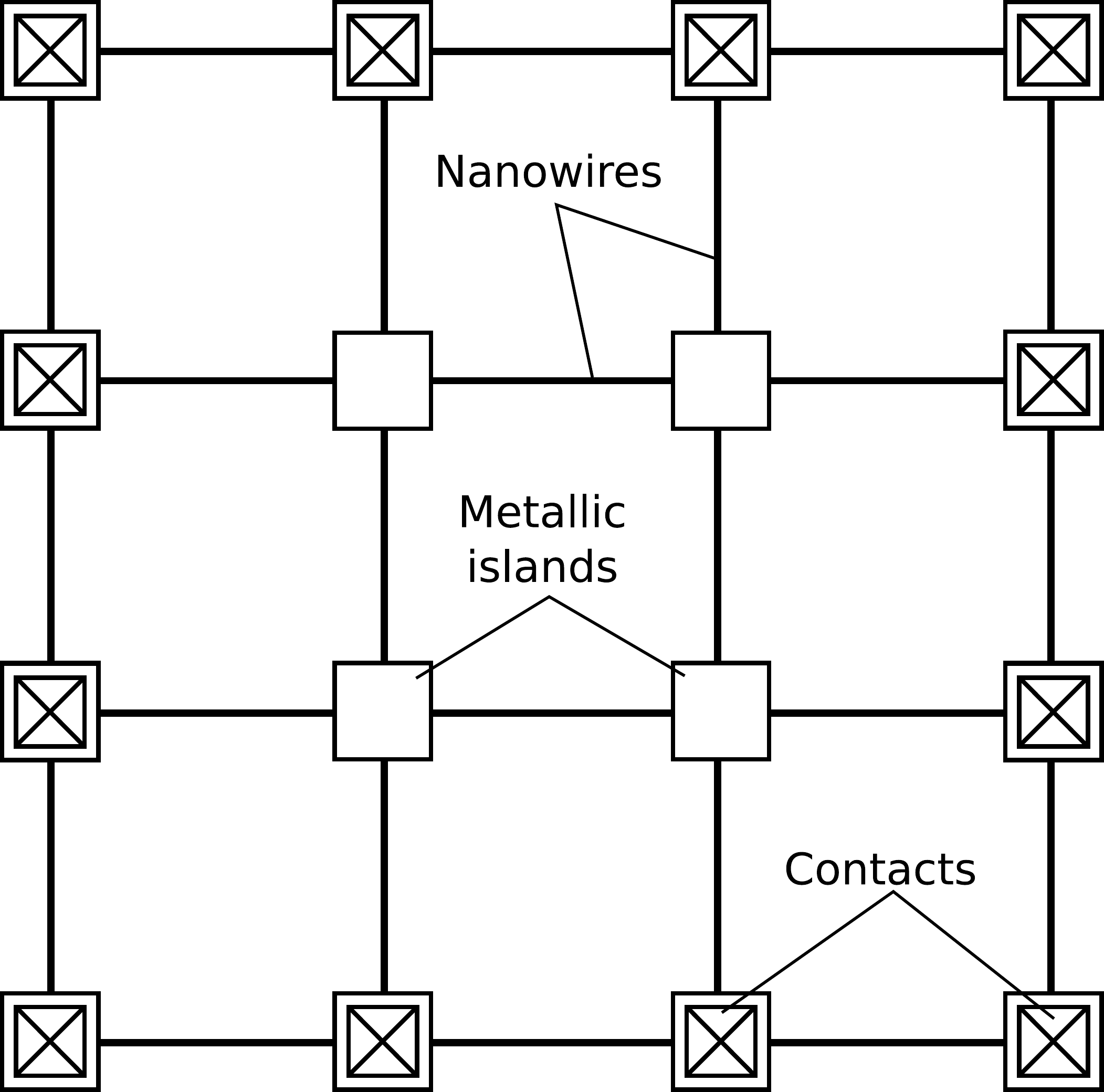}} \hspace{0.3cm}
    \subfigure[]{\includegraphics[width=0.45\linewidth]{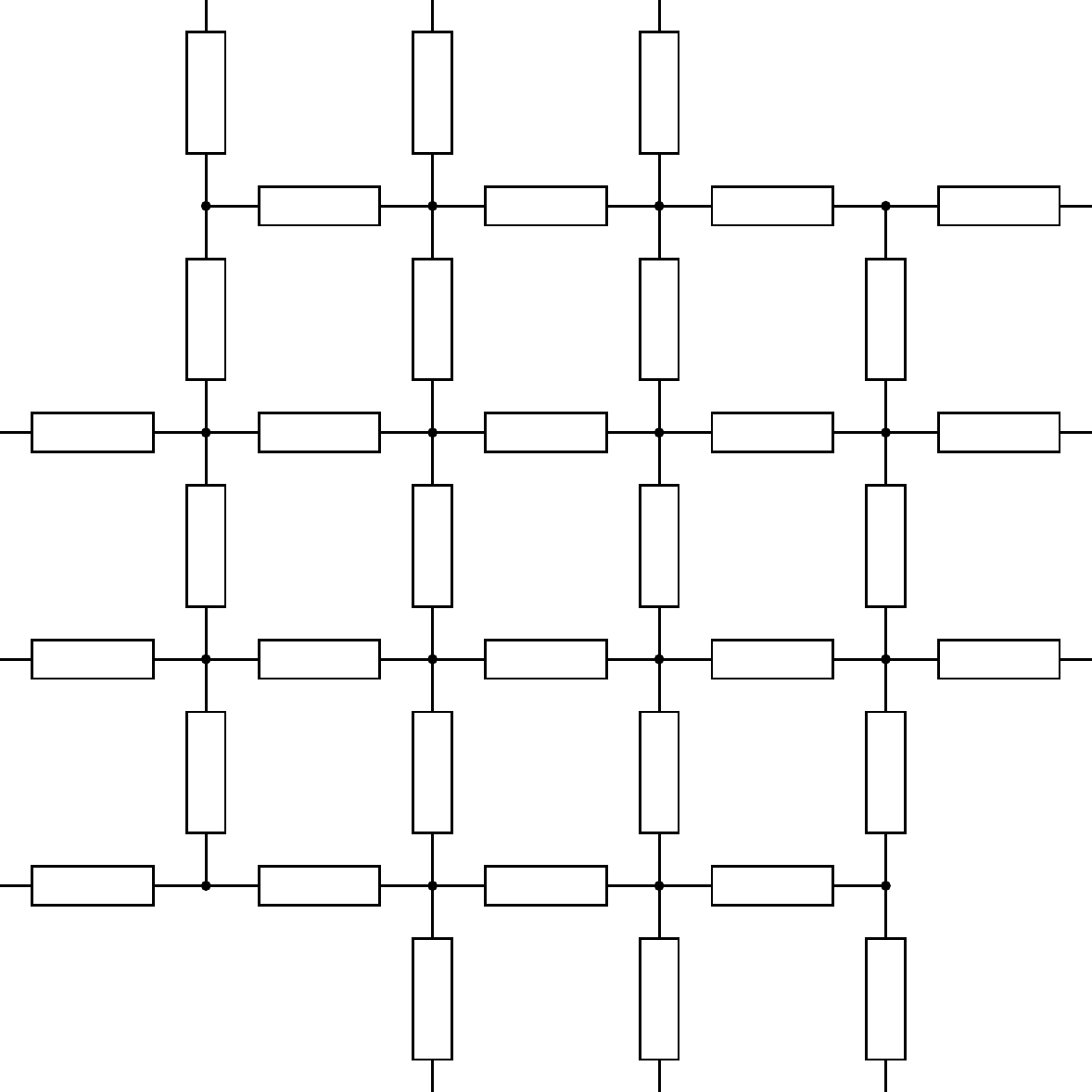}}
    \caption{a) Set-up for the nanoscale electrical impedance tomography of organic thin films and b) its equivalent circuit.  We show a 4x4 array of NW resistors only for illustrating purposes. In reality the number of the resistors determining the resolution of the measurement setup could be much larger.}
    \label{fig:impedance}
\end{figure}

As well as temporal resolution, the spatial NW layout allows the position of the mobile charges to be determined. The 1D array layout of silicon NWs shown in Fig. \ref{fig:intro} provides information on a linear distribution of charge carriers in the direction along the NW array. In order to access information on the 2D distribution of charge carriers in the organic semiconductor thin film, we propose a setup representing a grid of NW resistors shown in Fig.~\ref{fig:impedance}a. In principle, the resistance of each NW resistor can be read out with having contacts only at the edge nodes by means of electrical impedance tomography \cite{holder2004electrical}.

Electrical impedance tomography is a standard tool in modern medical imaging, and it has also recently found several applications for charge sensing of nanoengineered thin films \cite{tom, lynch}. In the equivalent circuit of the proposed device shown in Fig. \ref{fig:impedance}b all resistances are unknown while one has access to set up and/or to measure either the voltages on the edge nodes and/or the currents injected in the edge nodes. It has been proven that the problem of finding resistances for such an equivalent circuit has a unique solution \cite{holder2004electrical}. The technology relies on how fast this problem can be solved on the device level. The proposed setup requires additional analysis on the trade-offs between accuracy, resolution and bandwidth, which can be optimized for the parameters of given film to be studied.

\section{Summary}
In order to estimate the sensitivity of nanowire conductance detection, we consider silicon nanowires as a probe of charge states in tetracene molecules physisorbed on its surface as an example. We used the linear transport model and expressed the mean-free path determined by elastic scattering in terms of causal Green's functions. The Green's functions have been computed using the recursive algorithm and the semi-empirical tight-model with the reduced mode space transformation. The electrostatic potential of the molecules with excess charge has been computed using a combination  of  the  polarizable  continuum  model and density functional theory with the range-separated exchange-correlation functional.

The conductance change caused by elastic scattering on the electrostatic field of charged molecules decreases with increasing temperature and transverse sizes of nanowires. For a single molecule, ultra-thin silicon nanowires with characteristic sizes of the cross-section below 2 nm are characterized by the conductance change of about $0.1 \cdot 2e^2 / \hbar $ at room temperature. The conductance change grows with the number of charged molecules. Thus, the sub-4 nm nanowires are sensitive enough to detect several tens of charge carriers. At room temperature, only charge carriers in the first molecular monolayer are detectable. 

Our results show that silicon nanowires can be used to measure the concentration of charge carriers in the organic thin films assuming that their charge is localized within a single molecule. In addition the time evolution of molecular charge states leads to fluctuations of the NW conductance which, in turn, can be probed by noise spectroscopy. We propose using a grid of nanowire resistors to access the information on 2D spatial distribution of charge carriers in organic thin films, via electrical impedance tomography.

\section{Acknowledgments}
MK thanks Barbara Fresch and Francoise Remacle for fruitful discussions on DFT computations. MK and JC thank Pegah Maasoumi for discussions on relevant technological advances. The authors acknowledge support of the Australian Research Council through grant CE170100026. This research was undertaken with the assistance of resources and services from the National Computational Infrastructure, which is supported by the Australian Government.

\section{Appendix A: Device region Hamiltonian and leads self-energies}

In this appendix the band structure of the infinite ideal NW can be obtained from the solution of the following linear system of algebraic equations \cite{Wimmer}:

\begin{equation}
    \left( \begin{matrix} 
\mathbf{0} & \mathbf{I} \\
-\mathbf{h}_1^{-1} \mathbf{h}_{-1} & \mathbf{h}_1^{-1} \left(E\mathbf{I} - \mathbf{h}_0 \right)
\end{matrix}
\right)
\left( \begin{matrix} 
\mathbf{u}_n \\
\lambda_n \mathbf{u}_n
\end{matrix}
\right) =  \lambda_n \left( \begin{matrix} 
\mathbf{u}_n \\
\lambda_n \mathbf{u}_n
\end{matrix}
\right)
\label{lin}
\end{equation}
where $\mathbf{h}_0$ is the tight-binding Hamiltonian of the nanowire unit cell shown in Fig. \ref{fig:intro1}, $\mathbf{h}_{-1}$ and $\mathbf{h}_1$ are the Hamiltonians  describing couplings of the unit cell to the left and right neighbours correspondingly, $E$ is the energy set as a parameter and $\lambda_n = e^{-i k_n}$ is an eigenvalue expressed as an analytic function of the wave number $k$.

Following the procedure described in Ref. \onlinecite{Wimmer}, the eigenvectors $\mathbf{u}_n$ and eigenvalues $\lambda_n$ can be divided in two classes: the left-propagating modes, $\mathbf{u}_<$ and $\mathbf{\Lambda}_<$, and right-propagating modes, $\mathbf{u}_>$ and $\mathbf{\Lambda}_>$. The self-energies describing couplings to the semi-infinite leads can be expressed in terms of those modes \cite{Wimmer}:

\begin{equation}
    \mathbf{\Sigma}_L^r = \mathbf{h}_{1} \mathbf{u}_> \mathbf{\Lambda}_> \mathbf{u}_>^{-1}, \hspace{1cm} \mathbf{\Sigma}_R^r = \mathbf{h}_{-1} \mathbf{u}_< \mathbf{\Lambda}_> \mathbf{u}_<^{-1}
\end{equation}

The Hamiltonian for the device region, $\mathbf{H}$, in Eq. (\ref{eq:gf}) is composed of the Hamiltonians for the unit cells of the nanowire:

\begin{equation}
    \mathbf{H} = \left( 
    \begin{matrix} 
       \mathbf{h}_0 & \mathbf{h}_1  &  &  & \mathbf{0} \\
       \mathbf{h}_{-1} & \mathbf{h}_0 & \mathbf{h}_1 & & \\
       & \ddots & \ddots  & \ddots &  \\
       & & \mathbf{h}_{-1} & \mathbf{h}_0 & \mathbf{h}_1 \\
       \mathbf{0} & & & \mathbf{h}_{-1} & \mathbf{h}_0
       \end{matrix}
    \right)
\end{equation}

The number of blocks on the main diagonal is determined by the required length of the device region. The minimal length is determined by spreading of the electrostatic field of a single molecule with excess charge which spreads over approximately fifteen unit cells.

\section{Appendix B: Electrostatic fields of physisorbed tetracene molecules}

\begin{figure}[th]
    \centering
    \subfigure[]{\includegraphics[width=0.49\linewidth]{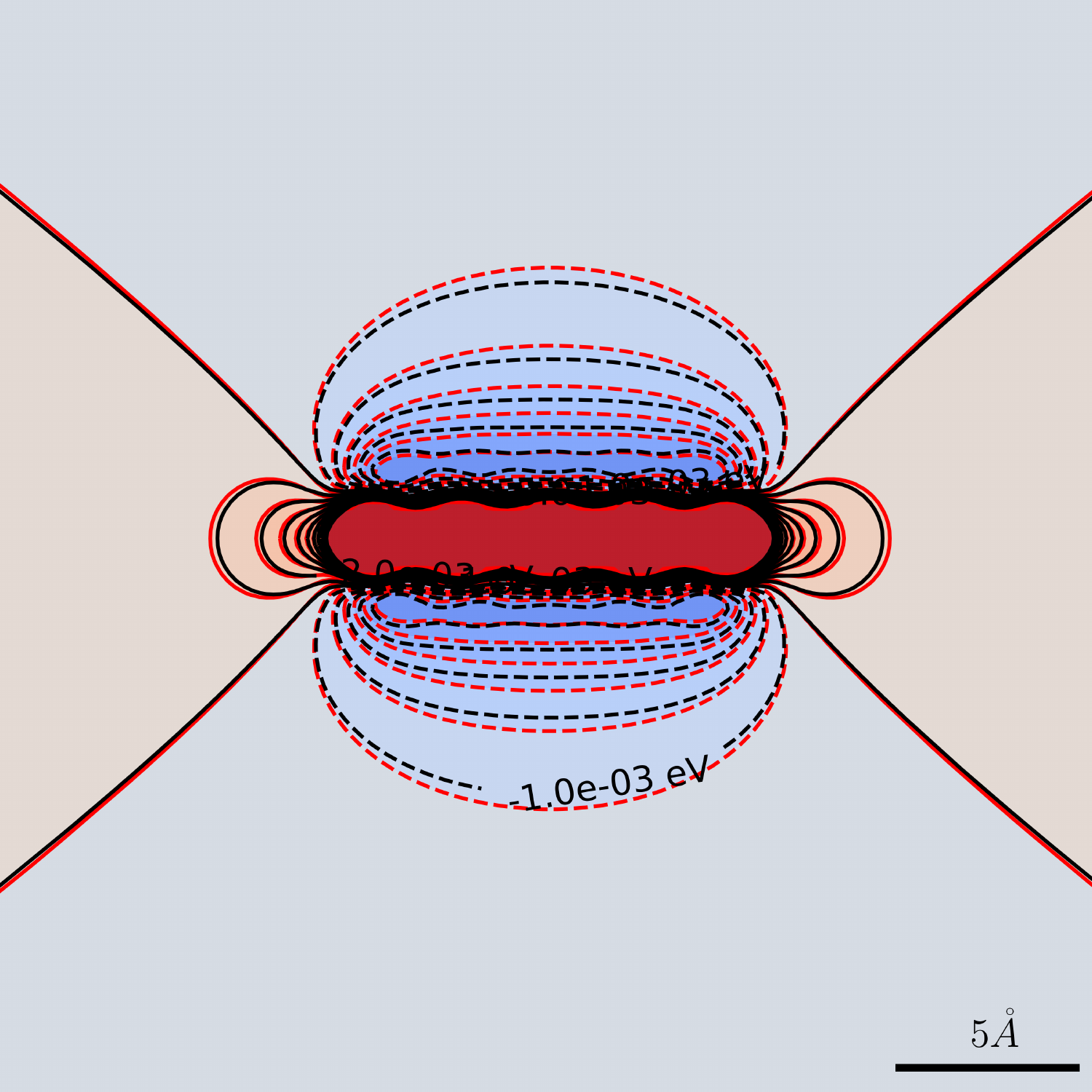}}
    \subfigure[]{\includegraphics[width=0.49\linewidth]{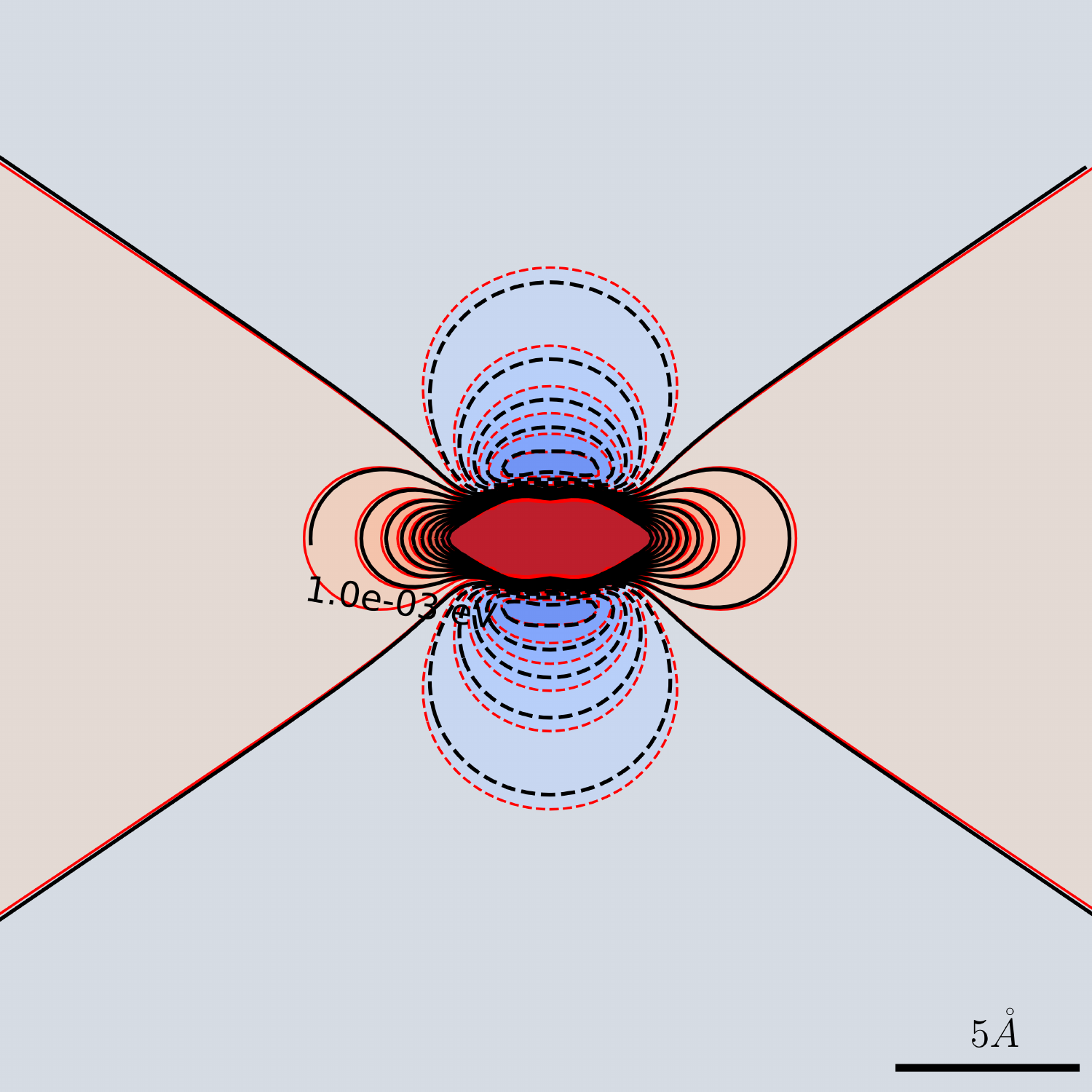}}
    \caption{Variations of electrostatic field of a neutral tetracene molecule in the near-field region computed with (black line contours) and without (red  line contours) optimizing the range-separation parameter for crystalline environment. The field distribution is shown for the planes perpendicular to the aromatic rings a) along the molecular axis $c$ and b) perpendicular the molecular axis $c$.}
    \label{fig:neutral}
\end{figure}

\begin{figure}[th]
    \centering
    \subfigure[]{\includegraphics[width=0.49\linewidth]{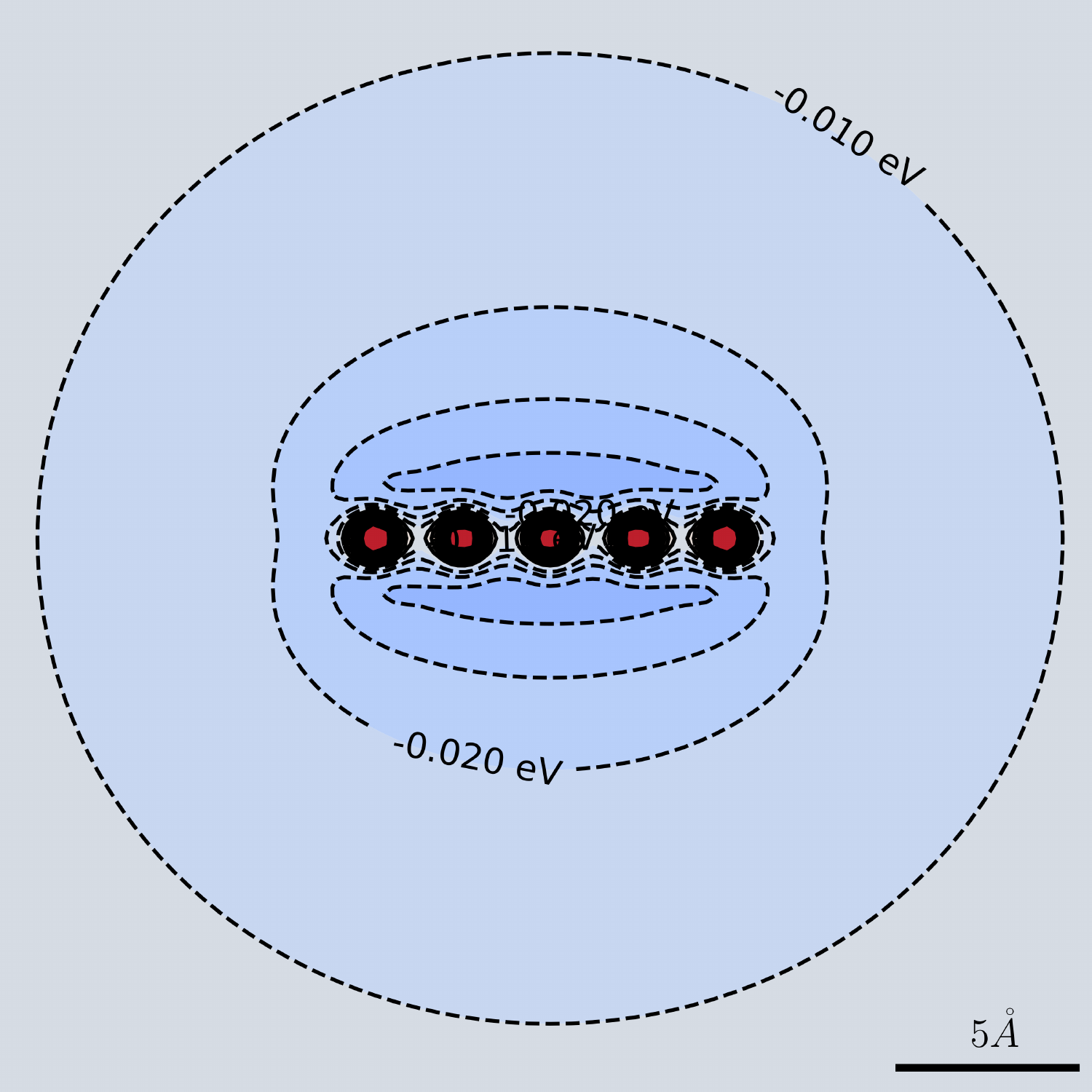}}
    \subfigure[]{\includegraphics[width=0.49\linewidth]{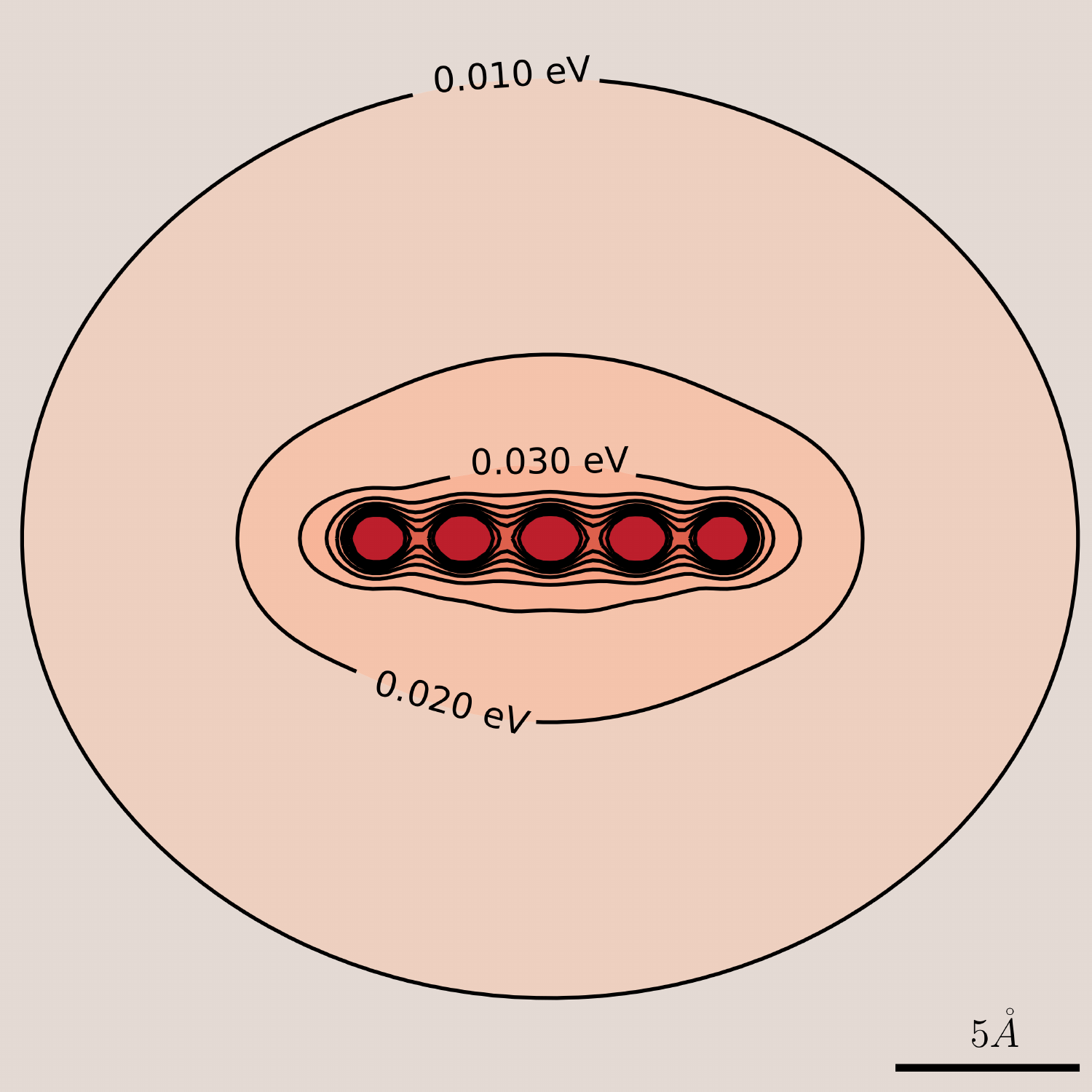}}
    \caption{Electrostatic potential of tetracene a) anion and b) cation computed in the homogeneous media approximation.}
    \label{fig:ions}
\end{figure}

In the homogeneous media approximation, the electrostatic potentials has been computed with the relative permittivity of $\epsilon_{Tc}=$3.8 to take into account dielectric screening caused by a polarizable environment.

 Results of DFT calculations shown in Fig. \ref{fig:neutral} suggest that the electrostatic field distribution of the neutral tetracene molecule is highly nonuniform in the near zone. Such field distribution is responsible for the herringbone structure of the tetracene crystal. The optimizing of the range-separation parameter $\omega$ in the exchange-correlation functional besides giving better agreement for ionization energy and electron affinity with experimental data, causes a slight difference in the electrostatic potential which is a result of a small redistribution of the electron density within a molecule due to interactions with surrounding molecules in the crystal.

The electrostatic potentials of the tetracene anion and cation shown in Fig. \ref{fig:ions} resemble the screened Coulomb potential at the distances larger than 15 \AA~from the molecular center of mass. At the shorter distances, electrostatic fields are characterized by a non-uniform angular dependence.

The dielectric permittivity of Si NWs are collected in Tab. \ref{tab:table1}. In order to take into account the dielectric mismatch between tetracene and the Si NW, we solve the Poisson equation: 
\begin{equation}
\nabla \left[ \epsilon(\mathbf{r}) \nabla  \phi(\mathbf{r}) \right] = - 4 \pi \rho_0    
\label{eq:poisson}
\end{equation}
for the charge density $\rho_0$ obtained in the homogeneous media approximation. The charge density can be extracted from the electrostatic potential, $\phi_0$, shown in Fig.~\ref{fig:ions} using the relationship $\rho_0 = - (\epsilon_{Tc} / 4 \pi) \nabla^2 \phi_0$.

\begin{table}[th]
\caption{\label{tab:table1}%
Relative dielectric permittivity of silicon NWs obtained from Penn's model
}
\begin{tabular}{c @{\qquad} c @{\qquad} c @{\qquad} c @{\qquad} c @{\qquad} c}
\colrule
NW width (nm) & 1.38 & 1.86 & 2.40 & 2.95 & 3.49\\
$\epsilon_{NW}$ & 8.11 & 9.26 & 9.97 & 10.38 & 10.62 \
\end{tabular}
\end{table}

Eq.~\ref{eq:poisson} can be rewritten in the form:
\begin{equation}
\nabla^2  \phi(\mathbf{r}) = - \frac{1}{\epsilon(\mathbf{r})} \left[ 4 \pi \rho_0 + \nabla \epsilon(\mathbf{r}) \nabla  \phi(\mathbf{r}) \right]
\label{eq:poisson1}
\end{equation}

The right hand side of the equation can be thought as an effective charge, $\rho'$, dependent on the electrostatic field. This is the essence of the so-called the induced charge computation method \cite{doi:10.1021/acsami.5b01606}. The equation can be solved iteratively assuming $\rho' = 4 \pi \rho_0 / \epsilon(\mathbf{r})$ as a first guess and updating charge $\rho'$ for each next step. The advantage of the iteration procedure is that, at each iteration step, we solve the standard Poisson equation with constant coefficients, for which a variety of standard Poisson solvers exists. In this work we use the so-called fast Poisson solver from GPAW package \cite{gpaw}. The solver uses a combination of Fourier and Fourier-sine transforms in combination with parallel array transposes. 

\begin{figure}[t]
    \centering
    \includegraphics[width=0.7\linewidth]{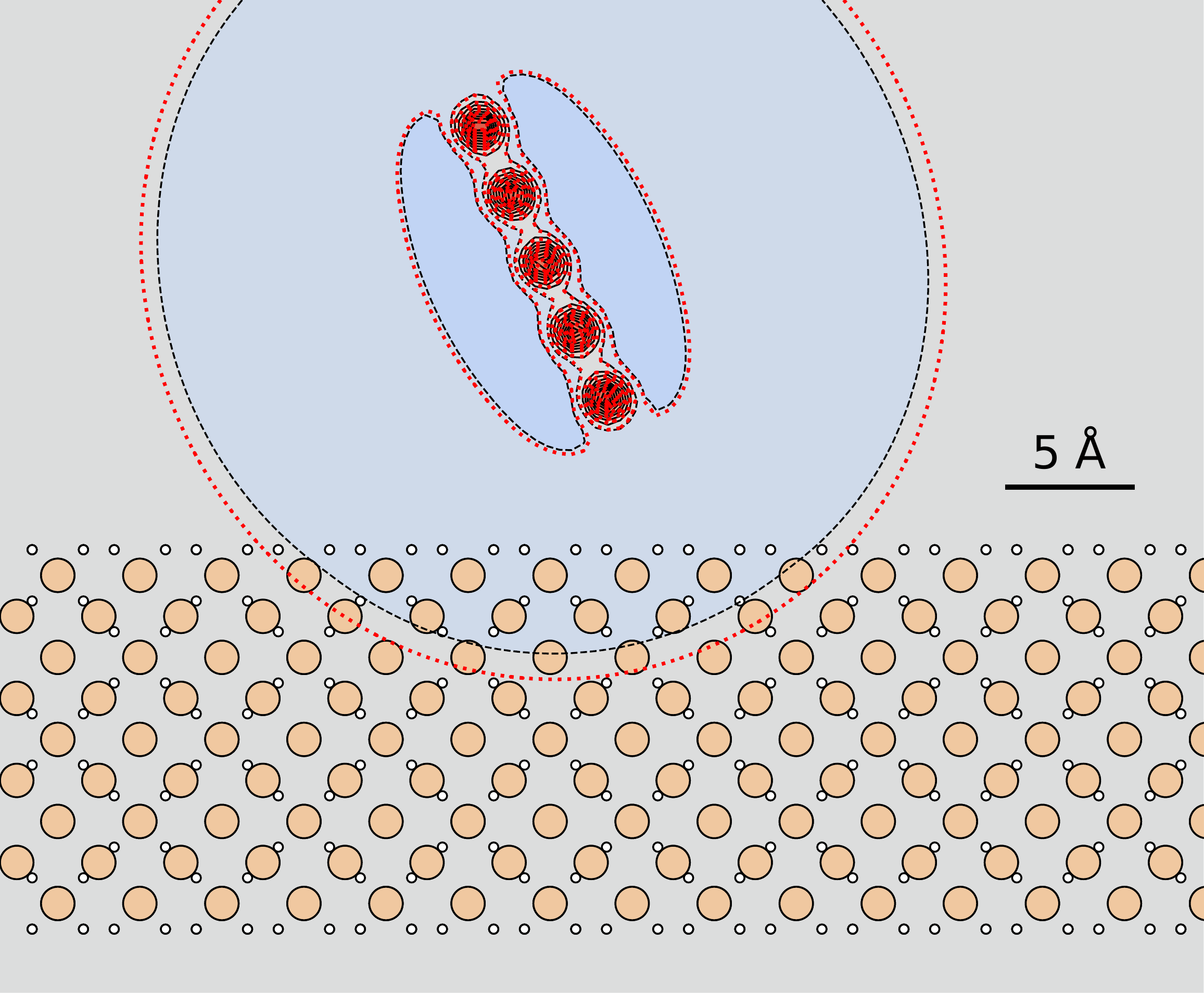}
    \caption{The electrostatic potential of tetracene anion computed with $\epsilon=3.8$ (red dotted contours) with the position-dependent dielectric permittivity (black dashed contours) taking into account the dielectric mismatch between silicon NW and tetracene crystal.}
    \label{fig:corrections}
\end{figure}

In Fig.~\ref{fig:corrections}, we show the effect of the dielectric mismatch on the electrostatic field of tetracene anion. The effect is pretty weak resulting in relative deviation of less than 5 \% in NW conductances.

\bibliographystyle{apsrev4-1}
\bibliography{bib}

\begin{thebibliography}{59}%
\makeatletter
\providecommand \@ifxundefined [1]{%
 \@ifx{#1\undefined}
}%
\providecommand \@ifnum [1]{%
 \ifnum #1\expandafter \@firstoftwo
 \else \expandafter \@secondoftwo
 \fi
}%
\providecommand \@ifx [1]{%
 \ifx #1\expandafter \@firstoftwo
 \else \expandafter \@secondoftwo
 \fi
}%
\providecommand \natexlab [1]{#1}%
\providecommand \enquote  [1]{``#1''}%
\providecommand \bibnamefont  [1]{#1}%
\providecommand \bibfnamefont [1]{#1}%
\providecommand \citenamefont [1]{#1}%
\providecommand \href@noop [0]{\@secondoftwo}%
\providecommand \href [0]{\begingroup \@sanitize@url \@href}%
\providecommand \@href[1]{\@@startlink{#1}\@@href}%
\providecommand \@@href[1]{\endgroup#1\@@endlink}%
\providecommand \@sanitize@url [0]{\catcode `\\12\catcode `\$12\catcode
  `\&12\catcode `\#12\catcode `\^12\catcode `\_12\catcode `\%12\relax}%
\providecommand \@@startlink[1]{}%
\providecommand \@@endlink[0]{}%
\providecommand \url  [0]{\begingroup\@sanitize@url \@url }%
\providecommand \@url [1]{\endgroup\@href {#1}{\urlprefix }}%
\providecommand \urlprefix  [0]{URL }%
\providecommand \Eprint [0]{\href }%
\providecommand \doibase [0]{http://dx.doi.org/}%
\providecommand \selectlanguage [0]{\@gobble}%
\providecommand \bibinfo  [0]{\@secondoftwo}%
\providecommand \bibfield  [0]{\@secondoftwo}%
\providecommand \translation [1]{[#1]}%
\providecommand \BibitemOpen [0]{}%
\providecommand \bibitemStop [0]{}%
\providecommand \bibitemNoStop [0]{.\EOS\space}%
\providecommand \EOS [0]{\spacefactor3000\relax}%
\providecommand \BibitemShut  [1]{\csname bibitem#1\endcsname}%
\let\auto@bib@innerbib\@empty
\bibitem [{\citenamefont {Moses}\ \emph {et~al.}(2006)\citenamefont {Moses},
  \citenamefont {Soci}, \citenamefont {Chi},\ and\ \citenamefont
  {Ramirez}}]{Moses}%
  \BibitemOpen
  \bibfield  {author} {\bibinfo {author} {\bibfnamefont {D.}~\bibnamefont
  {Moses}}, \bibinfo {author} {\bibfnamefont {C.}~\bibnamefont {Soci}},
  \bibinfo {author} {\bibfnamefont {X.}~\bibnamefont {Chi}}, \ and\ \bibinfo
  {author} {\bibfnamefont {A.~P.}\ \bibnamefont {Ramirez}},\ }\href {\doibase
  10.1103/PhysRevLett.97.067401} {\bibfield  {journal} {\bibinfo  {journal}
  {Phys. Rev. Lett.}\ }\textbf {\bibinfo {volume} {97}},\ \bibinfo {pages}
  {067401} (\bibinfo {year} {2006})}\BibitemShut {NoStop}%
\bibitem [{\citenamefont {Torsi}\ \emph {et~al.}(2013)\citenamefont {Torsi},
  \citenamefont {Magliulo}, \citenamefont {Manoli},\ and\ \citenamefont
  {Palazzo}}]{OFET}%
  \BibitemOpen
  \bibfield  {author} {\bibinfo {author} {\bibfnamefont {L.}~\bibnamefont
  {Torsi}}, \bibinfo {author} {\bibfnamefont {M.}~\bibnamefont {Magliulo}},
  \bibinfo {author} {\bibfnamefont {K.}~\bibnamefont {Manoli}}, \ and\ \bibinfo
  {author} {\bibfnamefont {G.}~\bibnamefont {Palazzo}},\ }\href {\doibase
  10.1039/C3CS60127G} {\bibfield  {journal} {\bibinfo  {journal} {Chem. Soc.
  Rev.}\ }\textbf {\bibinfo {volume} {42}},\ \bibinfo {pages} {8612} (\bibinfo
  {year} {2013})}\BibitemShut {NoStop}%
\bibitem [{\citenamefont {MacQueen}\ \emph {et~al.}(2018)\citenamefont
  {MacQueen}, \citenamefont {Liebhaber}, \citenamefont {Niederhausen},
  \citenamefont {Mews}, \citenamefont {Gersmann}, \citenamefont {Jäckle},
  \citenamefont {Jäger}, \citenamefont {Tayebjee}, \citenamefont {Schmidt},
  \citenamefont {Rech},\ and\ \citenamefont {Lips}}]{MacQueen}%
  \BibitemOpen
  \bibfield  {author} {\bibinfo {author} {\bibfnamefont {R.~W.}\ \bibnamefont
  {MacQueen}}, \bibinfo {author} {\bibfnamefont {M.}~\bibnamefont {Liebhaber}},
  \bibinfo {author} {\bibfnamefont {J.}~\bibnamefont {Niederhausen}}, \bibinfo
  {author} {\bibfnamefont {M.}~\bibnamefont {Mews}}, \bibinfo {author}
  {\bibfnamefont {C.}~\bibnamefont {Gersmann}}, \bibinfo {author}
  {\bibfnamefont {S.}~\bibnamefont {Jäckle}}, \bibinfo {author} {\bibfnamefont
  {K.}~\bibnamefont {Jäger}}, \bibinfo {author} {\bibfnamefont {M.~J.~Y.}\
  \bibnamefont {Tayebjee}}, \bibinfo {author} {\bibfnamefont {T.~W.}\
  \bibnamefont {Schmidt}}, \bibinfo {author} {\bibfnamefont {B.}~\bibnamefont
  {Rech}}, \ and\ \bibinfo {author} {\bibfnamefont {K.}~\bibnamefont {Lips}},\
  }\href {\doibase 10.1039/C8MH00853A} {\bibfield  {journal} {\bibinfo
  {journal} {Mater. Horiz.}\ }\textbf {\bibinfo {volume} {5}},\ \bibinfo
  {pages} {1065} (\bibinfo {year} {2018})}\BibitemShut {NoStop}%
\bibitem [{\citenamefont {Ruzicka}\ \emph {et~al.}(2012)\citenamefont
  {Ruzicka}, \citenamefont {Wang}, \citenamefont {Liu}, \citenamefont {Loh},
  \citenamefont {Wu},\ and\ \citenamefont {Zhao}}]{Ruzicka}%
  \BibitemOpen
  \bibfield  {author} {\bibinfo {author} {\bibfnamefont {B.~A.}\ \bibnamefont
  {Ruzicka}}, \bibinfo {author} {\bibfnamefont {S.}~\bibnamefont {Wang}},
  \bibinfo {author} {\bibfnamefont {J.}~\bibnamefont {Liu}}, \bibinfo {author}
  {\bibfnamefont {K.-P.}\ \bibnamefont {Loh}}, \bibinfo {author} {\bibfnamefont
  {J.~Z.}\ \bibnamefont {Wu}}, \ and\ \bibinfo {author} {\bibfnamefont
  {H.}~\bibnamefont {Zhao}},\ }\href {\doibase 10.1364/OME.2.000708} {\bibfield
   {journal} {\bibinfo  {journal} {Opt. Mater. Express}\ }\textbf {\bibinfo
  {volume} {2}},\ \bibinfo {pages} {708} (\bibinfo {year} {2012})}\BibitemShut
  {NoStop}%
\bibitem [{\citenamefont {Kumar}\ \emph {et~al.}(2011)\citenamefont {Kumar},
  \citenamefont {Ruzicka}, \citenamefont {Butch}, \citenamefont {Syers},
  \citenamefont {Kirshenbaum}, \citenamefont {Paglione},\ and\ \citenamefont
  {Zhao}}]{Kumar}%
  \BibitemOpen
  \bibfield  {author} {\bibinfo {author} {\bibfnamefont {N.}~\bibnamefont
  {Kumar}}, \bibinfo {author} {\bibfnamefont {B.~A.}\ \bibnamefont {Ruzicka}},
  \bibinfo {author} {\bibfnamefont {N.~P.}\ \bibnamefont {Butch}}, \bibinfo
  {author} {\bibfnamefont {P.}~\bibnamefont {Syers}}, \bibinfo {author}
  {\bibfnamefont {K.}~\bibnamefont {Kirshenbaum}}, \bibinfo {author}
  {\bibfnamefont {J.}~\bibnamefont {Paglione}}, \ and\ \bibinfo {author}
  {\bibfnamefont {H.}~\bibnamefont {Zhao}},\ }\href {\doibase
  10.1103/PhysRevB.83.235306} {\bibfield  {journal} {\bibinfo  {journal} {Phys.
  Rev. B}\ }\textbf {\bibinfo {volume} {83}},\ \bibinfo {pages} {235306}
  (\bibinfo {year} {2011})}\BibitemShut {NoStop}%
\bibitem [{\citenamefont {Cui}\ \emph {et~al.}(2001)\citenamefont {Cui},
  \citenamefont {Wei}, \citenamefont {Park},\ and\ \citenamefont
  {Lieber}}]{Cui1289}%
  \BibitemOpen
  \bibfield  {author} {\bibinfo {author} {\bibfnamefont {Y.}~\bibnamefont
  {Cui}}, \bibinfo {author} {\bibfnamefont {Q.}~\bibnamefont {Wei}}, \bibinfo
  {author} {\bibfnamefont {H.}~\bibnamefont {Park}}, \ and\ \bibinfo {author}
  {\bibfnamefont {C.~M.}\ \bibnamefont {Lieber}},\ }\href {\doibase
  10.1126/science.1062711} {\bibfield  {journal} {\bibinfo  {journal}
  {Science}\ }\textbf {\bibinfo {volume} {293}},\ \bibinfo {pages} {1289}
  (\bibinfo {year} {2001})},\ \Eprint
  {http://arxiv.org/abs/https://science.sciencemag.org/content/293/5533/1289.full.pdf}
  {https://science.sciencemag.org/content/293/5533/1289.full.pdf} \BibitemShut
  {NoStop}%
\bibitem [{\citenamefont {Li}\ \emph {et~al.}(2012)\citenamefont {Li},
  \citenamefont {Krali}, \citenamefont {Fobelets}, \citenamefont {Cheng},\ and\
  \citenamefont {Wang}}]{ammonia1}%
  \BibitemOpen
  \bibfield  {author} {\bibinfo {author} {\bibfnamefont {C.}~\bibnamefont
  {Li}}, \bibinfo {author} {\bibfnamefont {E.}~\bibnamefont {Krali}}, \bibinfo
  {author} {\bibfnamefont {K.}~\bibnamefont {Fobelets}}, \bibinfo {author}
  {\bibfnamefont {B.}~\bibnamefont {Cheng}}, \ and\ \bibinfo {author}
  {\bibfnamefont {Q.}~\bibnamefont {Wang}},\ }\href {\doibase
  10.1063/1.4768692} {\bibfield  {journal} {\bibinfo  {journal} {Applied
  Physics Letters}\ }\textbf {\bibinfo {volume} {101}},\ \bibinfo {pages}
  {222101} (\bibinfo {year} {2012})},\ \Eprint
  {http://arxiv.org/abs/https://doi.org/10.1063/1.4768692}
  {https://doi.org/10.1063/1.4768692} \BibitemShut {NoStop}%
\bibitem [{\citenamefont {Li}\ \emph {et~al.}(2013)\citenamefont {Li},
  \citenamefont {Zhang}, \citenamefont {Fobelets}, \citenamefont {Zheng},
  \citenamefont {Xue}, \citenamefont {Zuo}, \citenamefont {Cheng},\ and\
  \citenamefont {Wang}}]{ammonia2}%
  \BibitemOpen
  \bibfield  {author} {\bibinfo {author} {\bibfnamefont {C.}~\bibnamefont
  {Li}}, \bibinfo {author} {\bibfnamefont {C.}~\bibnamefont {Zhang}}, \bibinfo
  {author} {\bibfnamefont {K.}~\bibnamefont {Fobelets}}, \bibinfo {author}
  {\bibfnamefont {J.}~\bibnamefont {Zheng}}, \bibinfo {author} {\bibfnamefont
  {C.}~\bibnamefont {Xue}}, \bibinfo {author} {\bibfnamefont {Y.}~\bibnamefont
  {Zuo}}, \bibinfo {author} {\bibfnamefont {B.}~\bibnamefont {Cheng}}, \ and\
  \bibinfo {author} {\bibfnamefont {Q.}~\bibnamefont {Wang}},\ }\href {\doibase
  10.1063/1.4827184} {\bibfield  {journal} {\bibinfo  {journal} {Journal of
  Applied Physics}\ }\textbf {\bibinfo {volume} {114}},\ \bibinfo {pages}
  {173702} (\bibinfo {year} {2013})},\ \Eprint
  {http://arxiv.org/abs/https://doi.org/10.1063/1.4827184}
  {https://doi.org/10.1063/1.4827184} \BibitemShut {NoStop}%
\bibitem [{\citenamefont {Elzerman}\ \emph {et~al.}(2004)\citenamefont
  {Elzerman}, \citenamefont {Hanson}, \citenamefont {Willems~van Beveren},
  \citenamefont {Witkamp}, \citenamefont {Vandersypen},\ and\ \citenamefont
  {Kouwenhoven}}]{Elzerman2004}%
  \BibitemOpen
  \bibfield  {author} {\bibinfo {author} {\bibfnamefont {J.~M.}\ \bibnamefont
  {Elzerman}}, \bibinfo {author} {\bibfnamefont {R.}~\bibnamefont {Hanson}},
  \bibinfo {author} {\bibfnamefont {L.~H.}\ \bibnamefont {Willems~van
  Beveren}}, \bibinfo {author} {\bibfnamefont {B.}~\bibnamefont {Witkamp}},
  \bibinfo {author} {\bibfnamefont {L.~M.~K.}\ \bibnamefont {Vandersypen}}, \
  and\ \bibinfo {author} {\bibfnamefont {L.~P.}\ \bibnamefont {Kouwenhoven}},\
  }\href {\doibase 10.1038/nature02693} {\bibfield  {journal} {\bibinfo
  {journal} {Nature}\ }\textbf {\bibinfo {volume} {430}},\ \bibinfo {pages}
  {431} (\bibinfo {year} {2004})}\BibitemShut {NoStop}%
\bibitem [{\citenamefont {Schleser}\ \emph {et~al.}(2004)\citenamefont
  {Schleser}, \citenamefont {Ruh}, \citenamefont {Ihn}, \citenamefont
  {Ensslin}, \citenamefont {Driscoll},\ and\ \citenamefont {Gossard}}]{QPC}%
  \BibitemOpen
  \bibfield  {author} {\bibinfo {author} {\bibfnamefont {R.}~\bibnamefont
  {Schleser}}, \bibinfo {author} {\bibfnamefont {E.}~\bibnamefont {Ruh}},
  \bibinfo {author} {\bibfnamefont {T.}~\bibnamefont {Ihn}}, \bibinfo {author}
  {\bibfnamefont {K.}~\bibnamefont {Ensslin}}, \bibinfo {author} {\bibfnamefont
  {D.~C.}\ \bibnamefont {Driscoll}}, \ and\ \bibinfo {author} {\bibfnamefont
  {A.~C.}\ \bibnamefont {Gossard}},\ }\href {\doibase 10.1063/1.1784875}
  {\bibfield  {journal} {\bibinfo  {journal} {Applied Physics Letters}\
  }\textbf {\bibinfo {volume} {85}},\ \bibinfo {pages} {2005} (\bibinfo {year}
  {2004})},\ \Eprint {http://arxiv.org/abs/https://doi.org/10.1063/1.1784875}
  {https://doi.org/10.1063/1.1784875} \BibitemShut {NoStop}%
\bibitem [{\citenamefont {Feng}\ \emph {et~al.}(1986)\citenamefont {Feng},
  \citenamefont {Lee},\ and\ \citenamefont {Stone}}]{PhysRevLett.56.1960}%
  \BibitemOpen
  \bibfield  {author} {\bibinfo {author} {\bibfnamefont {S.}~\bibnamefont
  {Feng}}, \bibinfo {author} {\bibfnamefont {P.~A.}\ \bibnamefont {Lee}}, \
  and\ \bibinfo {author} {\bibfnamefont {A.~D.}\ \bibnamefont {Stone}},\ }\href
  {\doibase 10.1103/PhysRevLett.56.1960} {\bibfield  {journal} {\bibinfo
  {journal} {Phys. Rev. Lett.}\ }\textbf {\bibinfo {volume} {56}},\ \bibinfo
  {pages} {1960} (\bibinfo {year} {1986})}\BibitemShut {NoStop}%
\bibitem [{\citenamefont {Hershfield}(1988)}]{PhysRevB.37.8557}%
  \BibitemOpen
  \bibfield  {author} {\bibinfo {author} {\bibfnamefont {S.}~\bibnamefont
  {Hershfield}},\ }\href {\doibase 10.1103/PhysRevB.37.8557} {\bibfield
  {journal} {\bibinfo  {journal} {Phys. Rev. B}\ }\textbf {\bibinfo {volume}
  {37}},\ \bibinfo {pages} {8557} (\bibinfo {year} {1988})}\BibitemShut
  {NoStop}%
\bibitem [{\citenamefont {Meisenheimer}\ and\ \citenamefont
  {Giordano}(1989)}]{PhysRevB.39.9929}%
  \BibitemOpen
  \bibfield  {author} {\bibinfo {author} {\bibfnamefont {T.~L.}\ \bibnamefont
  {Meisenheimer}}\ and\ \bibinfo {author} {\bibfnamefont {N.}~\bibnamefont
  {Giordano}},\ }\href {\doibase 10.1103/PhysRevB.39.9929} {\bibfield
  {journal} {\bibinfo  {journal} {Phys. Rev. B}\ }\textbf {\bibinfo {volume}
  {39}},\ \bibinfo {pages} {9929} (\bibinfo {year} {1989})}\BibitemShut
  {NoStop}%
\bibitem [{\citenamefont {Washburn}\ and\ \citenamefont
  {Webb}(1992)}]{Washburn_1992}%
  \BibitemOpen
  \bibfield  {author} {\bibinfo {author} {\bibfnamefont {S.}~\bibnamefont
  {Washburn}}\ and\ \bibinfo {author} {\bibfnamefont {R.~A.}\ \bibnamefont
  {Webb}},\ }\href {\doibase 10.1088/0034-4885/55/8/004} {\bibfield  {journal}
  {\bibinfo  {journal} {Reports on Progress in Physics}\ }\textbf {\bibinfo
  {volume} {55}},\ \bibinfo {pages} {1311} (\bibinfo {year}
  {1992})}\BibitemShut {NoStop}%
\bibitem [{\citenamefont {Lundstrom}\ and\ \citenamefont
  {Jeong}(2012)}]{Lundstrom}%
  \BibitemOpen
  \bibfield  {author} {\bibinfo {author} {\bibfnamefont {M.}~\bibnamefont
  {Lundstrom}}\ and\ \bibinfo {author} {\bibfnamefont {C.}~\bibnamefont
  {Jeong}},\ }\href {https://books.google.com.au/books?id=Kyg8DQAAQBAJ} {\emph
  {\bibinfo {title} {Near-Equilibrium Transport: Fundamentals and
  Applications}}},\ Lessons from Nanoscience: A Lecture Notes Series\ (\bibinfo
   {publisher} {World Scientific Publishing Company},\ \bibinfo {year}
  {2012})\BibitemShut {NoStop}%
\bibitem [{\citenamefont {Markussen}\ \emph {et~al.}(2006)\citenamefont
  {Markussen}, \citenamefont {Rurali}, \citenamefont {Brandbyge},\ and\
  \citenamefont {Jauho}}]{Jauho}%
  \BibitemOpen
  \bibfield  {author} {\bibinfo {author} {\bibfnamefont {T.}~\bibnamefont
  {Markussen}}, \bibinfo {author} {\bibfnamefont {R.}~\bibnamefont {Rurali}},
  \bibinfo {author} {\bibfnamefont {M.}~\bibnamefont {Brandbyge}}, \ and\
  \bibinfo {author} {\bibfnamefont {A.-P.}\ \bibnamefont {Jauho}},\ }\href
  {\doibase 10.1103/PhysRevB.74.245313} {\bibfield  {journal} {\bibinfo
  {journal} {Phys. Rev. B}\ }\textbf {\bibinfo {volume} {74}},\ \bibinfo
  {pages} {245313} (\bibinfo {year} {2006})}\BibitemShut {NoStop}%
\bibitem [{\citenamefont {Mil'nikov}\ \emph {et~al.}(2012)\citenamefont
  {Mil'nikov}, \citenamefont {Mori},\ and\ \citenamefont
  {Kamakura}}]{PhysRevB.85.035317}%
  \BibitemOpen
  \bibfield  {author} {\bibinfo {author} {\bibfnamefont {G.}~\bibnamefont
  {Mil'nikov}}, \bibinfo {author} {\bibfnamefont {N.}~\bibnamefont {Mori}}, \
  and\ \bibinfo {author} {\bibfnamefont {Y.}~\bibnamefont {Kamakura}},\ }\href
  {\doibase 10.1103/PhysRevB.85.035317} {\bibfield  {journal} {\bibinfo
  {journal} {Phys. Rev. B}\ }\textbf {\bibinfo {volume} {85}},\ \bibinfo
  {pages} {035317} (\bibinfo {year} {2012})}\BibitemShut {NoStop}%
\bibitem [{\citenamefont {Huang}\ \emph {et~al.}(2018)\citenamefont {Huang},
  \citenamefont {Ilatikhameneh}, \citenamefont {Povolotskyi},\ and\
  \citenamefont {Klimeck}}]{Huang_Klimeck}%
  \BibitemOpen
  \bibfield  {author} {\bibinfo {author} {\bibfnamefont {J.~Z.}\ \bibnamefont
  {Huang}}, \bibinfo {author} {\bibfnamefont {H.}~\bibnamefont
  {Ilatikhameneh}}, \bibinfo {author} {\bibfnamefont {M.}~\bibnamefont
  {Povolotskyi}}, \ and\ \bibinfo {author} {\bibfnamefont {G.}~\bibnamefont
  {Klimeck}},\ }\href {\doibase 10.1063/1.5010238} {\bibfield  {journal}
  {\bibinfo  {journal} {Journal of Applied Physics}\ }\textbf {\bibinfo
  {volume} {123}},\ \bibinfo {pages} {044303} (\bibinfo {year} {2018})},\
  \Eprint {http://arxiv.org/abs/https://doi.org/10.1063/1.5010238}
  {https://doi.org/10.1063/1.5010238} \BibitemShut {NoStop}%
\bibitem [{\citenamefont {Sun}\ \emph {et~al.}(2016)\citenamefont {Sun},
  \citenamefont {Ryno}, \citenamefont {Zhong}, \citenamefont {Ravva},
  \citenamefont {Sun}, \citenamefont {Körzdörfer},\ and\ \citenamefont
  {Brédas}}]{pcm2}%
  \BibitemOpen
  \bibfield  {author} {\bibinfo {author} {\bibfnamefont {H.}~\bibnamefont
  {Sun}}, \bibinfo {author} {\bibfnamefont {S.}~\bibnamefont {Ryno}}, \bibinfo
  {author} {\bibfnamefont {C.}~\bibnamefont {Zhong}}, \bibinfo {author}
  {\bibfnamefont {M.~K.}\ \bibnamefont {Ravva}}, \bibinfo {author}
  {\bibfnamefont {Z.}~\bibnamefont {Sun}}, \bibinfo {author} {\bibfnamefont
  {T.}~\bibnamefont {Körzdörfer}}, \ and\ \bibinfo {author} {\bibfnamefont
  {J.-L.}\ \bibnamefont {Brédas}},\ }\href {\doibase 10.1021/acs.jctc.6b00225}
  {\bibfield  {journal} {\bibinfo  {journal} {Journal of Chemical Theory and
  Computation}\ }\textbf {\bibinfo {volume} {12}},\ \bibinfo {pages} {2906}
  (\bibinfo {year} {2016})},\ \bibinfo {note} {pMID: 27183355},\ \Eprint
  {http://arxiv.org/abs/https://doi.org/10.1021/acs.jctc.6b00225}
  {https://doi.org/10.1021/acs.jctc.6b00225} \BibitemShut {NoStop}%
\bibitem [{\citenamefont {Liu}\ \emph {et~al.}(1994)\citenamefont {Liu},
  \citenamefont {Biegelsen}, \citenamefont {Ponce}, \citenamefont {Johnson},\
  and\ \citenamefont {Pease}}]{doi:10.1063/1.111914}%
  \BibitemOpen
  \bibfield  {author} {\bibinfo {author} {\bibfnamefont {H.~I.}\ \bibnamefont
  {Liu}}, \bibinfo {author} {\bibfnamefont {D.~K.}\ \bibnamefont {Biegelsen}},
  \bibinfo {author} {\bibfnamefont {F.~A.}\ \bibnamefont {Ponce}}, \bibinfo
  {author} {\bibfnamefont {N.~M.}\ \bibnamefont {Johnson}}, \ and\ \bibinfo
  {author} {\bibfnamefont {R.~F.~W.}\ \bibnamefont {Pease}},\ }\href {\doibase
  10.1063/1.111914} {\bibfield  {journal} {\bibinfo  {journal} {Applied Physics
  Letters}\ }\textbf {\bibinfo {volume} {64}},\ \bibinfo {pages} {1383}
  (\bibinfo {year} {1994})},\ \Eprint
  {http://arxiv.org/abs/https://doi.org/10.1063/1.111914}
  {https://doi.org/10.1063/1.111914} \BibitemShut {NoStop}%
\bibitem [{\citenamefont {{Hiramoto}}\ \emph {et~al.}(2016)\citenamefont
  {{Hiramoto}}, \citenamefont {{Mizutani}}, \citenamefont {{Saraya}},
  \citenamefont {{Takeuchi}},\ and\ \citenamefont {{Kobayashi}}}]{7998896}%
  \BibitemOpen
  \bibfield  {author} {\bibinfo {author} {\bibfnamefont {T.}~\bibnamefont
  {{Hiramoto}}}, \bibinfo {author} {\bibfnamefont {T.}~\bibnamefont
  {{Mizutani}}}, \bibinfo {author} {\bibfnamefont {T.}~\bibnamefont
  {{Saraya}}}, \bibinfo {author} {\bibfnamefont {K.}~\bibnamefont
  {{Takeuchi}}}, \ and\ \bibinfo {author} {\bibfnamefont {M.}~\bibnamefont
  {{Kobayashi}}},\ }in\ \href {\doibase 10.1109/ICSICT.2016.7998896} {\emph
  {\bibinfo {booktitle} {2016 13th IEEE International Conference on Solid-State
  and Integrated Circuit Technology (ICSICT)}}}\ (\bibinfo {year} {2016})\ pp.\
  \bibinfo {pages} {272--274}\BibitemShut {NoStop}%
\bibitem [{\citenamefont {Weng}\ \emph {et~al.}(2018)\citenamefont {Weng},
  \citenamefont {Hennes}, \citenamefont {Coati}, \citenamefont {Vlad},
  \citenamefont {Garreau}, \citenamefont {Sauvage-Simkin}, \citenamefont
  {Fonda}, \citenamefont {Patriarche}, \citenamefont {Demaille}, \citenamefont
  {Vidal},\ and\ \citenamefont {Zheng}}]{PhysRevMaterials.2.106003}%
  \BibitemOpen
  \bibfield  {author} {\bibinfo {author} {\bibfnamefont {X.}~\bibnamefont
  {Weng}}, \bibinfo {author} {\bibfnamefont {M.}~\bibnamefont {Hennes}},
  \bibinfo {author} {\bibfnamefont {A.}~\bibnamefont {Coati}}, \bibinfo
  {author} {\bibfnamefont {A.}~\bibnamefont {Vlad}}, \bibinfo {author}
  {\bibfnamefont {Y.}~\bibnamefont {Garreau}}, \bibinfo {author} {\bibfnamefont
  {M.}~\bibnamefont {Sauvage-Simkin}}, \bibinfo {author} {\bibfnamefont
  {E.}~\bibnamefont {Fonda}}, \bibinfo {author} {\bibfnamefont
  {G.}~\bibnamefont {Patriarche}}, \bibinfo {author} {\bibfnamefont
  {D.}~\bibnamefont {Demaille}}, \bibinfo {author} {\bibfnamefont
  {F.}~\bibnamefont {Vidal}}, \ and\ \bibinfo {author} {\bibfnamefont
  {Y.}~\bibnamefont {Zheng}},\ }\href {\doibase
  10.1103/PhysRevMaterials.2.106003} {\bibfield  {journal} {\bibinfo  {journal}
  {Phys. Rev. Materials}\ }\textbf {\bibinfo {volume} {2}},\ \bibinfo {pages}
  {106003} (\bibinfo {year} {2018})}\BibitemShut {NoStop}%
\bibitem [{\citenamefont {Zhou}\ \emph {et~al.}(2015)\citenamefont {Zhou},
  \citenamefont {Liu}, \citenamefont {Gao}, \citenamefont {Chen}, \citenamefont
  {Senz}, \citenamefont {Zhang},\ and\ \citenamefont {Liu}}]{Zhou_2015}%
  \BibitemOpen
  \bibfield  {author} {\bibinfo {author} {\bibfnamefont {Q.}~\bibnamefont
  {Zhou}}, \bibinfo {author} {\bibfnamefont {L.}~\bibnamefont {Liu}}, \bibinfo
  {author} {\bibfnamefont {X.}~\bibnamefont {Gao}}, \bibinfo {author}
  {\bibfnamefont {L.}~\bibnamefont {Chen}}, \bibinfo {author} {\bibfnamefont
  {S.}~\bibnamefont {Senz}}, \bibinfo {author} {\bibfnamefont {Z.}~\bibnamefont
  {Zhang}}, \ and\ \bibinfo {author} {\bibfnamefont {J.}~\bibnamefont {Liu}},\
  }\href {\doibase 10.1088/0957-4484/26/7/075707} {\bibfield  {journal}
  {\bibinfo  {journal} {Nanotechnology}\ }\textbf {\bibinfo {volume} {26}},\
  \bibinfo {pages} {075707} (\bibinfo {year} {2015})}\BibitemShut {NoStop}%
\bibitem [{\citenamefont {Weber}\ \emph {et~al.}(2012)\citenamefont {Weber},
  \citenamefont {Mahapatra}, \citenamefont {Ryu}, \citenamefont {Lee},
  \citenamefont {Fuhrer}, \citenamefont {Reusch}, \citenamefont {Thompson},
  \citenamefont {Lee}, \citenamefont {Klimeck}, \citenamefont {Hollenberg},\
  and\ \citenamefont {Simmons}}]{Weber64}%
  \BibitemOpen
  \bibfield  {author} {\bibinfo {author} {\bibfnamefont {B.}~\bibnamefont
  {Weber}}, \bibinfo {author} {\bibfnamefont {S.}~\bibnamefont {Mahapatra}},
  \bibinfo {author} {\bibfnamefont {H.}~\bibnamefont {Ryu}}, \bibinfo {author}
  {\bibfnamefont {S.}~\bibnamefont {Lee}}, \bibinfo {author} {\bibfnamefont
  {A.}~\bibnamefont {Fuhrer}}, \bibinfo {author} {\bibfnamefont {T.~C.~G.}\
  \bibnamefont {Reusch}}, \bibinfo {author} {\bibfnamefont {D.~L.}\
  \bibnamefont {Thompson}}, \bibinfo {author} {\bibfnamefont {W.~C.~T.}\
  \bibnamefont {Lee}}, \bibinfo {author} {\bibfnamefont {G.}~\bibnamefont
  {Klimeck}}, \bibinfo {author} {\bibfnamefont {L.~C.~L.}\ \bibnamefont
  {Hollenberg}}, \ and\ \bibinfo {author} {\bibfnamefont {M.~Y.}\ \bibnamefont
  {Simmons}},\ }\href {\doibase 10.1126/science.1214319} {\bibfield  {journal}
  {\bibinfo  {journal} {Science}\ }\textbf {\bibinfo {volume} {335}},\ \bibinfo
  {pages} {64} (\bibinfo {year} {2012})},\ \Eprint
  {http://arxiv.org/abs/https://science.sciencemag.org/content/335/6064/64.full.pdf}
  {https://science.sciencemag.org/content/335/6064/64.full.pdf} \BibitemShut
  {NoStop}%
\bibitem [{\citenamefont {Meir}\ and\ \citenamefont
  {Wingreen}(1992)}]{Landauer}%
  \BibitemOpen
  \bibfield  {author} {\bibinfo {author} {\bibfnamefont {Y.}~\bibnamefont
  {Meir}}\ and\ \bibinfo {author} {\bibfnamefont {N.~S.}\ \bibnamefont
  {Wingreen}},\ }\href {\doibase 10.1103/PhysRevLett.68.2512} {\bibfield
  {journal} {\bibinfo  {journal} {Phys. Rev. Lett.}\ }\textbf {\bibinfo
  {volume} {68}},\ \bibinfo {pages} {2512} (\bibinfo {year}
  {1992})}\BibitemShut {NoStop}%
\bibitem [{\citenamefont {Todorov}(1996)}]{Todorov}%
  \BibitemOpen
  \bibfield  {author} {\bibinfo {author} {\bibfnamefont {T.~N.}\ \bibnamefont
  {Todorov}},\ }\href {\doibase 10.1103/PhysRevB.54.5801} {\bibfield  {journal}
  {\bibinfo  {journal} {Phys. Rev. B}\ }\textbf {\bibinfo {volume} {54}},\
  \bibinfo {pages} {5801} (\bibinfo {year} {1996})}\BibitemShut {NoStop}%
\bibitem [{\citenamefont {Lu}\ \emph {et~al.}(2005)\citenamefont {Lu},
  \citenamefont {Xiang}, \citenamefont {Timko}, \citenamefont {Wu},\ and\
  \citenamefont {Lieber}}]{Lu10046}%
  \BibitemOpen
  \bibfield  {author} {\bibinfo {author} {\bibfnamefont {W.}~\bibnamefont
  {Lu}}, \bibinfo {author} {\bibfnamefont {J.}~\bibnamefont {Xiang}}, \bibinfo
  {author} {\bibfnamefont {B.~P.}\ \bibnamefont {Timko}}, \bibinfo {author}
  {\bibfnamefont {Y.}~\bibnamefont {Wu}}, \ and\ \bibinfo {author}
  {\bibfnamefont {C.~M.}\ \bibnamefont {Lieber}},\ }\href {\doibase
  10.1073/pnas.0504581102} {\bibfield  {journal} {\bibinfo  {journal}
  {Proceedings of the National Academy of Sciences}\ }\textbf {\bibinfo
  {volume} {102}},\ \bibinfo {pages} {10046} (\bibinfo {year} {2005})},\
  \Eprint
  {http://arxiv.org/abs/https://www.pnas.org/content/102/29/10046.full.pdf}
  {https://www.pnas.org/content/102/29/10046.full.pdf} \BibitemShut {NoStop}%
\bibitem [{\citenamefont {Graf}\ and\ \citenamefont {Vogl}(1995)}]{Graf}%
  \BibitemOpen
  \bibfield  {author} {\bibinfo {author} {\bibfnamefont {M.}~\bibnamefont
  {Graf}}\ and\ \bibinfo {author} {\bibfnamefont {P.}~\bibnamefont {Vogl}},\
  }\href {\doibase 10.1103/PhysRevB.51.4940} {\bibfield  {journal} {\bibinfo
  {journal} {Phys. Rev. B}\ }\textbf {\bibinfo {volume} {51}},\ \bibinfo
  {pages} {4940} (\bibinfo {year} {1995})}\BibitemShut {NoStop}%
\bibitem [{\citenamefont {Jancu}\ \emph {et~al.}(1998)\citenamefont {Jancu},
  \citenamefont {Scholz}, \citenamefont {Beltram},\ and\ \citenamefont
  {Bassani}}]{PhysRevB.57.6493}%
  \BibitemOpen
  \bibfield  {author} {\bibinfo {author} {\bibfnamefont {J.-M.}\ \bibnamefont
  {Jancu}}, \bibinfo {author} {\bibfnamefont {R.}~\bibnamefont {Scholz}},
  \bibinfo {author} {\bibfnamefont {F.}~\bibnamefont {Beltram}}, \ and\
  \bibinfo {author} {\bibfnamefont {F.}~\bibnamefont {Bassani}},\ }\href
  {\doibase 10.1103/PhysRevB.57.6493} {\bibfield  {journal} {\bibinfo
  {journal} {Phys. Rev. B}\ }\textbf {\bibinfo {volume} {57}},\ \bibinfo
  {pages} {6493} (\bibinfo {year} {1998})}\BibitemShut {NoStop}%
\bibitem [{\citenamefont {Luisier}\ \emph {et~al.}(2006)\citenamefont
  {Luisier}, \citenamefont {Schenk}, \citenamefont {Fichtner},\ and\
  \citenamefont {Klimeck}}]{PhysRevB.74.205323}%
  \BibitemOpen
  \bibfield  {author} {\bibinfo {author} {\bibfnamefont {M.}~\bibnamefont
  {Luisier}}, \bibinfo {author} {\bibfnamefont {A.}~\bibnamefont {Schenk}},
  \bibinfo {author} {\bibfnamefont {W.}~\bibnamefont {Fichtner}}, \ and\
  \bibinfo {author} {\bibfnamefont {G.}~\bibnamefont {Klimeck}},\ }\href
  {\doibase 10.1103/PhysRevB.74.205323} {\bibfield  {journal} {\bibinfo
  {journal} {Phys. Rev. B}\ }\textbf {\bibinfo {volume} {74}},\ \bibinfo
  {pages} {205323} (\bibinfo {year} {2006})}\BibitemShut {NoStop}%
\bibitem [{\citenamefont {Podolskiy}\ and\ \citenamefont
  {Vogl}(2004)}]{PhysRevB.69.233101}%
  \BibitemOpen
  \bibfield  {author} {\bibinfo {author} {\bibfnamefont {A.~V.}\ \bibnamefont
  {Podolskiy}}\ and\ \bibinfo {author} {\bibfnamefont {P.}~\bibnamefont
  {Vogl}},\ }\href {\doibase 10.1103/PhysRevB.69.233101} {\bibfield  {journal}
  {\bibinfo  {journal} {Phys. Rev. B}\ }\textbf {\bibinfo {volume} {69}},\
  \bibinfo {pages} {233101} (\bibinfo {year} {2004})}\BibitemShut {NoStop}%
\bibitem [{nan(2018)}]{nanonet}%
  \BibitemOpen
  \href@noop {} {\enquote {\bibinfo {title} {Nanonet: extendable python
  framework for the electronic structure computations based on the
  tight-binding method},}\ }\bibinfo {howpublished}
  {\url{https://github.com/freude/NanoNet}} (\bibinfo {year}
  {2018})\BibitemShut {NoStop}%
\bibitem [{\citenamefont {Anantram}\ \emph {et~al.}(2008)\citenamefont
  {Anantram}, \citenamefont {Lundstrom},\ and\ \citenamefont
  {Nikonov}}]{gf_recursive}%
  \BibitemOpen
  \bibfield  {author} {\bibinfo {author} {\bibfnamefont {M.~P.}\ \bibnamefont
  {Anantram}}, \bibinfo {author} {\bibfnamefont {M.~S.}\ \bibnamefont
  {Lundstrom}}, \ and\ \bibinfo {author} {\bibfnamefont {D.~E.}\ \bibnamefont
  {Nikonov}},\ }\href {\doibase 10.1109/JPROC.2008.927355} {\bibfield
  {journal} {\bibinfo  {journal} {Proceedings of the IEEE}\ }\textbf {\bibinfo
  {volume} {96}},\ \bibinfo {pages} {1511} (\bibinfo {year}
  {2008})}\BibitemShut {NoStop}%
\bibitem [{\citenamefont {Wimmer}(2009)}]{Wimmer}%
  \BibitemOpen
  \bibfield  {author} {\bibinfo {author} {\bibfnamefont {M.}~\bibnamefont
  {Wimmer}},\ }\href {https://books.google.com.au/books?id=WQtZngEACAAJ} {\emph
  {\bibinfo {title} {Quantum Transport in Nanostructures: From Computational
  Concepts to Spintronics in Graphene and Magnetic Tunnel Junctions}}},\
  Dissertationsreihe der Fakult{\"a}t f{\"u}r Physik der Universit{\"a}t
  Regensburg\ (\bibinfo  {publisher} {Univ.-Verlag Regensburg},\ \bibinfo
  {year} {2009})\BibitemShut {NoStop}%
\bibitem [{\citenamefont {Tersigni}\ \emph {et~al.}(2006)\citenamefont
  {Tersigni}, \citenamefont {Shi}, \citenamefont {Jiang},\ and\ \citenamefont
  {Qin}}]{structure}%
  \BibitemOpen
  \bibfield  {author} {\bibinfo {author} {\bibfnamefont {A.}~\bibnamefont
  {Tersigni}}, \bibinfo {author} {\bibfnamefont {J.}~\bibnamefont {Shi}},
  \bibinfo {author} {\bibfnamefont {D.~T.}\ \bibnamefont {Jiang}}, \ and\
  \bibinfo {author} {\bibfnamefont {X.~R.}\ \bibnamefont {Qin}},\ }\href
  {\doibase 10.1103/PhysRevB.74.205326} {\bibfield  {journal} {\bibinfo
  {journal} {Phys. Rev. B}\ }\textbf {\bibinfo {volume} {74}},\ \bibinfo
  {pages} {205326} (\bibinfo {year} {2006})}\BibitemShut {NoStop}%
\bibitem [{\citenamefont {Ortmann}\ \emph {et~al.}(2009)\citenamefont
  {Ortmann}, \citenamefont {Bechstedt},\ and\ \citenamefont
  {Hannewald}}]{PhysRevB.79.235206}%
  \BibitemOpen
  \bibfield  {author} {\bibinfo {author} {\bibfnamefont {F.}~\bibnamefont
  {Ortmann}}, \bibinfo {author} {\bibfnamefont {F.}~\bibnamefont {Bechstedt}},
  \ and\ \bibinfo {author} {\bibfnamefont {K.}~\bibnamefont {Hannewald}},\
  }\href {\doibase 10.1103/PhysRevB.79.235206} {\bibfield  {journal} {\bibinfo
  {journal} {Phys. Rev. B}\ }\textbf {\bibinfo {volume} {79}},\ \bibinfo
  {pages} {235206} (\bibinfo {year} {2009})}\BibitemShut {NoStop}%
\bibitem [{\citenamefont {Hannewald}\ \emph {et~al.}(2004)\citenamefont
  {Hannewald}, \citenamefont {Stojanovi\ifmmode~\acute{c}\else \'{c}\fi{}},
  \citenamefont {Schellekens}, \citenamefont {Bobbert}, \citenamefont
  {Kresse},\ and\ \citenamefont {Hafner}}]{PhysRevB.69.075211}%
  \BibitemOpen
  \bibfield  {author} {\bibinfo {author} {\bibfnamefont {K.}~\bibnamefont
  {Hannewald}}, \bibinfo {author} {\bibfnamefont {V.~M.}\ \bibnamefont
  {Stojanovi\ifmmode~\acute{c}\else \'{c}\fi{}}}, \bibinfo {author}
  {\bibfnamefont {J.~M.~T.}\ \bibnamefont {Schellekens}}, \bibinfo {author}
  {\bibfnamefont {P.~A.}\ \bibnamefont {Bobbert}}, \bibinfo {author}
  {\bibfnamefont {G.}~\bibnamefont {Kresse}}, \ and\ \bibinfo {author}
  {\bibfnamefont {J.}~\bibnamefont {Hafner}},\ }\href {\doibase
  10.1103/PhysRevB.69.075211} {\bibfield  {journal} {\bibinfo  {journal} {Phys.
  Rev. B}\ }\textbf {\bibinfo {volume} {69}},\ \bibinfo {pages} {075211}
  (\bibinfo {year} {2004})}\BibitemShut {NoStop}%
\bibitem [{\citenamefont {Limketkai}\ \emph {et~al.}(2007)\citenamefont
  {Limketkai}, \citenamefont {Jadhav},\ and\ \citenamefont
  {Baldo}}]{PhysRevB.75.113203}%
  \BibitemOpen
  \bibfield  {author} {\bibinfo {author} {\bibfnamefont {B.~N.}\ \bibnamefont
  {Limketkai}}, \bibinfo {author} {\bibfnamefont {P.}~\bibnamefont {Jadhav}}, \
  and\ \bibinfo {author} {\bibfnamefont {M.~A.}\ \bibnamefont {Baldo}},\ }\href
  {\doibase 10.1103/PhysRevB.75.113203} {\bibfield  {journal} {\bibinfo
  {journal} {Phys. Rev. B}\ }\textbf {\bibinfo {volume} {75}},\ \bibinfo
  {pages} {113203} (\bibinfo {year} {2007})}\BibitemShut {NoStop}%
\bibitem [{\citenamefont {Pope}\ and\ \citenamefont
  {Swenberg}(1999)}]{book:pope}%
  \BibitemOpen
  \bibfield  {author} {\bibinfo {author} {\bibfnamefont {M.}~\bibnamefont
  {Pope}}\ and\ \bibinfo {author} {\bibfnamefont {C.}~\bibnamefont
  {Swenberg}},\ }\href {https://books.google.com.au/books?id=AZVUAAAAMAAJ}
  {\emph {\bibinfo {title} {Electronic Processes in Organic Crystals and
  Polymers}}},\ Monographs on the physics and chemistry of materials\ (\bibinfo
   {publisher} {Oxford University Press},\ \bibinfo {year} {1999})\BibitemShut
  {NoStop}%
\bibitem [{\citenamefont {Andersen}\ \emph {et~al.}(2015)\citenamefont
  {Andersen}, \citenamefont {Latini},\ and\ \citenamefont {Thygesen}}]{genome}%
  \BibitemOpen
  \bibfield  {author} {\bibinfo {author} {\bibfnamefont {K.}~\bibnamefont
  {Andersen}}, \bibinfo {author} {\bibfnamefont {S.}~\bibnamefont {Latini}}, \
  and\ \bibinfo {author} {\bibfnamefont {K.~S.}\ \bibnamefont {Thygesen}},\
  }\href {\doibase 10.1021/acs.nanolett.5b01251} {\bibfield  {journal}
  {\bibinfo  {journal} {Nano Letters}\ }\textbf {\bibinfo {volume} {15}},\
  \bibinfo {pages} {4616} (\bibinfo {year} {2015})},\ \bibinfo {note} {pMID:
  26047386},\ \Eprint
  {http://arxiv.org/abs/https://doi.org/10.1021/acs.nanolett.5b01251}
  {https://doi.org/10.1021/acs.nanolett.5b01251} \BibitemShut {NoStop}%
\bibitem [{\citenamefont {Tomasi}\ \emph {et~al.}(2005)\citenamefont {Tomasi},
  \citenamefont {Mennucci},\ and\ \citenamefont {Cammi}}]{pcm}%
  \BibitemOpen
  \bibfield  {author} {\bibinfo {author} {\bibfnamefont {J.}~\bibnamefont
  {Tomasi}}, \bibinfo {author} {\bibfnamefont {B.}~\bibnamefont {Mennucci}}, \
  and\ \bibinfo {author} {\bibfnamefont {R.}~\bibnamefont {Cammi}},\ }\href
  {\doibase 10.1021/cr9904009} {\bibfield  {journal} {\bibinfo  {journal}
  {Chemical Reviews}\ }\textbf {\bibinfo {volume} {105}},\ \bibinfo {pages}
  {2999} (\bibinfo {year} {2005})},\ \bibinfo {note} {pMID: 16092826},\ \Eprint
  {http://arxiv.org/abs/https://doi.org/10.1021/cr9904009}
  {https://doi.org/10.1021/cr9904009} \BibitemShut {NoStop}%
\bibitem [{\citenamefont {Refaely-Abramson}\ \emph {et~al.}(2013)\citenamefont
  {Refaely-Abramson}, \citenamefont {Sharifzadeh}, \citenamefont {Jain},
  \citenamefont {Baer}, \citenamefont {Neaton},\ and\ \citenamefont
  {Kronik}}]{pcm1}%
  \BibitemOpen
  \bibfield  {author} {\bibinfo {author} {\bibfnamefont {S.}~\bibnamefont
  {Refaely-Abramson}}, \bibinfo {author} {\bibfnamefont {S.}~\bibnamefont
  {Sharifzadeh}}, \bibinfo {author} {\bibfnamefont {M.}~\bibnamefont {Jain}},
  \bibinfo {author} {\bibfnamefont {R.}~\bibnamefont {Baer}}, \bibinfo {author}
  {\bibfnamefont {J.~B.}\ \bibnamefont {Neaton}}, \ and\ \bibinfo {author}
  {\bibfnamefont {L.}~\bibnamefont {Kronik}},\ }\href {\doibase
  10.1103/PhysRevB.88.081204} {\bibfield  {journal} {\bibinfo  {journal} {Phys.
  Rev. B}\ }\textbf {\bibinfo {volume} {88}},\ \bibinfo {pages} {081204}
  (\bibinfo {year} {2013})}\BibitemShut {NoStop}%
\bibitem [{\citenamefont {Sebastian}\ \emph {et~al.}(1981)\citenamefont
  {Sebastian}, \citenamefont {Weiser},\ and\ \citenamefont
  {Bässler}}]{tet_eps}%
  \BibitemOpen
  \bibfield  {author} {\bibinfo {author} {\bibfnamefont {L.}~\bibnamefont
  {Sebastian}}, \bibinfo {author} {\bibfnamefont {G.}~\bibnamefont {Weiser}}, \
  and\ \bibinfo {author} {\bibfnamefont {H.}~\bibnamefont {Bässler}},\ }\href
  {\doibase https://doi.org/10.1016/0301-0104(81)85055-0} {\bibfield  {journal}
  {\bibinfo  {journal} {Chemical Physics}\ }\textbf {\bibinfo {volume} {61}},\
  \bibinfo {pages} {125 } (\bibinfo {year} {1981})}\BibitemShut {NoStop}%
\bibitem [{\citenamefont {Frisch}\ \emph {et~al.}(2009)\citenamefont {Frisch},
  \citenamefont {Trucks}, \citenamefont {Schlegel}, \citenamefont {Scuseria},
  \citenamefont {Robb}, \citenamefont {Cheeseman}, \citenamefont {Scalmani},
  \citenamefont {Barone}, \citenamefont {Mennucci}, \citenamefont {Petersson},
  \citenamefont {Nakatsuji}, \citenamefont {Caricato}, \citenamefont {Li},
  \citenamefont {Hratchian}, \citenamefont {Izmaylov}, \citenamefont {Bloino},
  \citenamefont {Zheng}, \citenamefont {Sonnenberg}, \citenamefont {Hada},
  \citenamefont {Ehara}, \citenamefont {Toyota}, \citenamefont {Fukuda},
  \citenamefont {Hasegawa}, \citenamefont {Ishida}, \citenamefont {Nakajima},
  \citenamefont {Honda}, \citenamefont {Kitao}, \citenamefont {Nakai},
  \citenamefont {Vreven}, \citenamefont {Montgomery}, \citenamefont {Peralta},
  \citenamefont {Ogliaro}, \citenamefont {Bearpark}, \citenamefont {Heyd},
  \citenamefont {Brothers}, \citenamefont {Kudin}, \citenamefont {Staroverov},
  \citenamefont {Kobayashi}, \citenamefont {Normand}, \citenamefont
  {Raghavachari}, \citenamefont {Rendell}, \citenamefont {Burant},
  \citenamefont {Iyengar}, \citenamefont {Tomasi}, \citenamefont {Cossi},
  \citenamefont {Rega}, \citenamefont {Millam}, \citenamefont {Klene},
  \citenamefont {Knox}, \citenamefont {Cross}, \citenamefont {Bakken},
  \citenamefont {Adamo}, \citenamefont {Jaramillo}, \citenamefont {Gomperts},
  \citenamefont {Stratmann}, \citenamefont {Yazyev}, \citenamefont {Austin},
  \citenamefont {Cammi}, \citenamefont {Pomelli}, \citenamefont {Ochterski},
  \citenamefont {Martin}, \citenamefont {Morokuma}, \citenamefont {Zakrzewski},
  \citenamefont {Voth}, \citenamefont {Salvador}, \citenamefont {Dannenberg},
  \citenamefont {Dapprich}, \citenamefont {Daniels}, \citenamefont {Farkas},
  \citenamefont {Foresman}, \citenamefont {Ortiz}, \citenamefont {Cioslowski},\
  and\ \citenamefont {Fox}}]{g09}%
  \BibitemOpen
  \bibfield  {author} {\bibinfo {author} {\bibfnamefont {M.~J.}\ \bibnamefont
  {Frisch}}, \bibinfo {author} {\bibfnamefont {G.~W.}\ \bibnamefont {Trucks}},
  \bibinfo {author} {\bibfnamefont {H.~B.}\ \bibnamefont {Schlegel}}, \bibinfo
  {author} {\bibfnamefont {G.~E.}\ \bibnamefont {Scuseria}}, \bibinfo {author}
  {\bibfnamefont {M.~A.}\ \bibnamefont {Robb}}, \bibinfo {author}
  {\bibfnamefont {J.~R.}\ \bibnamefont {Cheeseman}}, \bibinfo {author}
  {\bibfnamefont {G.}~\bibnamefont {Scalmani}}, \bibinfo {author}
  {\bibfnamefont {V.}~\bibnamefont {Barone}}, \bibinfo {author} {\bibfnamefont
  {B.}~\bibnamefont {Mennucci}}, \bibinfo {author} {\bibfnamefont {G.~A.}\
  \bibnamefont {Petersson}}, \bibinfo {author} {\bibfnamefont {H.}~\bibnamefont
  {Nakatsuji}}, \bibinfo {author} {\bibfnamefont {M.}~\bibnamefont {Caricato}},
  \bibinfo {author} {\bibfnamefont {X.}~\bibnamefont {Li}}, \bibinfo {author}
  {\bibfnamefont {H.~P.}\ \bibnamefont {Hratchian}}, \bibinfo {author}
  {\bibfnamefont {A.~F.}\ \bibnamefont {Izmaylov}}, \bibinfo {author}
  {\bibfnamefont {J.}~\bibnamefont {Bloino}}, \bibinfo {author} {\bibfnamefont
  {G.}~\bibnamefont {Zheng}}, \bibinfo {author} {\bibfnamefont {J.~L.}\
  \bibnamefont {Sonnenberg}}, \bibinfo {author} {\bibfnamefont
  {M.}~\bibnamefont {Hada}}, \bibinfo {author} {\bibfnamefont {M.}~\bibnamefont
  {Ehara}}, \bibinfo {author} {\bibfnamefont {K.}~\bibnamefont {Toyota}},
  \bibinfo {author} {\bibfnamefont {R.}~\bibnamefont {Fukuda}}, \bibinfo
  {author} {\bibfnamefont {J.}~\bibnamefont {Hasegawa}}, \bibinfo {author}
  {\bibfnamefont {M.}~\bibnamefont {Ishida}}, \bibinfo {author} {\bibfnamefont
  {T.}~\bibnamefont {Nakajima}}, \bibinfo {author} {\bibfnamefont
  {Y.}~\bibnamefont {Honda}}, \bibinfo {author} {\bibfnamefont
  {O.}~\bibnamefont {Kitao}}, \bibinfo {author} {\bibfnamefont
  {H.}~\bibnamefont {Nakai}}, \bibinfo {author} {\bibfnamefont
  {T.}~\bibnamefont {Vreven}}, \bibinfo {author} {\bibfnamefont {J.~A.}\
  \bibnamefont {Montgomery}, \bibfnamefont {{Jr.}}}, \bibinfo {author}
  {\bibfnamefont {J.~E.}\ \bibnamefont {Peralta}}, \bibinfo {author}
  {\bibfnamefont {F.}~\bibnamefont {Ogliaro}}, \bibinfo {author} {\bibfnamefont
  {M.}~\bibnamefont {Bearpark}}, \bibinfo {author} {\bibfnamefont {J.~J.}\
  \bibnamefont {Heyd}}, \bibinfo {author} {\bibfnamefont {E.}~\bibnamefont
  {Brothers}}, \bibinfo {author} {\bibfnamefont {K.~N.}\ \bibnamefont {Kudin}},
  \bibinfo {author} {\bibfnamefont {V.~N.}\ \bibnamefont {Staroverov}},
  \bibinfo {author} {\bibfnamefont {R.}~\bibnamefont {Kobayashi}}, \bibinfo
  {author} {\bibfnamefont {J.}~\bibnamefont {Normand}}, \bibinfo {author}
  {\bibfnamefont {K.}~\bibnamefont {Raghavachari}}, \bibinfo {author}
  {\bibfnamefont {A.}~\bibnamefont {Rendell}}, \bibinfo {author} {\bibfnamefont
  {J.~C.}\ \bibnamefont {Burant}}, \bibinfo {author} {\bibfnamefont {S.~S.}\
  \bibnamefont {Iyengar}}, \bibinfo {author} {\bibfnamefont {J.}~\bibnamefont
  {Tomasi}}, \bibinfo {author} {\bibfnamefont {M.}~\bibnamefont {Cossi}},
  \bibinfo {author} {\bibfnamefont {N.}~\bibnamefont {Rega}}, \bibinfo {author}
  {\bibfnamefont {J.~M.}\ \bibnamefont {Millam}}, \bibinfo {author}
  {\bibfnamefont {M.}~\bibnamefont {Klene}}, \bibinfo {author} {\bibfnamefont
  {J.~E.}\ \bibnamefont {Knox}}, \bibinfo {author} {\bibfnamefont {J.~B.}\
  \bibnamefont {Cross}}, \bibinfo {author} {\bibfnamefont {V.}~\bibnamefont
  {Bakken}}, \bibinfo {author} {\bibfnamefont {C.}~\bibnamefont {Adamo}},
  \bibinfo {author} {\bibfnamefont {J.}~\bibnamefont {Jaramillo}}, \bibinfo
  {author} {\bibfnamefont {R.}~\bibnamefont {Gomperts}}, \bibinfo {author}
  {\bibfnamefont {R.~E.}\ \bibnamefont {Stratmann}}, \bibinfo {author}
  {\bibfnamefont {O.}~\bibnamefont {Yazyev}}, \bibinfo {author} {\bibfnamefont
  {A.~J.}\ \bibnamefont {Austin}}, \bibinfo {author} {\bibfnamefont
  {R.}~\bibnamefont {Cammi}}, \bibinfo {author} {\bibfnamefont
  {C.}~\bibnamefont {Pomelli}}, \bibinfo {author} {\bibfnamefont {J.~W.}\
  \bibnamefont {Ochterski}}, \bibinfo {author} {\bibfnamefont {R.~L.}\
  \bibnamefont {Martin}}, \bibinfo {author} {\bibfnamefont {K.}~\bibnamefont
  {Morokuma}}, \bibinfo {author} {\bibfnamefont {V.~G.}\ \bibnamefont
  {Zakrzewski}}, \bibinfo {author} {\bibfnamefont {G.~A.}\ \bibnamefont
  {Voth}}, \bibinfo {author} {\bibfnamefont {P.}~\bibnamefont {Salvador}},
  \bibinfo {author} {\bibfnamefont {J.~J.}\ \bibnamefont {Dannenberg}},
  \bibinfo {author} {\bibfnamefont {S.}~\bibnamefont {Dapprich}}, \bibinfo
  {author} {\bibfnamefont {A.~D.}\ \bibnamefont {Daniels}}, \bibinfo {author}
  {\bibfnamefont {{\"O}.}~\bibnamefont {Farkas}}, \bibinfo {author}
  {\bibfnamefont {J.~B.}\ \bibnamefont {Foresman}}, \bibinfo {author}
  {\bibfnamefont {J.~V.}\ \bibnamefont {Ortiz}}, \bibinfo {author}
  {\bibfnamefont {J.}~\bibnamefont {Cioslowski}}, \ and\ \bibinfo {author}
  {\bibfnamefont {D.~J.}\ \bibnamefont {Fox}},\ }\href@noop {} {\enquote
  {\bibinfo {title} {Gaussian~09 {R}evision {E}.01},}\ } (\bibinfo {year}
  {2009}),\ \bibinfo {note} {gaussian Inc. Wallingford CT}\BibitemShut
  {NoStop}%
\bibitem [{\citenamefont {Chai}\ and\ \citenamefont
  {Head-Gordon}(2008)}]{wB97XD}%
  \BibitemOpen
  \bibfield  {author} {\bibinfo {author} {\bibfnamefont {J.-D.}\ \bibnamefont
  {Chai}}\ and\ \bibinfo {author} {\bibfnamefont {M.}~\bibnamefont
  {Head-Gordon}},\ }\href {\doibase 10.1039/B810189B} {\bibfield  {journal}
  {\bibinfo  {journal} {Phys. Chem. Chem. Phys.}\ }\textbf {\bibinfo {volume}
  {10}},\ \bibinfo {pages} {6615} (\bibinfo {year} {2008})}\BibitemShut
  {NoStop}%
\bibitem [{\citenamefont {Penn}(1962)}]{Penn}%
  \BibitemOpen
  \bibfield  {author} {\bibinfo {author} {\bibfnamefont {D.~R.}\ \bibnamefont
  {Penn}},\ }\href {\doibase 10.1103/PhysRev.128.2093} {\bibfield  {journal}
  {\bibinfo  {journal} {Phys. Rev.}\ }\textbf {\bibinfo {volume} {128}},\
  \bibinfo {pages} {2093} (\bibinfo {year} {1962})}\BibitemShut {NoStop}%
\bibitem [{\citenamefont {Wang}\ and\ \citenamefont {Zunger}(1994)}]{Zunger}%
  \BibitemOpen
  \bibfield  {author} {\bibinfo {author} {\bibfnamefont {L.-W.}\ \bibnamefont
  {Wang}}\ and\ \bibinfo {author} {\bibfnamefont {A.}~\bibnamefont {Zunger}},\
  }\href {\doibase 10.1103/PhysRevLett.73.1039} {\bibfield  {journal} {\bibinfo
   {journal} {Phys. Rev. Lett.}\ }\textbf {\bibinfo {volume} {73}},\ \bibinfo
  {pages} {1039} (\bibinfo {year} {1994})}\BibitemShut {NoStop}%
\bibitem [{\citenamefont {Tsu}\ \emph {et~al.}(1997)\citenamefont {Tsu},
  \citenamefont {Babić},\ and\ \citenamefont {Ioriatti}}]{Tsu}%
  \BibitemOpen
  \bibfield  {author} {\bibinfo {author} {\bibfnamefont {R.}~\bibnamefont
  {Tsu}}, \bibinfo {author} {\bibfnamefont {D.}~\bibnamefont {Babić}}, \ and\
  \bibinfo {author} {\bibfnamefont {L.}~\bibnamefont {Ioriatti}},\ }\href
  {\doibase 10.1063/1.365762} {\bibfield  {journal} {\bibinfo  {journal}
  {Journal of Applied Physics}\ }\textbf {\bibinfo {volume} {82}},\ \bibinfo
  {pages} {1327} (\bibinfo {year} {1997})},\ \Eprint
  {http://arxiv.org/abs/https://doi.org/10.1063/1.365762}
  {https://doi.org/10.1063/1.365762} \BibitemShut {NoStop}%
\bibitem [{\citenamefont {Kasap}\ and\ \citenamefont
  {Capper}(2017)}]{handbook}%
  \BibitemOpen
  \bibfield  {author} {\bibinfo {author} {\bibfnamefont {S.}~\bibnamefont
  {Kasap}}\ and\ \bibinfo {author} {\bibfnamefont {P.}~\bibnamefont {Capper}},\
  }\href {https://books.google.com.au/books?id=3JQ4DwAAQBAJ} {\emph {\bibinfo
  {title} {Springer Handbook of Electronic and Photonic Materials}}},\ Springer
  Handbooks\ (\bibinfo  {publisher} {Springer International Publishing},\
  \bibinfo {year} {2017})\BibitemShut {NoStop}%
\bibitem [{\citenamefont {Sherkar}\ and\ \citenamefont
  {Koster}(2015)}]{doi:10.1021/acsami.5b01606}%
  \BibitemOpen
  \bibfield  {author} {\bibinfo {author} {\bibfnamefont {T.~S.}\ \bibnamefont
  {Sherkar}}\ and\ \bibinfo {author} {\bibfnamefont {L.~J.~A.}\ \bibnamefont
  {Koster}},\ }\href {\doibase 10.1021/acsami.5b01606} {\bibfield  {journal}
  {\bibinfo  {journal} {ACS Applied Materials \& Interfaces}\ }\textbf
  {\bibinfo {volume} {7}},\ \bibinfo {pages} {11881} (\bibinfo {year}
  {2015})},\ \bibinfo {note} {pMID: 25989847},\ \Eprint
  {http://arxiv.org/abs/https://doi.org/10.1021/acsami.5b01606}
  {https://doi.org/10.1021/acsami.5b01606} \BibitemShut {NoStop}%
\bibitem [{\citenamefont {Campbell}\ \emph {et~al.}(1962)\citenamefont
  {Campbell}, \citenamefont {Robertson},\ and\ \citenamefont
  {Trotter}}]{Campbell:a03426}%
  \BibitemOpen
  \bibfield  {author} {\bibinfo {author} {\bibfnamefont {R.~B.}\ \bibnamefont
  {Campbell}}, \bibinfo {author} {\bibfnamefont {J.~M.}\ \bibnamefont
  {Robertson}}, \ and\ \bibinfo {author} {\bibfnamefont {J.}~\bibnamefont
  {Trotter}},\ }\href {\doibase 10.1107/S0365110X62000699} {\bibfield
  {journal} {\bibinfo  {journal} {Acta Crystallographica}\ }\textbf {\bibinfo
  {volume} {15}},\ \bibinfo {pages} {289} (\bibinfo {year} {1962})}\BibitemShut
  {NoStop}%
\bibitem [{\citenamefont {Vitusevich}\ and\ \citenamefont
  {Zadorozhnyi}(2017)}]{Vitusevich}%
  \BibitemOpen
  \bibfield  {author} {\bibinfo {author} {\bibfnamefont {S.}~\bibnamefont
  {Vitusevich}}\ and\ \bibinfo {author} {\bibfnamefont {I.}~\bibnamefont
  {Zadorozhnyi}},\ }\href {\doibase 10.1088/1361-6641/aa5cf3} {\bibfield
  {journal} {\bibinfo  {journal} {Semiconductor Science and Technology}\
  }\textbf {\bibinfo {volume} {32}},\ \bibinfo {pages} {043002} (\bibinfo
  {year} {2017})}\BibitemShut {NoStop}%
\bibitem [{\citenamefont {Kogan}(2008)}]{kogan2008}%
  \BibitemOpen
  \bibfield  {author} {\bibinfo {author} {\bibfnamefont {S.}~\bibnamefont
  {Kogan}},\ }\href {https://books.google.com.au/books?id=s5tupGCMzBYC} {\emph
  {\bibinfo {title} {Electronic Noise and Fluctuations in Solids}}}\ (\bibinfo
  {publisher} {Cambridge University Press},\ \bibinfo {year}
  {2008})\BibitemShut {NoStop}%
\bibitem [{\citenamefont {Constantin}\ and\ \citenamefont
  {Yu}(2007)}]{PhysRevLett.99.207001}%
  \BibitemOpen
  \bibfield  {author} {\bibinfo {author} {\bibfnamefont {M.}~\bibnamefont
  {Constantin}}\ and\ \bibinfo {author} {\bibfnamefont {C.~C.}\ \bibnamefont
  {Yu}},\ }\href {\doibase 10.1103/PhysRevLett.99.207001} {\bibfield  {journal}
  {\bibinfo  {journal} {Phys. Rev. Lett.}\ }\textbf {\bibinfo {volume} {99}},\
  \bibinfo {pages} {207001} (\bibinfo {year} {2007})}\BibitemShut {NoStop}%
\bibitem [{\citenamefont {{Kamioka}}\ \emph {et~al.}(2012)\citenamefont
  {{Kamioka}}, \citenamefont {{Imai}}, \citenamefont {{Kamakura}},
  \citenamefont {{Ohmori}}, \citenamefont {{Shiraishi}}, \citenamefont
  {{Niwa}}, \citenamefont {{Yamada}},\ and\ \citenamefont
  {{Watanabe}}}]{Watanabe}%
  \BibitemOpen
  \bibfield  {author} {\bibinfo {author} {\bibfnamefont {T.}~\bibnamefont
  {{Kamioka}}}, \bibinfo {author} {\bibfnamefont {H.}~\bibnamefont {{Imai}}},
  \bibinfo {author} {\bibfnamefont {Y.}~\bibnamefont {{Kamakura}}}, \bibinfo
  {author} {\bibfnamefont {K.}~\bibnamefont {{Ohmori}}}, \bibinfo {author}
  {\bibfnamefont {K.}~\bibnamefont {{Shiraishi}}}, \bibinfo {author}
  {\bibfnamefont {M.}~\bibnamefont {{Niwa}}}, \bibinfo {author} {\bibfnamefont
  {K.}~\bibnamefont {{Yamada}}}, \ and\ \bibinfo {author} {\bibfnamefont
  {T.}~\bibnamefont {{Watanabe}}},\ }in\ \href {\doibase
  10.1109/IEDM.2012.6479058} {\emph {\bibinfo {booktitle} {2012 International
  Electron Devices Meeting}}}\ (\bibinfo {year} {2012})\ pp.\ \bibinfo {pages}
  {17.2.1--17.2.4}\BibitemShut {NoStop}%
\bibitem [{\citenamefont {Holder}(2004)}]{holder2004electrical}%
  \BibitemOpen
  \bibfield  {author} {\bibinfo {author} {\bibfnamefont {D.}~\bibnamefont
  {Holder}},\ }\href {https://books.google.com.au/books?id=cjcRd4m\_jUQC}
  {\emph {\bibinfo {title} {Electrical Impedance Tomography: Methods, History
  and Applications}}},\ Series in Medical Physics and Biomedical Engineering\
  (\bibinfo  {publisher} {CRC Press},\ \bibinfo {year} {2004})\BibitemShut
  {NoStop}%
\bibitem [{\citenamefont {{Loyola}}\ \emph {et~al.}(2013)\citenamefont
  {{Loyola}}, \citenamefont {{Saponara}}, \citenamefont {{Loh}}, \citenamefont
  {{Briggs}}, \citenamefont {{O'Bryan}},\ and\ \citenamefont
  {{Skinner}}}]{tom}%
  \BibitemOpen
  \bibfield  {author} {\bibinfo {author} {\bibfnamefont {B.~R.}\ \bibnamefont
  {{Loyola}}}, \bibinfo {author} {\bibfnamefont {V.~L.}\ \bibnamefont
  {{Saponara}}}, \bibinfo {author} {\bibfnamefont {K.~J.}\ \bibnamefont
  {{Loh}}}, \bibinfo {author} {\bibfnamefont {T.~M.}\ \bibnamefont {{Briggs}}},
  \bibinfo {author} {\bibfnamefont {G.}~\bibnamefont {{O'Bryan}}}, \ and\
  \bibinfo {author} {\bibfnamefont {J.~L.}\ \bibnamefont {{Skinner}}},\ }\href
  {\doibase 10.1109/JSEN.2013.2253456} {\bibfield  {journal} {\bibinfo
  {journal} {IEEE Sensors Journal}\ }\textbf {\bibinfo {volume} {13}},\
  \bibinfo {pages} {2357} (\bibinfo {year} {2013})}\BibitemShut {NoStop}%
\bibitem [{\citenamefont {Lynch}\ \emph {et~al.}(2012)\citenamefont {Lynch},
  \citenamefont {Huo}, \citenamefont {Kotov}, \citenamefont {Wong Shi~Kam},\
  and\ \citenamefont {Loh}}]{lynch}%
  \BibitemOpen
  \bibfield  {author} {\bibinfo {author} {\bibfnamefont {J.~P.}\ \bibnamefont
  {Lynch}}, \bibinfo {author} {\bibfnamefont {T.-C.}\ \bibnamefont {Huo}},
  \bibinfo {author} {\bibfnamefont {N.~A.}\ \bibnamefont {Kotov}}, \bibinfo
  {author} {\bibfnamefont {N.}~\bibnamefont {Wong Shi~Kam}}, \ and\ \bibinfo
  {author} {\bibfnamefont {K.~J.}\ \bibnamefont {Loh}},\ }\href@noop {}
  {\enquote {\bibinfo {title} {Electrical impedance tomography of
  nanoengineered thin films},}\ } (\bibinfo {year} {2012})\BibitemShut
  {NoStop}%
\bibitem [{\citenamefont {Enkovaara}\ \emph {et~al.}(2010)\citenamefont
  {Enkovaara}, \citenamefont {Rostgaard}, \citenamefont {Mortensen},
  \citenamefont {Chen}, \citenamefont {Du{\l}ak}, \citenamefont {Ferrighi},
  \citenamefont {Gavnholt}, \citenamefont {Glinsvad}, \citenamefont {Haikola},
  \citenamefont {Hansen}, \citenamefont {Kristoffersen}, \citenamefont
  {Kuisma}, \citenamefont {Larsen}, \citenamefont {Lehtovaara}, \citenamefont
  {Ljungberg}, \citenamefont {Lopez-Acevedo}, \citenamefont {Moses},
  \citenamefont {Ojanen}, \citenamefont {Olsen}, \citenamefont {Petzold},
  \citenamefont {Romero}, \citenamefont {Stausholm-M{\o}ller}, \citenamefont
  {Strange}, \citenamefont {Tritsaris}, \citenamefont {Vanin}, \citenamefont
  {Walter}, \citenamefont {Hammer}, \citenamefont {Häkkinen}, \citenamefont
  {Madsen}, \citenamefont {Nieminen}, \citenamefont {N{\o}rskov}, \citenamefont
  {Puska}, \citenamefont {Rantala}, \citenamefont {Schi{\o}tz}, \citenamefont
  {Thygesen},\ and\ \citenamefont {Jacobsen}}]{gpaw}%
  \BibitemOpen
  \bibfield  {author} {\bibinfo {author} {\bibfnamefont {J.}~\bibnamefont
  {Enkovaara}}, \bibinfo {author} {\bibfnamefont {C.}~\bibnamefont
  {Rostgaard}}, \bibinfo {author} {\bibfnamefont {J.~J.}\ \bibnamefont
  {Mortensen}}, \bibinfo {author} {\bibfnamefont {J.}~\bibnamefont {Chen}},
  \bibinfo {author} {\bibfnamefont {M.}~\bibnamefont {Du{\l}ak}}, \bibinfo
  {author} {\bibfnamefont {L.}~\bibnamefont {Ferrighi}}, \bibinfo {author}
  {\bibfnamefont {J.}~\bibnamefont {Gavnholt}}, \bibinfo {author}
  {\bibfnamefont {C.}~\bibnamefont {Glinsvad}}, \bibinfo {author}
  {\bibfnamefont {V.}~\bibnamefont {Haikola}}, \bibinfo {author} {\bibfnamefont
  {H.~A.}\ \bibnamefont {Hansen}}, \bibinfo {author} {\bibfnamefont {H.~H.}\
  \bibnamefont {Kristoffersen}}, \bibinfo {author} {\bibfnamefont
  {M.}~\bibnamefont {Kuisma}}, \bibinfo {author} {\bibfnamefont {A.~H.}\
  \bibnamefont {Larsen}}, \bibinfo {author} {\bibfnamefont {L.}~\bibnamefont
  {Lehtovaara}}, \bibinfo {author} {\bibfnamefont {M.}~\bibnamefont
  {Ljungberg}}, \bibinfo {author} {\bibfnamefont {O.}~\bibnamefont
  {Lopez-Acevedo}}, \bibinfo {author} {\bibfnamefont {P.~G.}\ \bibnamefont
  {Moses}}, \bibinfo {author} {\bibfnamefont {J.}~\bibnamefont {Ojanen}},
  \bibinfo {author} {\bibfnamefont {T.}~\bibnamefont {Olsen}}, \bibinfo
  {author} {\bibfnamefont {V.}~\bibnamefont {Petzold}}, \bibinfo {author}
  {\bibfnamefont {N.~A.}\ \bibnamefont {Romero}}, \bibinfo {author}
  {\bibfnamefont {J.}~\bibnamefont {Stausholm-M{\o}ller}}, \bibinfo {author}
  {\bibfnamefont {M.}~\bibnamefont {Strange}}, \bibinfo {author} {\bibfnamefont
  {G.~A.}\ \bibnamefont {Tritsaris}}, \bibinfo {author} {\bibfnamefont
  {M.}~\bibnamefont {Vanin}}, \bibinfo {author} {\bibfnamefont
  {M.}~\bibnamefont {Walter}}, \bibinfo {author} {\bibfnamefont
  {B.}~\bibnamefont {Hammer}}, \bibinfo {author} {\bibfnamefont
  {H.}~\bibnamefont {Häkkinen}}, \bibinfo {author} {\bibfnamefont {G.~K.~H.}\
  \bibnamefont {Madsen}}, \bibinfo {author} {\bibfnamefont {R.~M.}\
  \bibnamefont {Nieminen}}, \bibinfo {author} {\bibfnamefont {J.~K.}\
  \bibnamefont {N{\o}rskov}}, \bibinfo {author} {\bibfnamefont
  {M.}~\bibnamefont {Puska}}, \bibinfo {author} {\bibfnamefont {T.~T.}\
  \bibnamefont {Rantala}}, \bibinfo {author} {\bibfnamefont {J.}~\bibnamefont
  {Schi{\o}tz}}, \bibinfo {author} {\bibfnamefont {K.~S.}\ \bibnamefont
  {Thygesen}}, \ and\ \bibinfo {author} {\bibfnamefont {K.~W.}\ \bibnamefont
  {Jacobsen}},\ }\href {\doibase 10.1088/0953-8984/22/25/253202} {\bibfield
  {journal} {\bibinfo  {journal} {Journal of Physics: Condensed Matter}\
  }\textbf {\bibinfo {volume} {22}},\ \bibinfo {pages} {253202} (\bibinfo
  {year} {2010})}\BibitemShut {NoStop}%
\end{thebibliography}%

\onecolumngrid

\section{Supplementary information}

The content of the Gaussian90 input file needed to reproduce the computational results for molecular electrostatic potentials. The code listing below is for the tetracene anion. Input files for the cation and neutral molecule differ only by the eighths line that contains the charge and spin multiplicity numbers. 

\begin{lstlisting}
%mem=32Gb
%NprocShared=16
%chk=chk_file.chk
# opt=tight uwb97xd/6-31++g(d,p) iop(3/107=0032000000, 3/108=0032000000) scrf=(pcm,read) scf=qc geom=connectivity

tetracene anion optimization

-1 2
 C                 -4.42200000   -0.63200000   11.62800000
 C                 -4.37500000   -0.89400000   10.25400000
 C                 -3.67300000    0.00200000    9.37300000
 H                 -4.00300000    0.30100000   11.99100000
 H                 -3.33700000    0.89800000    9.78400000
 C                  0.46800000    1.30700000    7.61800000
 C                  0.46300000    0.41900000    8.66800000
 C                 -0.17900000    0.81400000    9.87400000
 C                 -0.25100000   -0.06100000   10.97600000
 C                 -0.91600000    0.25400000   12.19700000
 C                 -1.40400000    2.45100000   11.24100000
 C                 -0.76600000    2.11900000   10.04000000
 C                 -0.69100000    2.95100000    8.96900000
 C                 -0.08600000    2.66600000    7.79200000
 H                  0.93100000    0.97100000    6.69400000
 H                  0.89600000   -0.55500000    8.52100000
 H                  0.17900000   -1.06000000   10.86300000
 H                 -1.82800000    3.40800000   11.37800000
 H                 -1.13900000    3.92800000    9.15900000
 H                 -0.04600000    3.34100000    7.05000000
 C                 -5.03100000   -1.46800000   12.51900000
 C                 -4.91200000   -2.09100000    9.76000000
 C                 -3.61300000   -0.25300000    7.99000000
 C                 -0.96700000   -0.59300000   13.28100000
 C                 -1.45500000    1.60400000   12.32500000
 C                 -5.62900000   -2.69100000   12.00200000
 C                 -5.11700000   -1.22000000   13.92800000
 C                 -4.85500000   -2.35100000    8.31400000
 C                 -5.54300000   -2.93900000   10.59300000
 C                 -4.21300000   -1.47000000    7.49100000
 H                 -3.17600000    0.43100000    7.35600000
 C                 -1.60400000   -0.26000000   14.48100000
 H                 -0.54300000   -1.54900000   13.14400000
 C                 -2.12000000    1.91900000   13.54600000
 C                 -6.23800000   -3.52800000   12.89300000
 C                 -5.74800000   -2.06800000   14.76200000
 H                 -4.68100000   -0.29300000   14.28400000
 H                 -5.26800000   -3.26100000    7.98200000
 H                 -5.97900000   -3.86700000   10.23800000
 H                 -4.19100000   -1.70400000    6.49800000
 C                 -2.19200000    1.04400000   14.64800000
 C                 -1.67900000   -1.09300000   15.55300000
 H                 -2.55000000    2.91900000   13.65900000
 H                 -6.65700000   -4.46000000   12.53100000
 C                 -6.28500000   -3.26500000   14.26800000
 C                 -5.80500000   -1.80800000   16.20800000
 C                 -2.83400000    1.44000000   15.85300000
 H                 -1.23200000   -2.07000000   15.36300000
 C                 -2.28500000   -0.80700000   16.73000000
 C                 -6.98700000   -4.16200000   15.14800000
 C                 -6.44700000   -2.68900000   17.03000000
 H                 -5.39200000   -0.89800000   16.54000000
 C                 -2.83900000    0.55100000   16.90400000
 H                 -3.26700000    2.41300000   16.00000000
 H                 -2.32500000   -1.48300000   17.47200000
 H                 -7.32300000   -5.05700000   14.73700000
 C                 -7.04700000   -3.90600000   16.53100000
 H                 -6.46900000   -2.45500000   18.02300000
 H                 -3.30200000    0.88800000   17.82700000
 H                 -7.48400000   -4.59000000   17.16500000

 1 21 2.0 2 1.5 4 1.0
 2 22 1.5 3 1.5
 3 23 1.5 5 1.0
 4
 5
 6 7 2.0 14 1.0 15 1.0
 7 16 1.0 8 1.5
 8 9 1.5 12 1.5
 9 10 1.5 17 1.0
 10 24 2.0 25 1.0
 11 12 1.5 18 1.0 25 2.0
 12 13 2.0
 13 14 2.0 19 1.0
 14 20 1.0
 15
 16
 17
 18
 19
 20
 21 26 1.0 27 1.5
 22 28 1.0 29 2.0
 23 30 1.5 31 1.0
 24 32 1.5 33 1.0
 25 34 1.5
 26 29 1.5 35 2.0
 27 36 2.0 37 1.0
 28 38 1.0 30 2.0
 29 39 1.0
 30 40 1.0
 31
 32 41 1.5 42 2.0
 33
 34 41 1.5 43 1.0
 35 44 1.0 45 1.5
 36 45 1.5 46 1.0
 37
 38
 39
 40
 41 47 1.5
 42 48 1.0 49 2.0
 43
 44
 45 50 1.5
 46 51 2.0 52 1.0
 47 53 2.0 54 1.0
 48
 49 53 1.0 55 1.0
 50 56 1.0 57 1.5
 51 57 1.5 58 1.0
 52
 53 59 1.0
 54
 55
 56
 57 60 1.0
 58
 59
 60

eps=3.8
\end{lstlisting}



\end{document}